\documentclass{aastex62}
\usepackage{amsmath}
\usepackage{graphicx}
\usepackage{array}
\DeclareGraphicsExtensions{.png,.eps} % use png if it exists rather than eps
\newcommand{\rocke}{ROCKE-3D}
\submitjournal{ApJS}

\begin{document}

\title{Climates of Warm Earth-like Planets I: 3-D Model Simulations}

\correspondingauthor{M.J. Way}

\author[0000-0003-3728-0475]{M.J. Way}
\affil{NASA Goddard Institute for Space Studies, 2880 Broadway, New York, NY, 10025, USA}
\affil{Department of Physics and Astronomy, Uppsala University, Uppsala, 75120, Sweden}

\author{Anthony D. Del Genio}
\affil{NASA Goddard Institute for Space Studies, 2880 Broadway, New York, NY, 10025, USA}
\author{Igor Aleinov}
\affil{NASA Goddard Institute for Space Studies, 2880 Broadway, New York, NY, 10025, USA}
\affil{Center for Climate Systems Research, Columbia University, New York, NY 10025, USA}
\author{Thomas L. Clune}
\affil{Global Modeling and Assimilation Office, NASA Goddard Space Flight Center, USA}
\author{Maxwell Kelley}
\affil{NASA Goddard Institute for Space Studies, 2880 Broadway, New York, NY, 10025, USA}
\author{Nancy Y. Kiang}
\affil{NASA Goddard Institute for Space Studies, 2880 Broadway, New York, NY, 10025, USA}

\email{Michael.J.Way@nasa.gov}

\begin{abstract}
We  present  a  large  ensemble  of  simulations  of  an  Earth-like  world
with increasing insolation and rotation rate.  Unlike previous work utilizing
idealized aquaplanet configurations we focus our simulations on modern
Earth-like topography.  The orbital period is the same as modern Earth, but with
zero obliquity and eccentricity.  The atmosphere is 1 bar N$_{2}$-dominated with
CO$_{2}$=400 ppmv and CH$_{4}$=1 ppmv.  The simulations include two types of
oceans; one without ocean heat transport (OHT) between grid cells as has been
commonly used in the exoplanet literature, while the other is a fully coupled
dynamic bathtub type ocean.  The dynamical regime transitions that occur as day
length increases induce climate feedbacks producing cooler temperatures, first
via the reduction of water vapor with increasing rotation period despite
decreasing shortwave cooling by clouds, and then via decreasing water vapor and
increasing shortwave cloud cooling, except at the highest insolations.
Simulations without OHT are more sensitive to insolation changes  for  fast
rotations  while  slower  rotations  are relatively  insensitive  to ocean choice.
OHT  runs  with  faster  rotations tend  to be similar with  gyres transporting heat
poleward making them warmer than those without OHT. For slower rotations OHT is
directed equator-ward and no high latitude gyres are apparent.  Uncertainties in
cloud parameterization preclude a precise determination of habitability but do
not affect robust aspects of exoplanet climate sensitivity.  This is the first
paper in a series that will investigate aspects of habitability in the
simulations presented herein. The datasets from this study are opensource and
publicly available.
\end{abstract}

\section{Introduction}\label{sec:intro}

In recent years studies of the liquid water habitable zone for terrestrial
exoplanets have mostly moved from the realm of 1-D
\citep[e.g.][]{kasting1993,Forget1998,Selsis2007,pierrehumbert2011b,Kopparapu2013,Popp2013,Kopparapu2014,Ramirez2014b}
to fully 3-D coupled Atmophere and Ocean General Circulation Models
\citep[e.g.][]{merlisschneider2010,Edson2011,WolfToon2013,Yang2013,leconte2013a,Shields2014,Yang2014,godolt2015,Kopparapu2016,Popp2016,Popp2017,Noda2017,Checlair2017,Boutle2017,Salameh2018}.
In general this has shown the value that 3-D modeling can bring to fully
characterizing the climate state of a given world, the importance of cloud and
ice albedo feedbacks and how different stellar types effect the habitable zone,
among many other effects. On the other hand, because of computational
limitations the majority of such 3-D studies have neglected the effect of
lateral ocean heat transport (OHT). Recent exceptions include the work of
\cite{Yang2013,cullum2014,huyang2014,Ferreira2014,Way2015,Way2016,Fujii2017,Way2017b,DelGenio2018,Kilic2018}
who utilized a fully coupled ocean, \cite{Charnay2013,Charnay2017} who use a
2-layer ocean \citep{Codron2012} that mimics some aspects of a fully coupled
ocean, and \cite{godolt2015,Edson2012,Kilic2017} who specified fixed lateral
ocean heat transports perhaps first applied by \cite{russell1985} for Earth
climate studies.  One of the more stark representations of the effect that OHT
has on such 3-D simulations was presented in the work of \cite{huyang2014}.
Initial studies of tidally locked aquaplanet simulations around M-dwarfs
demonstrated an ``eye-ball" state where open water would appear in a circular
region at the substellar point, but the rest of the planet would be covered in
ice \citep[e.g.][]{pierrehumbert2011} even for high CO$_2$ concentrations.
However, it is known that sea ice dynamics are intimately tied to OHT and has
had an effect in simulations of snowball Earth studies
\citep[e.g.][]{Yang2012a,Yang2012b}.  \cite{huyang2014} showed the ``eyeball"
state to be fictional and that a ``lobster" state is a better reflection of
reality, but it does require OHT.

To better understand the differences that OHT can have on climate dynamics we
have analyzed a suite of three-dimensional (3D) General Circulation Model (GCM)
simulations of an Earth-like planet on which two parameters, insolation and
length of day, are varied over ranges suitable for surface habitability. The
study includes a suite of simulations that both include and neglect OHT.
Previous studies exploring insolation and rotation have been shown to have a
range of effects on exoplanet climates
\citep{Yang2014,Kopparapu2016,Noda2017,Salameh2018}, but at the same time OHT
has been neglected in these studies.  Our use of \rocke{} \citep{Way2017} with
OHT and its different treatment of clouds, convection, radiative transfer,
dynamical core, and a host of other parameterized physics should help establish
the generality of results where they overlap. However, direct comparison with
previous work is also limited not only by the different treatment of OHT, but
also because most previous work relied upon an aquaplanet setup, whereas we use
an Earth-like land/sea mask.  One may even consider a planet with Earth's
present-day or past land-ocean distribution as a possibly more useful,
demonstrably habitable, template for the rise of life as compared with
aquaplanets. There is also an on-going debate in the community where some
\citep{Abbot2012} argue that it is not possible for an aquaplanet to support a
climate stabilizing carbonate-silicate cycle, and those \citep{Charnay2017} who
argue otherwise. Until the debate is settled we believe it is a good idea to
continue modeling the climates of both types of worlds.

\section{Methods}\label{sec:Methods}
\subsection{Experimental Setup}\label{sec:Experimental_Setup}

All simulations herein utilize \rocke{} \citep{Way2017}. \rocke{} is a
generalized version of the Goddard Institute for Space Studies (GISS) ModelE2
\citep{schmidt2014} General Circulation Model (GCM). \rocke{} has been
developed to allow a larger range of input variables than ModelE2 such as
higher and lower atmospheric temperatures, rotation rates, atmospheric
compositions and pressures. 

We use a baseline version of \rocke{} known as Planet\_1.0 which is described
in detail in \cite{Way2017}. In summary, \rocke{} is a Cartesian gridpoint
model run at 4$\degr$ $\times$ 5$\degr$ latitude $\times$ longitude resolution
with 40 atmospheric layers and a top at 0.1 hPa. For this study the model is
coupled to two different types of ocean with the same resolution as the
atmosphere. The first is a ``Q-flux" ocean \citep{russell1985}, i.e., a
thermodynamic mixed layer whose temperature is determined by radiative and
turbulent energy exchange with the atmosphere.  Lateral ocean heat transport is
set to zero (Q-flux=0). The second ocean is a fully coupled dynamic ocean with
13 vertical layers down to a possible depth of 5000 m \citep{russell1995}.
However, the ocean is limited to a depth of 1360 meters for this study, which
is the bottom of the 10th ocean layer in \rocke{}. This allows the model to
come into radiative equilibrium faster than it would with the full-depth 5000m
ModelE2 default Earth ocean.

The experiments use a baseline model that is similar to that used by ModelE2
for modern Earth climate studies, but differs from that model in several ways.
At model start the following properties were specified: 

\begin{itemize}
\item Planetary obliquity and eccentricity are zero. This avoids the additional
complications of unraveling effects related to seasonal cycles as we see on
Earth and Mars.

\item Continental layout and land topography are roughly that of modern Earth.
Some shallow ocean, seas and lake regions were replaced with land at locations
that tend to freeze to the bottom at lower insolations. For example, land
replaces the Baltic, Mediterranean, Red and Black Seas, and the Hudson Bay. The
higher latitude seas freeze to the bottom at low insolations as a consequence
of the zero obliquity used. Currently \rocke{} is unable to interactively
change ocean gridboxes that freeze to the bottom to land ice.\footnote{The
model will crash when this happens.} The height of the replacement land is set
to the average height of the neighboring land grid cells. As well, some land is
replaced by ocean where it may also cause problems in neighboring shallow
continental shelf areas (e.g. between Australia and Indonesia). See Figure
\ref{fig:bathtub}. A number of the default Earth boundary condition files were
modified for these purposes (see Table \ref{tab:boundarycondfiles}), especially
the river directions file which now contains river runoff direction for every
grid cell. Runoff was also allowed for the Caspian Sea region, which otherwise
does not have a runoff direction in the default Earth river directions file.

\item The dynamic fully coupled ocean is a bathtub type ocean. Continental
shelf regions are set to 591 meters in depth (ocean model level 8 of 13), while
the rest of the ocean is set to 1360 meters (ocean model level 10 of 13). See
Section \ref{sec:Sensitivity_to_Ocean_Model} and Figure~\ref{fig:bathtub}.

\item Land albedo is set to 0.2 everywhere. The value is chosen to be the same
as that used in the few non-aquaplanet simulations in \cite{Yang2014} to enable
some inter-comparison. Those authors chose this value as a rough average albedo
of a clay and sand mix. There is no glacial ice at model start, but snow is
allowed to accumulate on land, which changes the albedo. Greenland and
Antarctica maintain their present day topographies though they have no land-ice
at model start. Note that the default land surface albedo in \rocke{} is
spectrally flat.

\item Lakes are allowed to shrink and expand, but excess run-off is transported
downhill when the water exceeds the pre-defined lake height (also known as
`sill depth').

\item Soil texture was chosen to be comparable to a ``clay and sand" mix as in
\cite{Yang2014} for the majority of simulations herein.  The fractions for each
particle class were not explicitly defined in that study, so we set them to 0.5
each.  However, this corresponds to a soil texture classified as a sandy clay
on the soil texture triangle \citep{Jury1991}, an arbitrary choice with regard
to soil physical properties of porosity (saturated capacity) and water holding
capacity.  The latter is also known as ``available water" after soil has
drained from saturation, excluding water too tightly bound to soil particles to
be available for life.  For plants, that lower threshold of unavailable water
is generally considered to be the ``wilting point", $\sim$ --15 bar matric
potential, though in nature it may be more extreme, with microbes also able to
extract water at lower matric potentials.  At the hygroscopic point water is
bound to soil particles by molecular forces and cannot be evaporated.  To set
lower and upper bounds on the potential availability of soil water (see Paper
III\footnote{\citet{Kiang2018}} for details), we observed which simulations
were very dry and very wet with the 50/50 clay/sand soil, and then we added to
these end points more simulations with all sand soil (61D in Table
\ref{tab:dyn:simulations}) and all silt soil (22D \& 72D in Table
\ref{tab:dyn:simulations}), which have, respectively, the lowest and highest
water holding capacities.  Texture-dependent soil physical properties are
generally only estimated empirically, and representations differ among GCMs.
Those used in \rocke{}'s ground hydrology scheme for our chosen soil textures
are listed in Table~\ref{table:Soil_textures}.

\item No vegetation is included. 

\item Maximum land soil depth is 3.5 m uniformly around the globe, which is a
restriction of the current ground hydrology scheme.  Subsurface run-off is
transported directly to the ocean. Solar radiative heating is allowed to
penetrate to a soil depth of 3.5 meters. This choice is appropriate for the
length of day on modern Earth. However, for the slower rotation periods that we
simulate, a variable penetration depth consistent with the length of the day
would be appropriate to implement in future versions of \rocke{}.

\item Since the canonical habitable zone is the region around a star where an
orbiting planet with an Earth-like atmosphere (i.e. N$_{2}$, CO$_{2}$,
CH$_{4}$, H$_{2}$O) could maintain surface liquid water
\citep[e.g.][]{hart1979,kasting1993} we prescribe present day Earth surface
atmospheric pressure (984 millibars), with N$_{2}$ the major constituent and
prescribed spatially uniform concentrations of the minor radiatively active
constituents CO$_{2}$ = 400 ppmv\footnote{Parts Per Million by Volume.} and
CH$_{4}$ = 1 ppmv and a modern Earth water vapor profile at model start.
O$_{3}$ is set to zero in line with most previous exoplanet studies, with the
exception of \cite{godolt2015} who retain O$_{3}$ in most of their simulations.
Aerosols (i.e., hazes from photochemical, volcanic, or dust- or ocean-stirring
processes) are also not explicitly included, although their presence as nuclei
for cloud formation is implicit.

\item H$_{2}$O is the only other radiatively active constituent besides
CO$_{2}$ and CH$_{4}$. Its concentration varies in space and time in accordance
with the parameterized convection, condensation and evaporation physics and
dynamical transport in the model. The effect of variable H$_{2}$O on
atmospheric mass is neglected. The maximum monthly mean gridbox specific
humidity (mass concentration of H$_{2}$O) in the lowest altitude model layer in
the runs with the warmest surface temperatures were found to be $\lesssim$0.1
in any gridbox. Thus, our neglect of H$_{2}$O mass is a $\sim$10\% or smaller
effect in our simulations.

\end{itemize}

\rocke{} was run first at Earth's present day sidereal rotation period
t$_{rot}$ (X001) = 1 and insolation (S0) = 1360.67 W m$^{-2}$, and then in a
series of experiments at higher insolation and longer sidereal days; see Tables
\ref{tableqflux:rot-sox}, \ref{tabledynocn:rot-sox}).  Simulations were
generally conducted for surface temperatures that remain lower than 400 K. This
is close to the upper temperature limit for accurate calculations with our
radiation scheme SOCRATES \citep{Edwards1996a,Edwards1996b} which uses
HITRAN2012 \citep{HITRAN2013Rothman}. See \cite{Goldblatt2013} for details on
the accuracy of HITRAN2008 versus HITEMP2010, although our use of HITRAN 2012
may imply even smaller differences with the HITEMP database.

The suite of simulations are summarized in Tables \ref{tableqflux:rot-sox} and \ref{tabledynocn:rot-sox}.

\begin{deluxetable}{|l|c|c|c|}
\tablecaption{Soil textures used in experiments \label{table:Soil_textures}}
\tablehead{
\multicolumn{1}{|l|}{Soil Type} & \multicolumn{1}{|c|}{Soil Water Content\tablenotemark{a}} & \multicolumn{1}{|c|}{Hygroscopic water\tablenotemark{b}} & \multicolumn{1}{|c|}{Available Water\tablenotemark{c}}}
\startdata
\hline
50/50\% clay/sand & 1699.062238324 & 377.958232212 & 1322.8538127422\\
100\% sand        & 1378.847624922 & 108.488011098 & 1270.3596138239\\
100\% silt        & 1879.292321277 & 237.973701763 & 1641.3186195135\\
\hline
\enddata
\tablenotetext{a}{\rocke{} saturated water content for 3.5 m deep soil in units of kg m$^{-2}$. The large number of significant digits is required for accuracy when you consider that the ice-free land surface area is $\sim$1.30577 x 10$^{14}$ m$^{2}$.\hfill}
\tablenotetext{b}{\rocke{} hygroscopic water for 3.5 m deep soil in kg m$^{-2}$.}
\tablenotetext{c}{\rocke{} maximum available water soil water for 3.5 m deep soil in kg m$^{-2}$.}
\end{deluxetable}

\begin{deluxetable}{|c|ccccccccc|}
\tablecaption{Q-flux=0 (zero ocean heat transport) simulation summary\tablenotemark{1} \label{tableqflux:rot-sox}}
\tablehead{\multicolumn{1}{|c|}{Insolation\tablenotemark{2}} & \multicolumn{9}{|c|}{Rotation Period(d)\tablenotemark{3}}}
\startdata
\hline
S0X & X001 & X002 & X004 & X008 &X016 &X032 &X064 &X128&X256 \\
\hline
1.0/1360.67 & x,t,c&  x   &  x   &  x   &  x  &  x  &  x  & x  & x\\
1.1/1501.80 & x,t  &  x   &  x   &  x   &  x,c&  x  &  x  & x  & x\\
1.2/1638.40 & x,t  &  x   &  x   &  x   &  x  &  x  &  x,c& x  & x\\
1.3/1774.90 & n,t  &  x   &  x   &  x   &  x  &  x  &  x  & x,c& x\\
1.4/1911.40 &      &  x   &  n   &  x   &  x  &  x  &  x  & x  & x\\
1.5/2047.90 &      &      &      &      &  n  &  x  &  x  & x  & x\\
1.7/2321.00 &      &      &      &      &     &  x  &  x  & x  & x\\
1.9/2594.00 &      &      &      &      &     &  x  &  x  & x  & x\\
2.1/2867.10 &      &      &      &      &     &  n  &  x  & x  & x\\
2.3/3140.00 &      &      &      &      &     &     &  x  & x  & x\\
2.5/3413.25 &      &      &      &      &     &     &  x  & x  & x\\
2.7/3686.30 &      &      &      &      &     &     &  x  & x  & n\\
2.9/3959.37 &      &      &      &      &     &     &  x  & n  &  \\
3.1/4232.40 &      &      &      &      &     &     &  n  & n  &  \\
\hline
\enddata
\tablenotetext{1}{x: Run completed and is in net radiative and hydrological
balance. n: Ran for several model years, but crashed before reaching net
radiative balance.  c: Run with all convective cloud condensate converted to
precipitation.  t: Run with different cloud tuning parameters.
See Section \ref{cloudparam} for details on `c' and `t.'}
\tablenotetext{2}{Present day Earth solar insolation (S0X=1.0) is set to
1360.67 W m$^{-2}$. Subsequent S0X use S0=1365.3 W m$^{-2}$ multiplied by X.
For example, S0X=1.1 $\times$ 1365.3 = 1501.8}
\tablenotetext{3}{Rotation: values are in multiples of Earth length sidereal
days (d). Sidereal=Solar Day equivalents are 1=1, 16=16.7, 32=34.97, 64=76.6, 128=191, 256=848.}
\end{deluxetable}

\begin{deluxetable}{|c|ccccccccc|}
\tablecaption{Fully coupled dynamic ocean simulation summary\tablenotemark{1} \label{tabledynocn:rot-sox}}
\tablehead{\multicolumn{1}{|c|}{Insolation} & \multicolumn{9}{|c|}{Rotation Period(d)}}
\startdata
\hline
S0X & X001 & X002 & X004 & X008 &X016 &X032 &X064 &X128&X256 \\
\hline
1.0 &  x,c &  x   &  x   &  x   &  x  &  x  &  x  &  x & x \\
1.1 &   x  &  x   &  x   &  x   & x,c &  x  &  x  &  x & x \\
1.2 &   x  &  x   &  x   &  x   &  x  &  x  & x,c &  x & x \\
1.3 &   n  &  n   &  x   &  x   &  x  &  x  &  x  & x,c& x \\
1.4 &      &      &      &  x   &  x  &  x  &  x  &  x & x \\
1.5 &      &      &      &  n   &  n  &  x  &  x  &  x & x \\
1.7 &      &      &      &      &     &  x  &  x  &  x & x \\
1.9 &      &      &      &      &     &  x  &  x  &  x & x \\
2.1 &      &      &      &      &     &  x  &  x  &  x & x \\
2.3 &      &      &      &      &     &     &  x  &  x & x \\
2.5 &      &      &      &      &     &     &  x  &  x & x \\
2.7 &      &      &      &      &     &     &  x  &  x & n \\
2.9 &      &      &      &      &     &     &  x  &  n & n \\
3.1 &      &      &      &      &     &     &  n  &  n & n \\
\hline
\enddata
\tablenotetext{1}{See Table \ref{tableqflux:rot-sox} caption for column and row descriptions.}
\end{deluxetable}

\section{Results}
\subsection{Sensitivity to Ocean Model}\label{sec:Sensitivity_to_Ocean_Model}

Most previous exoplanet GCM studies have utilized a static thermodynamic ocean
with no ocean heat transport, because of its computational efficiency
\citep[e.g.][]{Yang2014,merlisschneider2010,Shields2013,WolfToon2015,
Turbet2016}.  In this section we examine the effect that a dynamic fully
coupled ocean with ocean heat transport has on planetary climate and
impressions of liquid water habitable zone extent.

% Figure 1
\begin{figure}[!htb]
\includegraphics[scale=0.35]{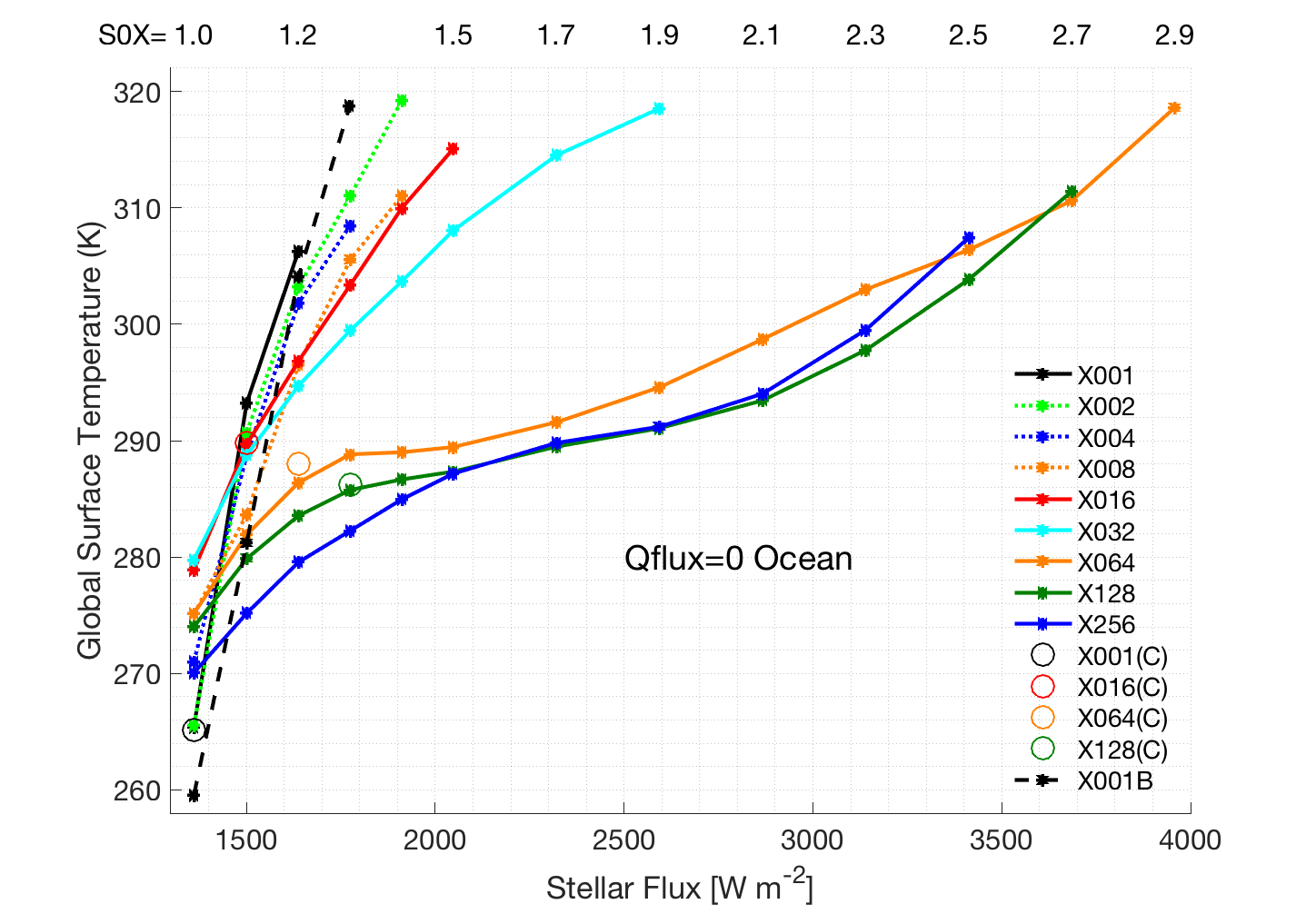}
\includegraphics[scale=0.35]{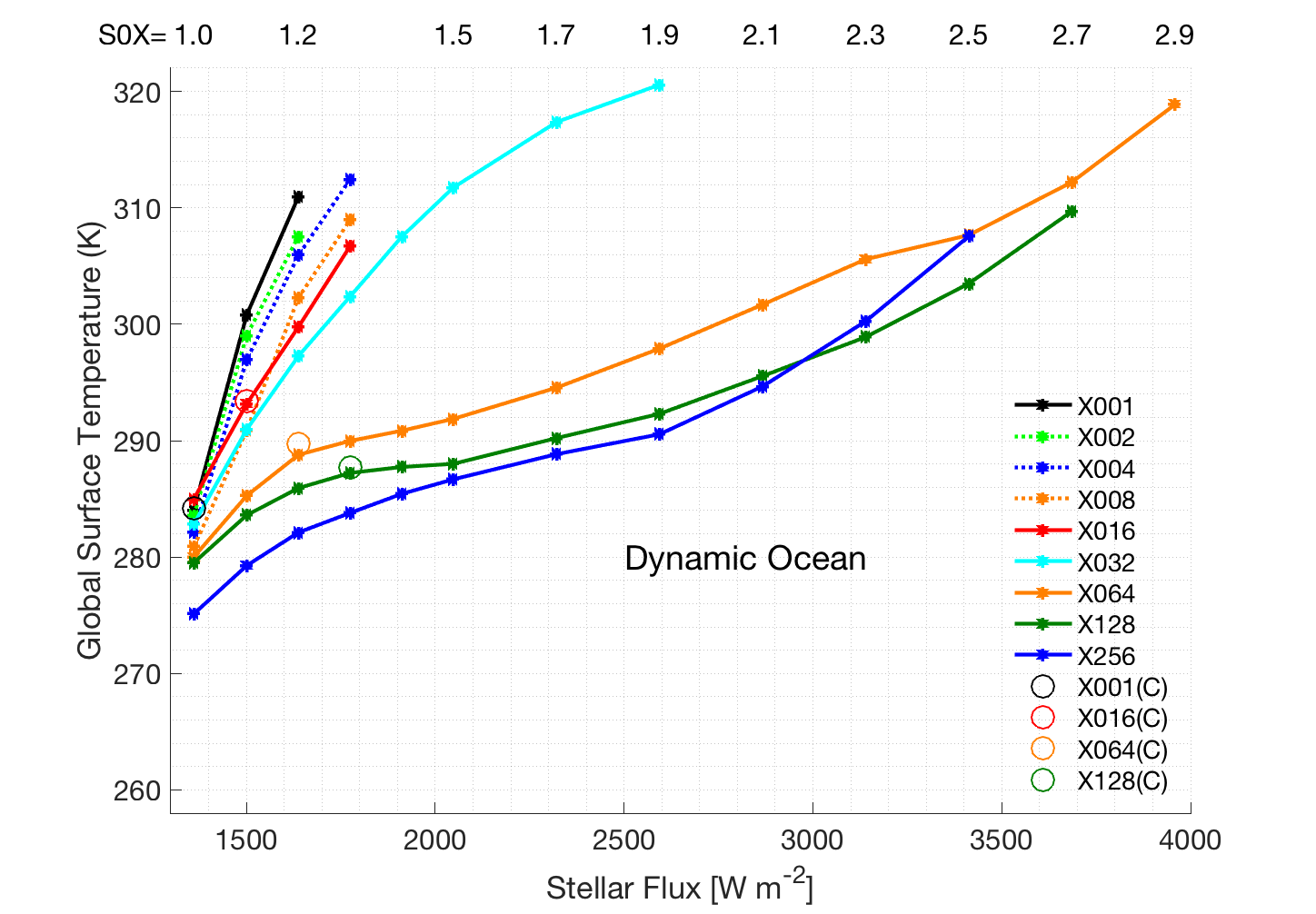}
\caption{\small Global mean surface temperature for \rocke{} as a function of
insolation at different rotation periods for thermodynamic ocean simulations
(a, left) and dynamic fully coupled ocean (b, right). Solid curves and dots
show results from the baseline model at 10\% increments above the  modern Earth
insolation value and 20\% increments above S0X=1.5. Open circles are for
sensitivity tests with no vertical transport of convective condensate. The
dashed curve is for an alternate model version with the same physics but using
different settings of free parameters to adjust the model to global radiative
balance.} \label{fig:s0xtsurf}
\end{figure}

% Figure 2
\begin{figure}[!htb]
\includegraphics[scale=0.6]{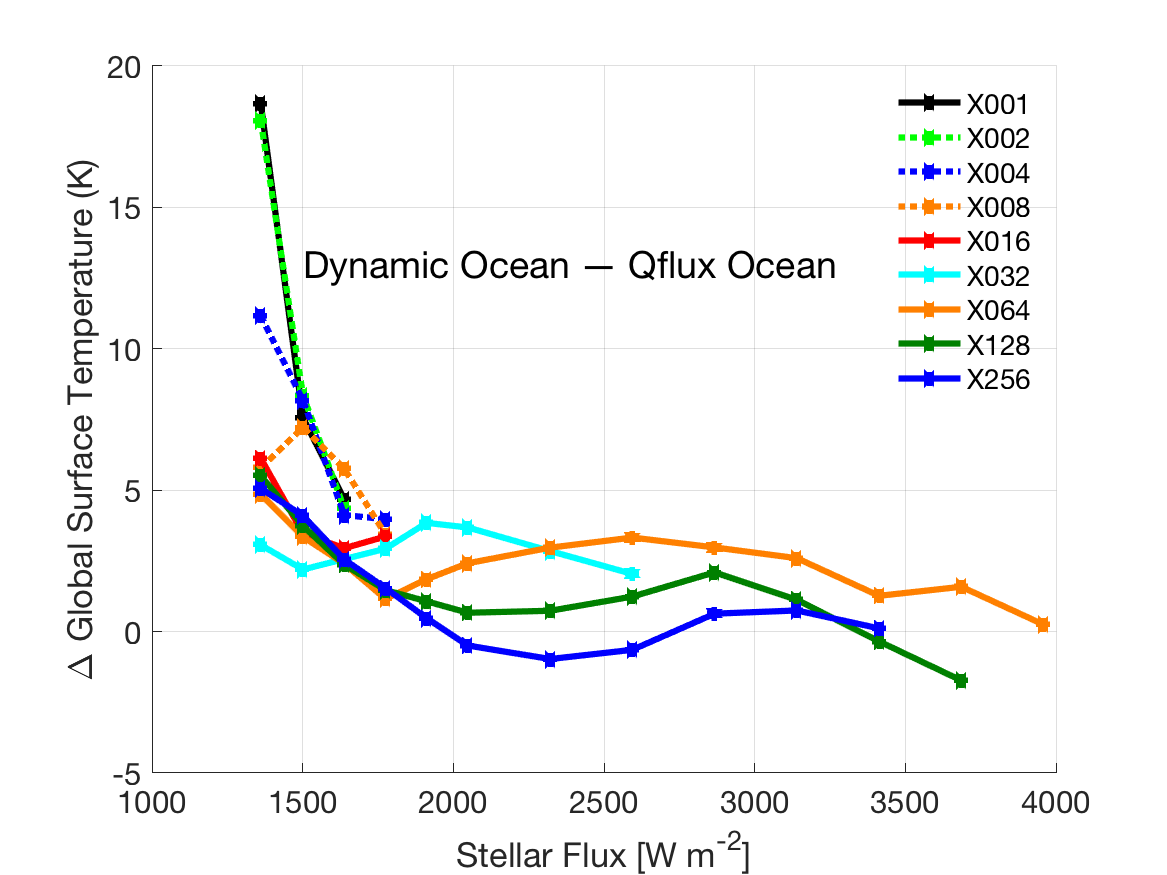}
\caption{\small Global mean surface temperature differences when subtracting
the Q-flux=0 mean temperature from that of the dynamic ocean.}
\label{fig:s0xdiff}
\end{figure}

% Figure 3
\begin{figure}[!htb]
\includegraphics[scale=0.35]{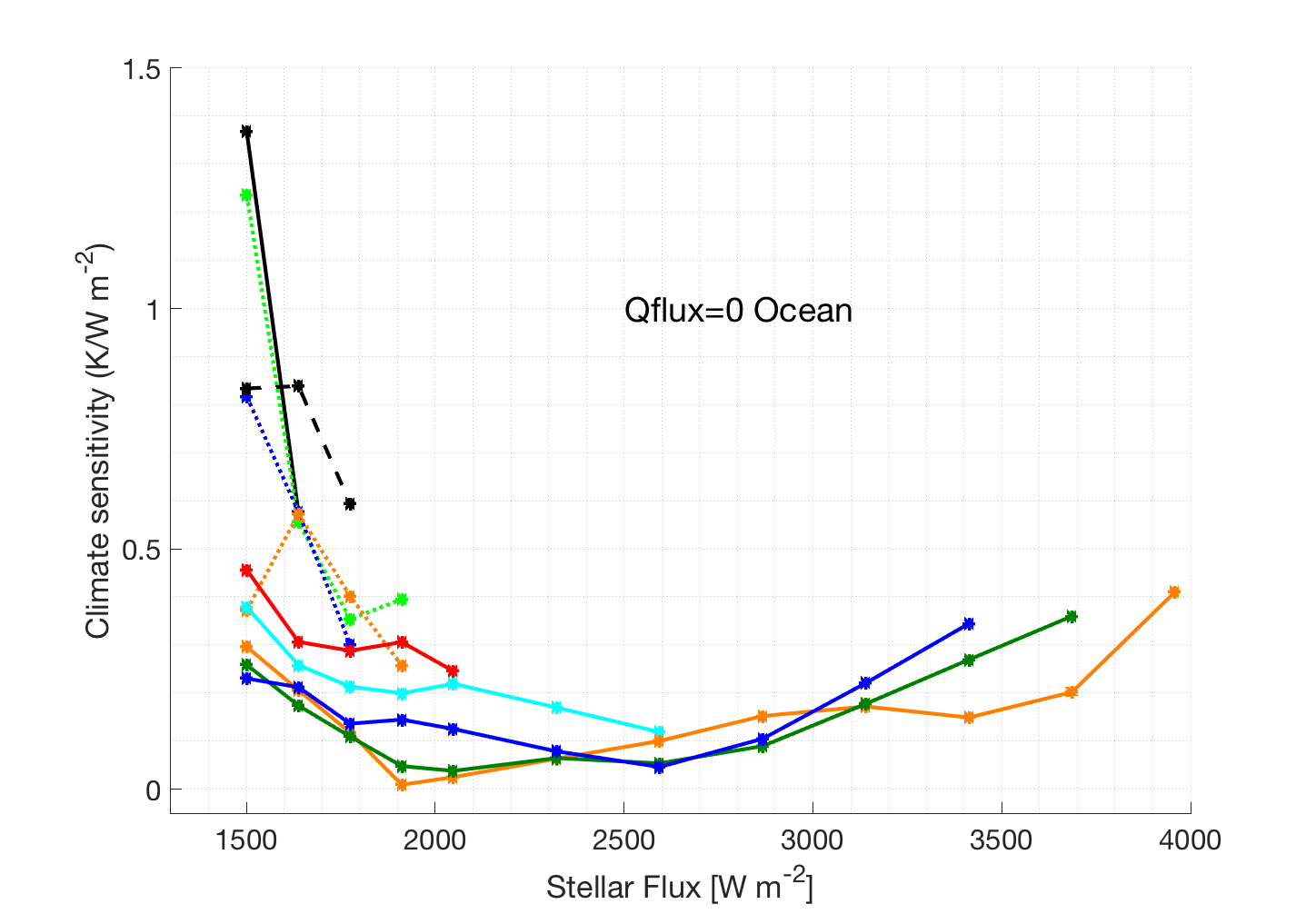}
\includegraphics[scale=0.35]{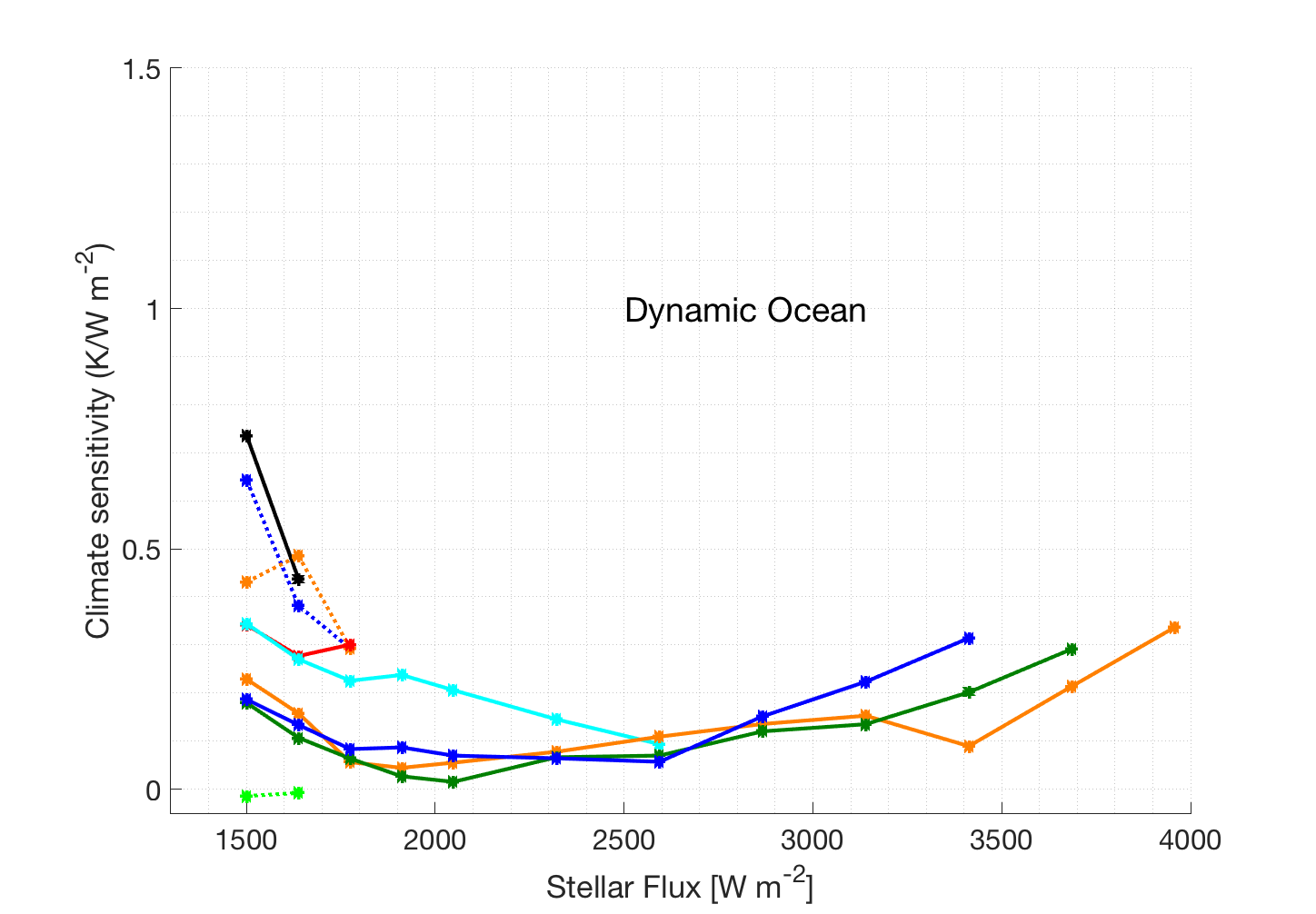}
\caption{\small Climate Sensitivity for Q-flux=0 (a, left) and Dynamic Ocean runs (b, right).}
\label{fig:s0xclimate}
\end{figure}

Figure \ref{fig:s0xtsurf}a shows the global mean surface temperature of
\rocke{} as a function of insolation for sidereal day lengths of 1 Earth day to
256 Earth days for a 100 meter deep thermodynamic Q-flux=0 ocean.  Our results
are qualitatively similar to those of \cite{Yang2014} who use the NCAR
CAM3\footnote{National Center for Atmospheric Research Community Atmospheric
Model version 3} in an aquaplanet configuration (their Figure 1). Recall again
that we use a modern Earth-like topography, not an aquaplanet. Mean surface
temperature increases sharply with insolation at present day Earth's day length
(X001) but is less sensitive as the day length increases. There are several
quantitative differences between \rocke{} and the CAM models used by
\cite{Yang2014} as far as one can compare our Earth-like land/sea mask to their
aquaplanet configuration. For the Q-flux=0 runs \rocke{} is more sensitive to
increasing insolation at present day Earth's day length (X001) and at 16
sidereal days (X016). The trend reverses for day lengths of 64 d (X064) and
larger.  Consequently, the \rocke{} simulations divide clearly into two
classes: moderate to rapidly rotating planets with high sensitivity to
insolation, and slowly rotating planets with low sensitivity, as opposed to the
CAM simulations (Fig. 1 of \cite{Yang2014}) for which sensitivity is a more
continuously decreasing function of rotation period.  We return to this
question below. 

Our results for Earth's rotation period (X001) can also be compared with those
in Figure 1 of \cite{WolfToon2015} who plot climate sensitivity for their CAM3
\& CAM4 results, but also for the LMDG\footnote{Laboratoire de M\'et\'eorologie
Dynamique Generic} results of \cite{leconte2013b}.  \rocke{} is much more
sensitive than the CAM3 and CAM4 results to a 10\% increase in insolation
(S0X=1.1), but much less sensitive than the LMDG GCM, which has a large warming
for insolation increases of 5-10\%. For a 20\% increase \citep[S0X=1.2, the
only other simulation comparable in][]{WolfToon2015}
\rocke{} is less sensitive than CAM4 because CAM4 warms dramatically at 11-12\%
insolation increase while \rocke{} does not. 

Figure \ref{fig:s0xtsurf}b is identical to Figure \ref{fig:s0xtsurf}a but for a
dynamic fully coupled ocean. The results for day lengths of 64 d (X064) and
longer are very similar to those for the thermodynamic ocean case. However, for
the X001 day lengths the dynamic ocean model is significantly warmer (see
Figure \ref{fig:s0xdiff}), and considerably less sensitive to insolation
increases.

The climate sensitivity to changes in sunlight is usually expressed as the
response of surface temperature to the absorbed sunlight rather than the
insolation itself (see, e.g. Eqn. 2 of \citealt{WolfToon2015}). This quantity
is shown in Figure \ref{fig:s0xclimate} for both types of ocean.  Climate
sensitivity is highest for the fastest rotators and most weakly irradiated
planets and is very small for the slow rotators regardless of absorbed
sunlight.  The dynamic ocean simulations have a markedly lower climate
sensitivity than the Q-flux=0 simulations for fast rotation periods.

To explain these differences among models, Figures \ref{fig:s0xAlbOice} \&
\ref{fig:s0xClouds} shows the extent to which the two different ocean types
affect aspects of the planetary (Bond) albedo, ocean ice fractional extent, and
cloudiness at different levels of the model. The upper panels of Figure
\ref{fig:s0xAlbOice} show that at the lower insolation values, the rapidly and
slowly rotating planets behave in opposite fashion: planetary albedo initially
decreases with insolation in the shorter day length simulations (more so for
the Q-flux=0 ocean) but increases with insolation for longer day lengths.  At
higher insolations, all simulations exhibit increasing planetary albedo as
insolation increases. This is partly explained by the sensitivity of the ocean
ice fraction to insolation, rotation, and ocean dynamics. The zero obliquity of
this Earth-like world allows ocean ice to grow considerably at higher latitudes
when the insolation is low, but the ice melts (and thus the surface albedo
decreases) when either insolation increases or the length of day increases. The
latter dependence occurs because heat is more efficiently transported poleward
on slowly rotating planets with broader Hadley cells \citep{delgenio1987}. The
higher ocean ice fractions for the Q-flux=0 versus dynamic ocean at lower
insolations clearly points to the role that a dynamic ocean plays in equator to
pole heat transport.

% Figure 4
\begin{figure}[!htb]
\includegraphics[scale=0.23]{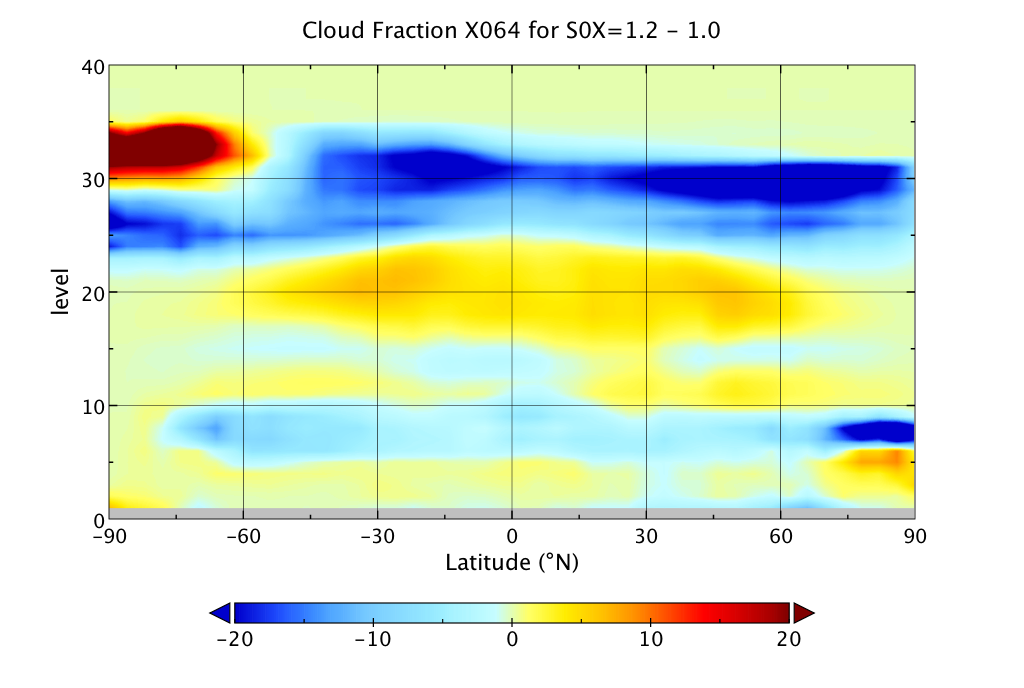}
\includegraphics[scale=0.23]{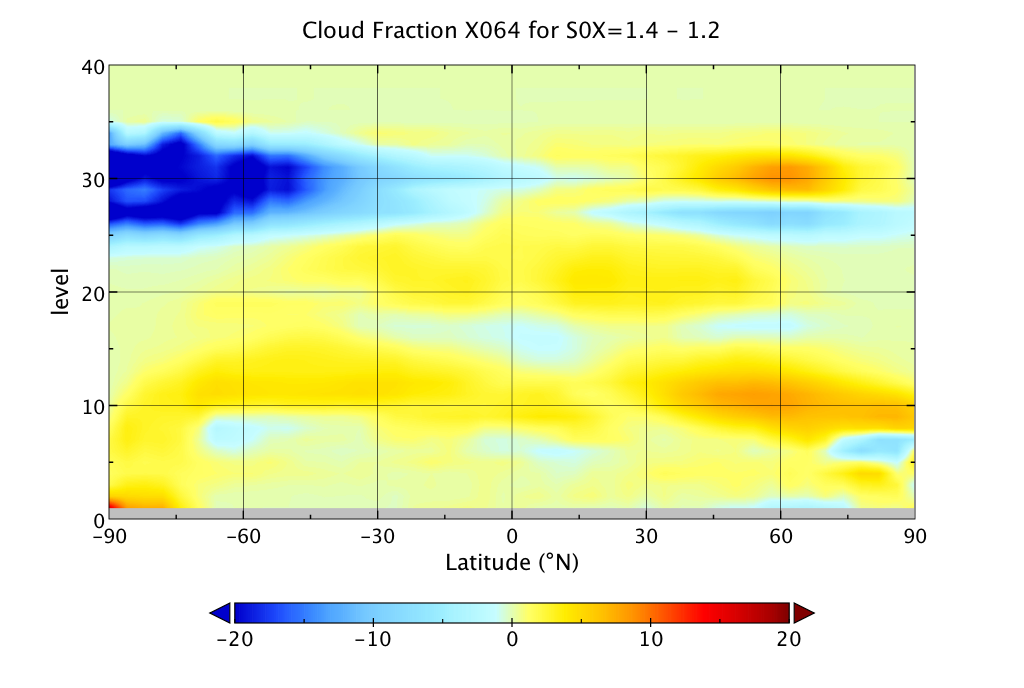}\\
\includegraphics[scale=0.23]{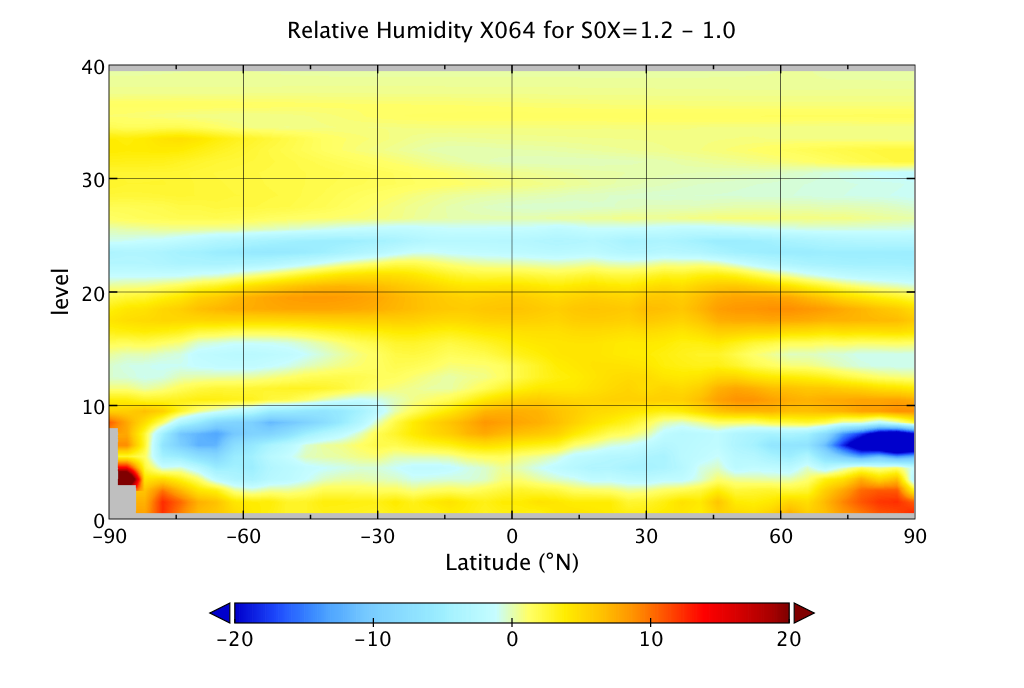}
\includegraphics[scale=0.23]{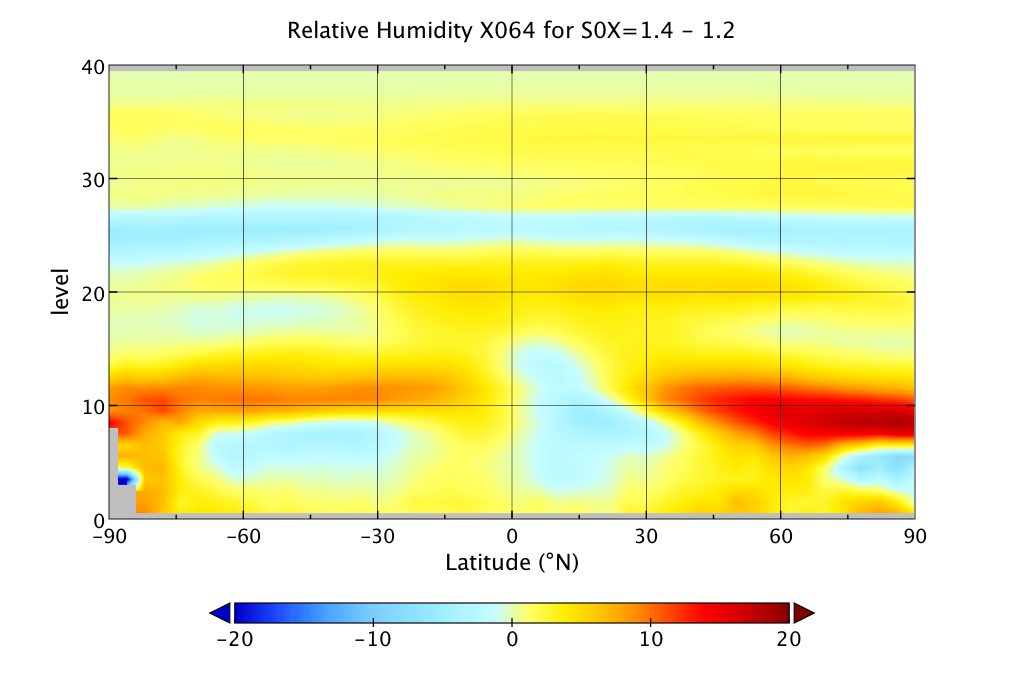}
\caption{\small Total cloud fraction (top) and relative humidity (bottom) for
the difference between two low insolation runs (left) and higher insolation
runs (right). These are all from a single rotation period of X064. X-axes are
all in degrees of latitude, while the y-axes are all in units of atmospheric
levels (1-40).} \label{fig:cfrh}
\end{figure}

The increase in planetary albedo with insolation for the longer day lengths is
due to the now well-documented substellar cloud bank as shown in previous 3D
model simulations \cite[e.g.][]{Yang2014}. This occurs because moist convection
at low latitudes increasingly transports water vapor and cloud particles upward
as insolation increases. The convection itself occupies a small area but
produces more extensive anvil clouds at the levels where water vapor and cloud
particles detrain into the environment \citep{Fu1990}. On Earth, the anvil
clouds are primarily a feature of the upper troposphere. This is also true in
the current experiments at low insolation. Interestingly, though, as insolation
increases convection depth seems to decrease, because high cloud cover
decreases as middle level cloud cover increases in Figure \ref{fig:s0xClouds}.
The latter clouds apparently dominate the optical thickness and thus the
dependence of planetary albedo on insolation. Low level clouds decrease with
increasing insolation at low insolation values, similar to how 3D Earth climate
models respond to 21st Century increases in anthropogenic greenhouse gas
concentrations \citep{Klein2017}. However, at higher insolation low cloud cover
tends to increase with insolation instead. For example, in Figure
\ref{fig:cfrh} the decrease in low clouds from insolation values of 1.0 to 1.2
and the increase from 1.2 to 1.4, are both related to relative humidity, mostly
at high latitudes.  This make sense because there is less shallow convection at
high latitudes.

% Figure 5
\begin{figure}[!htb]
\includegraphics[scale=0.35]{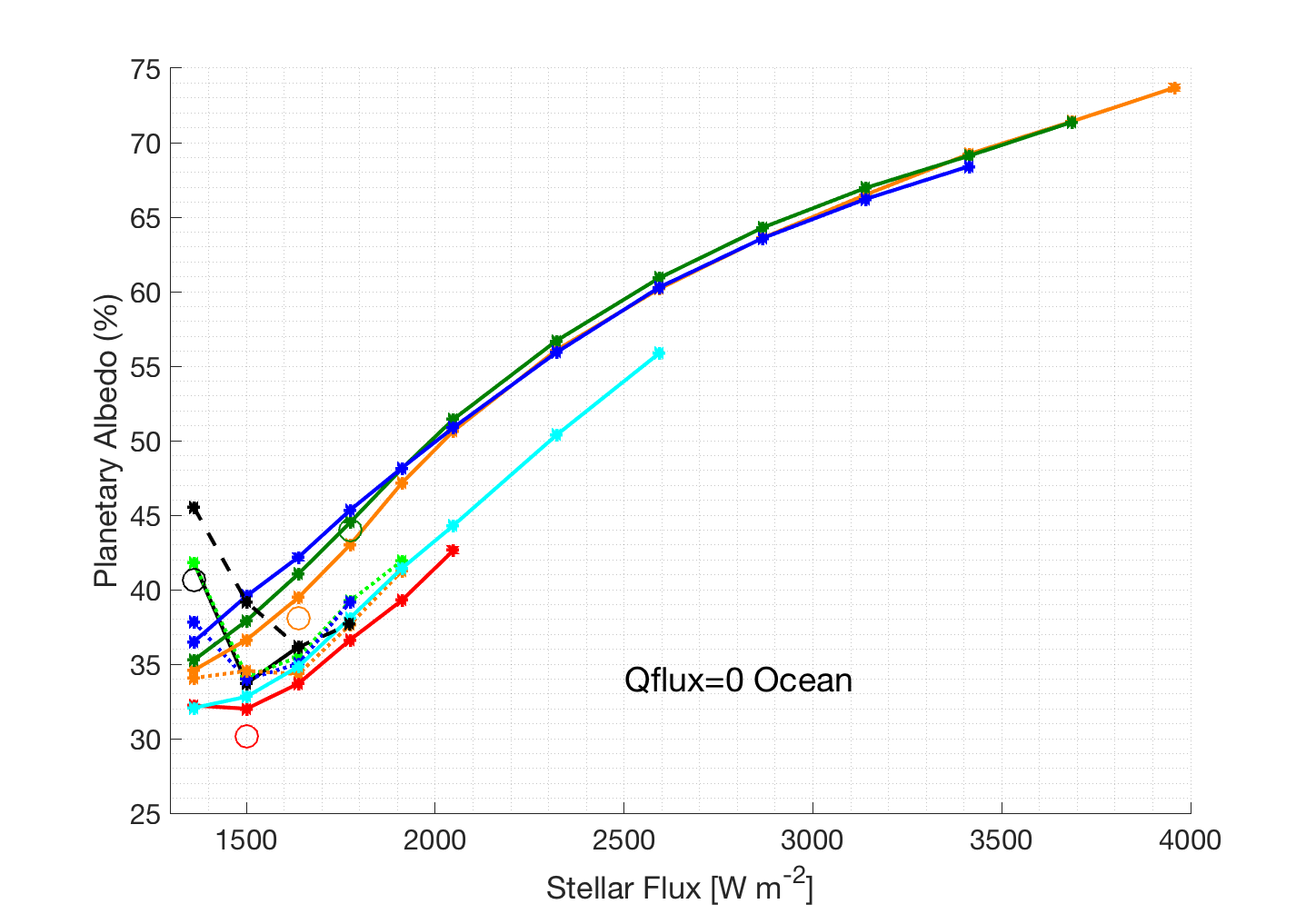}
\includegraphics[scale=0.35]{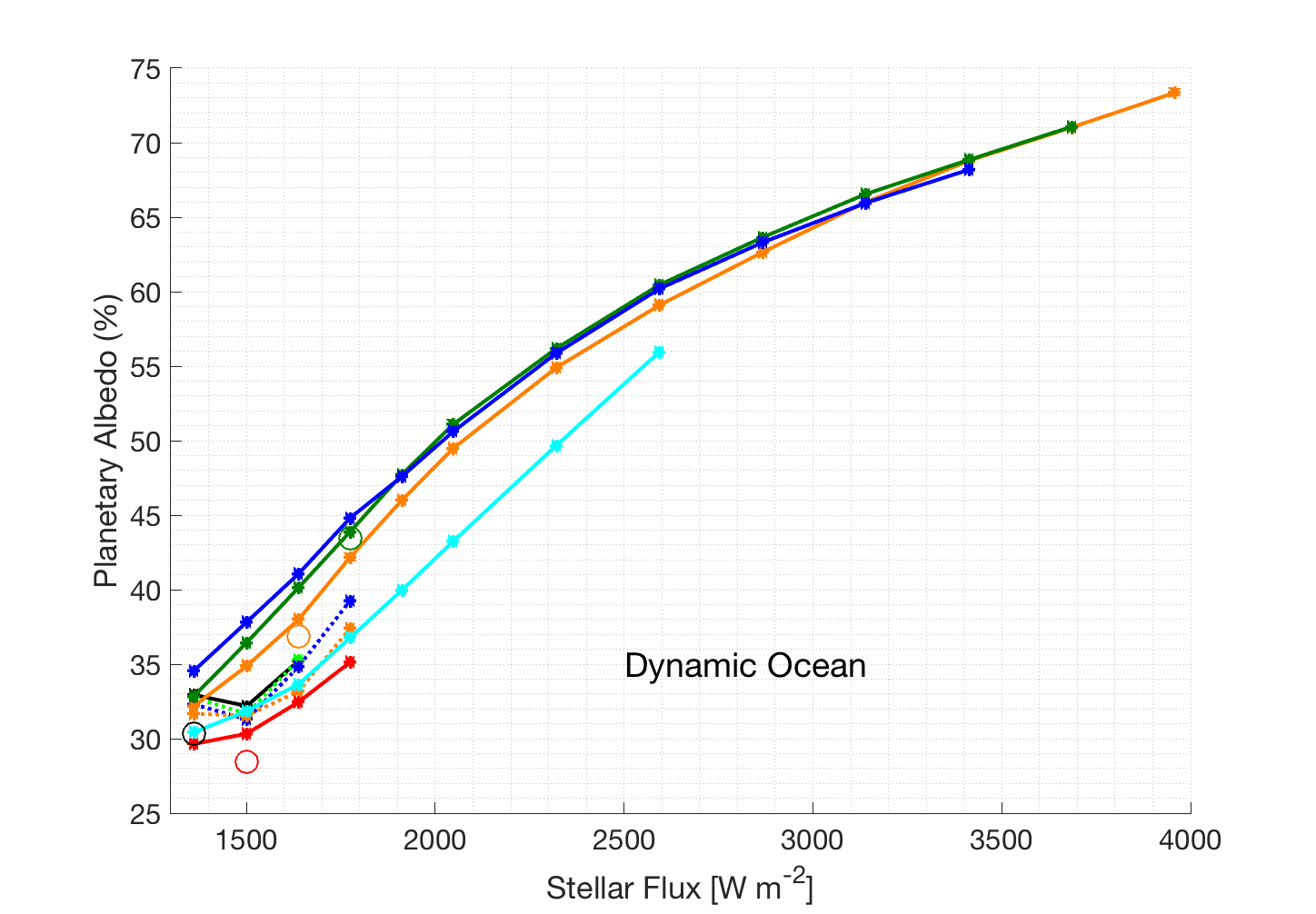}\\
\includegraphics[scale=0.35]{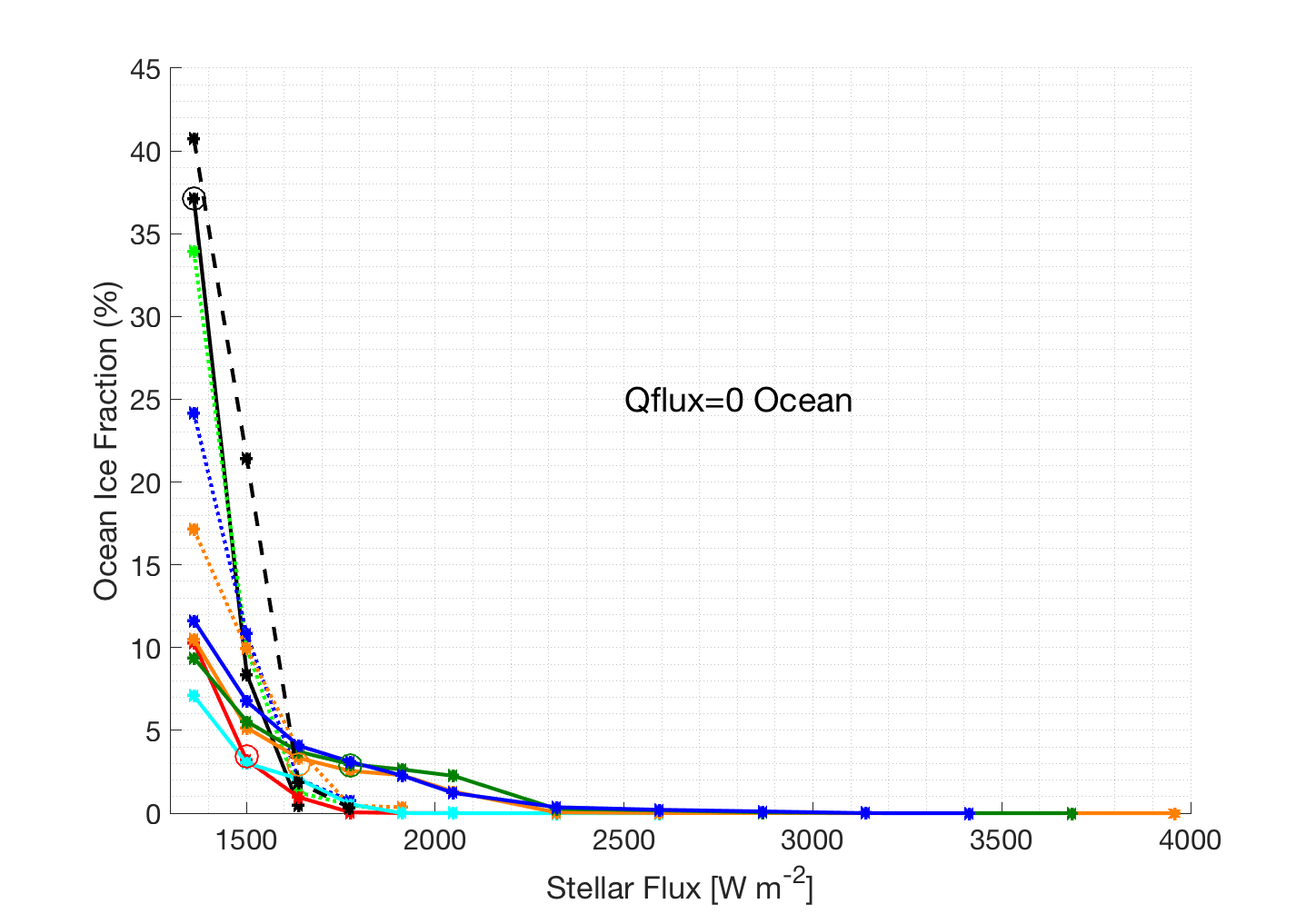}
\includegraphics[scale=0.35]{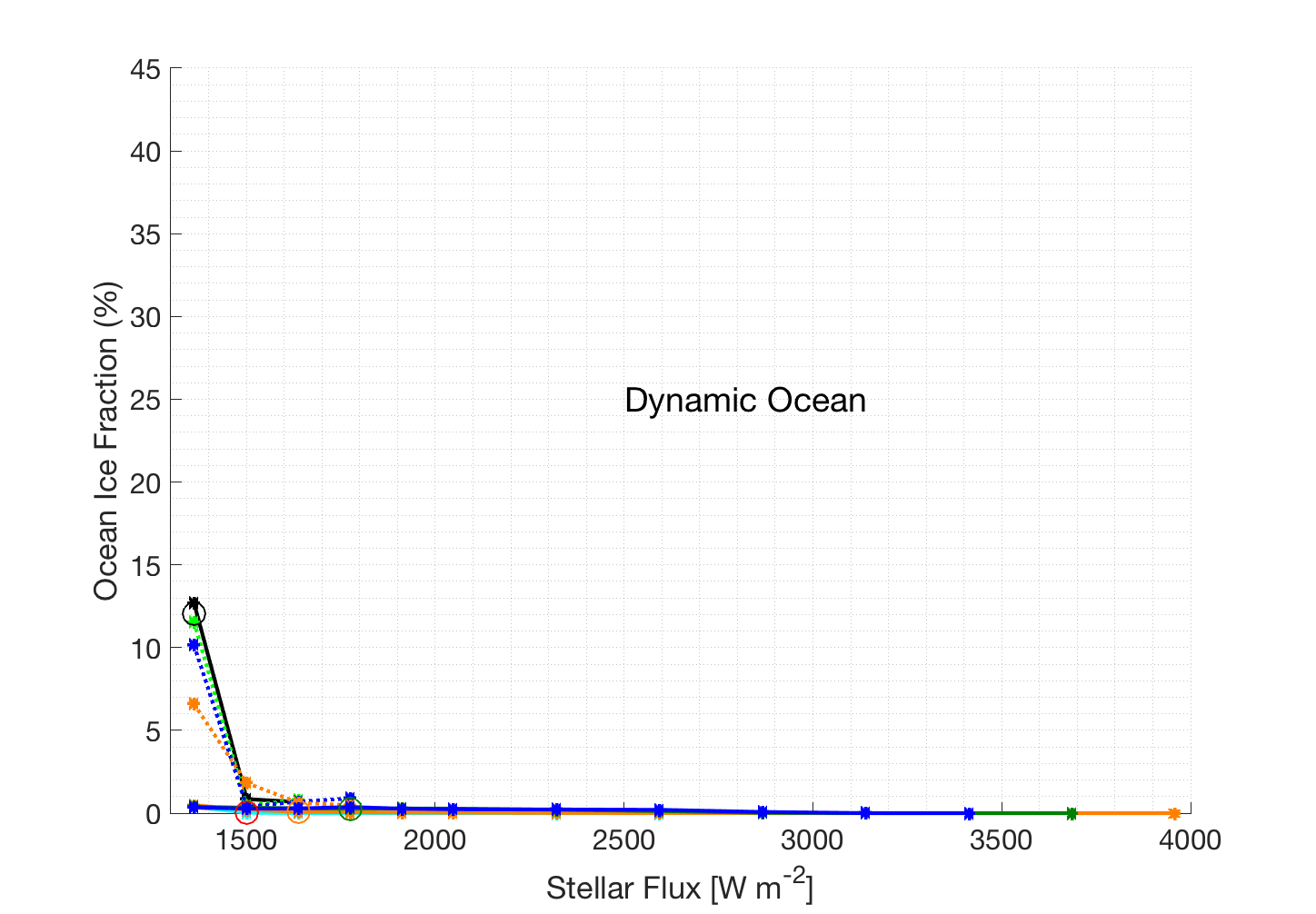}
\caption{\small Planetary Albedo (top) and Ocean Ice Fraction (bottom) for
Q-flux=0 (left) and Dynamic Ocean runs (right).} \label{fig:s0xAlbOice}
\end{figure}

% Figure 6
\begin{figure}[!htb]
\includegraphics[scale=0.35]{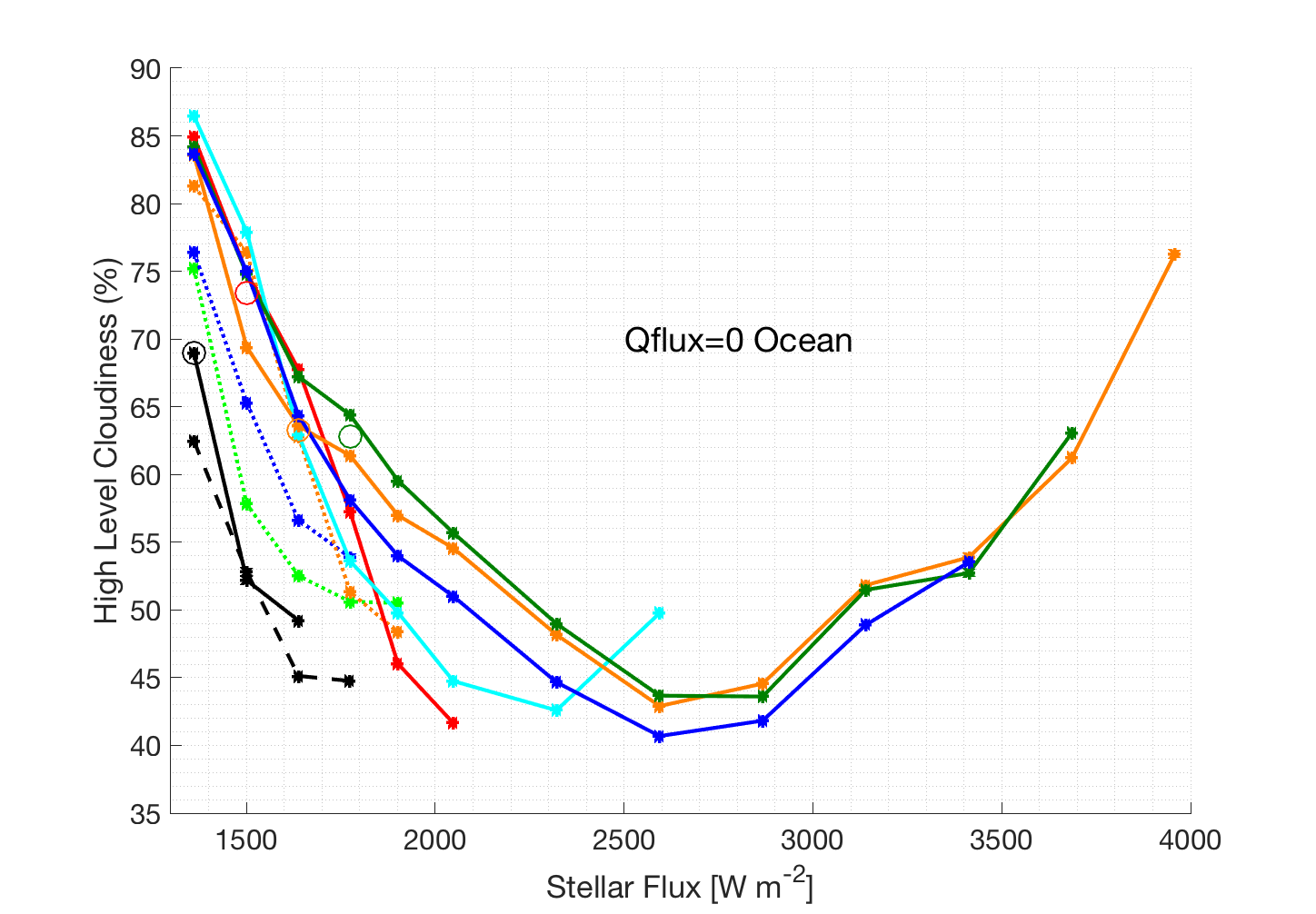}
\includegraphics[scale=0.35]{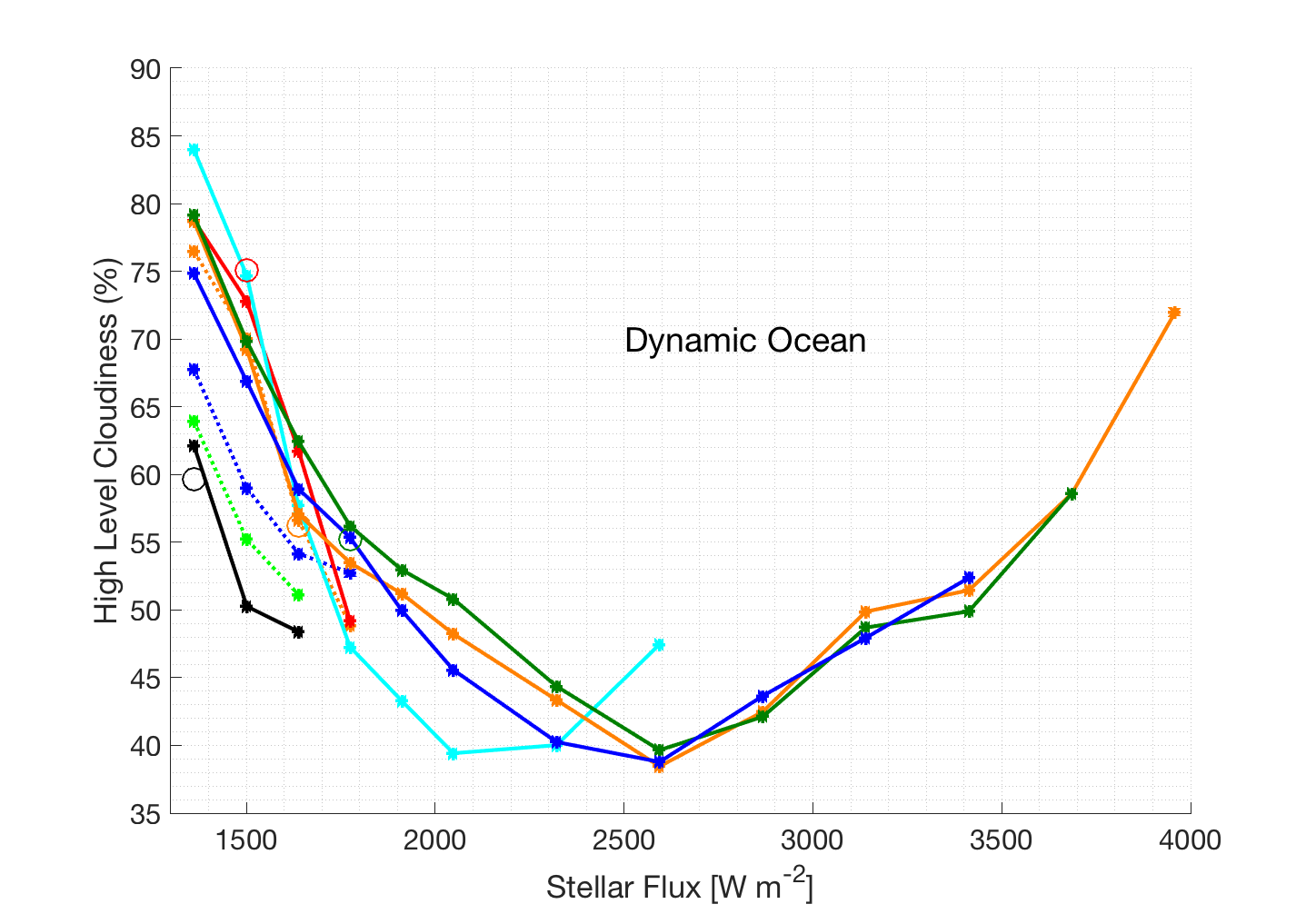}\\
\includegraphics[scale=0.35]{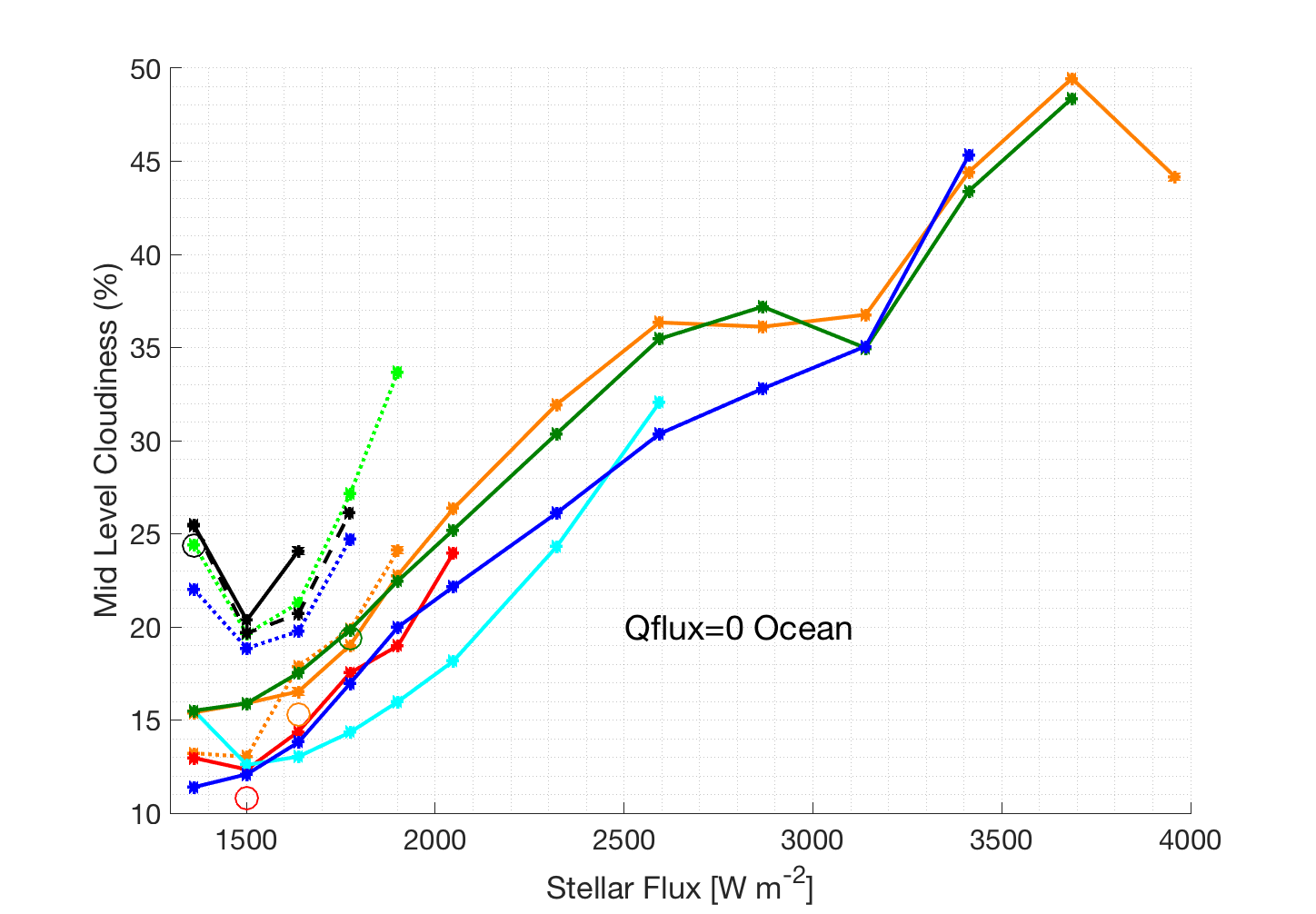}
\includegraphics[scale=0.35]{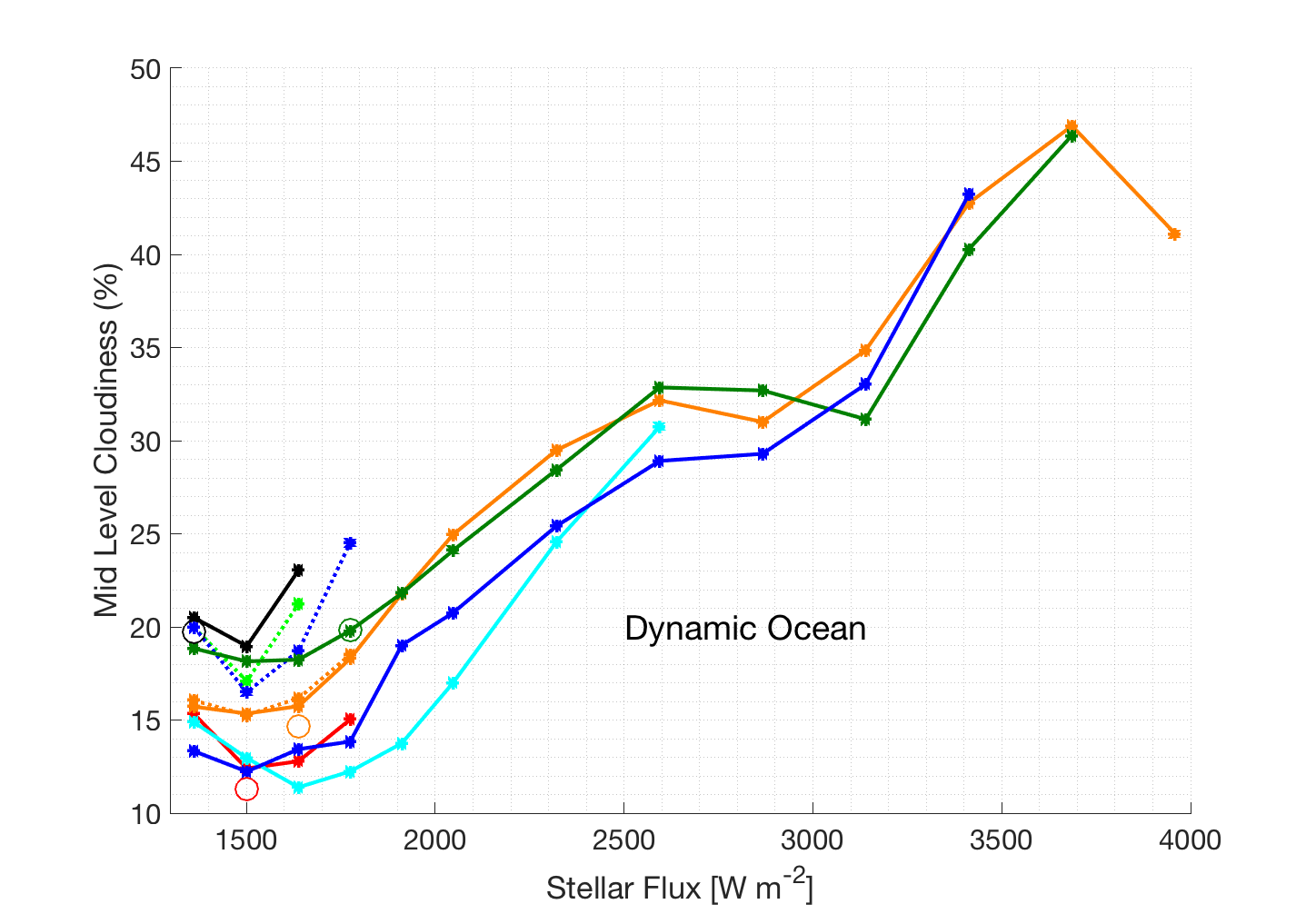}\\
\includegraphics[scale=0.35]{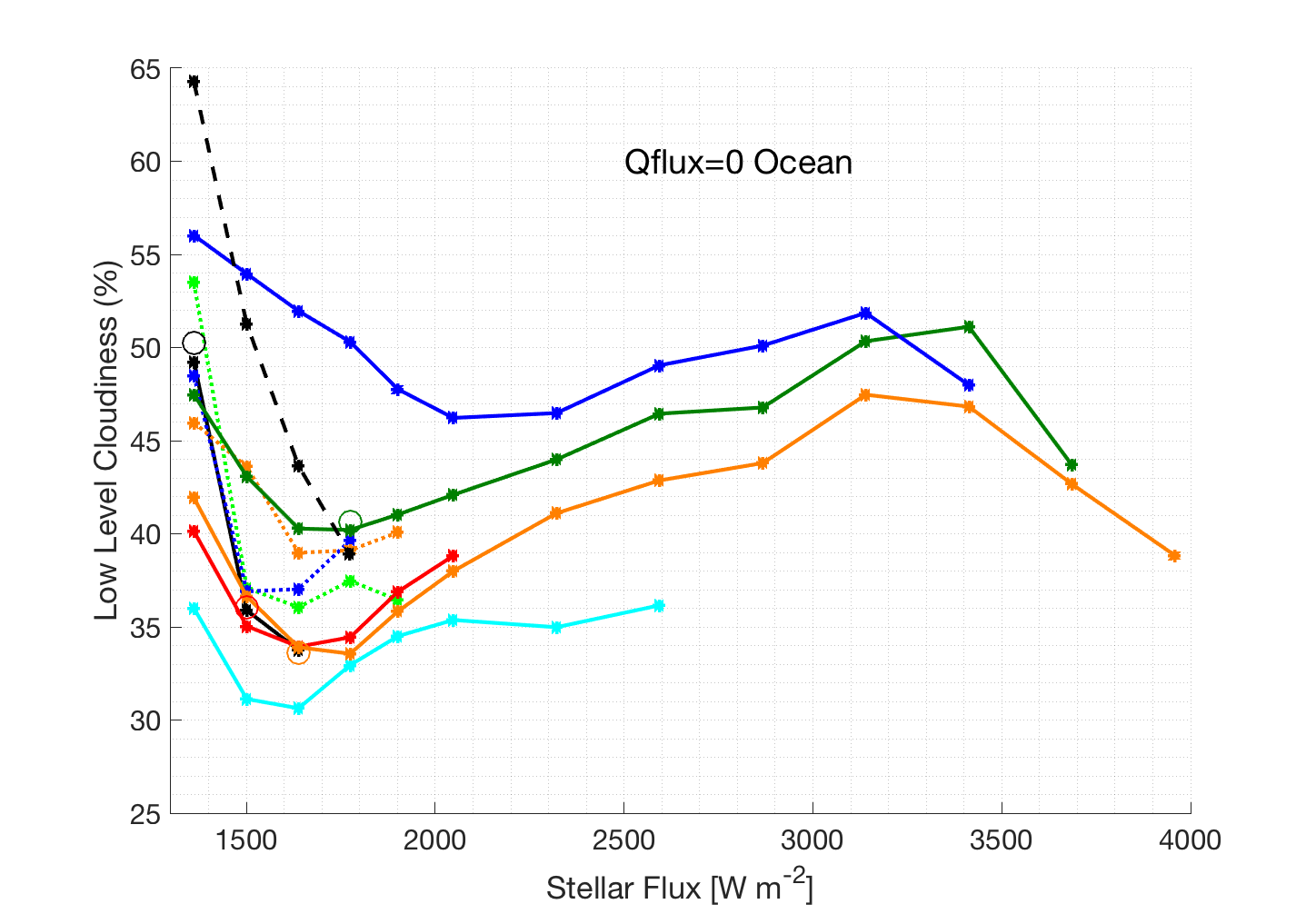}
\includegraphics[scale=0.35]{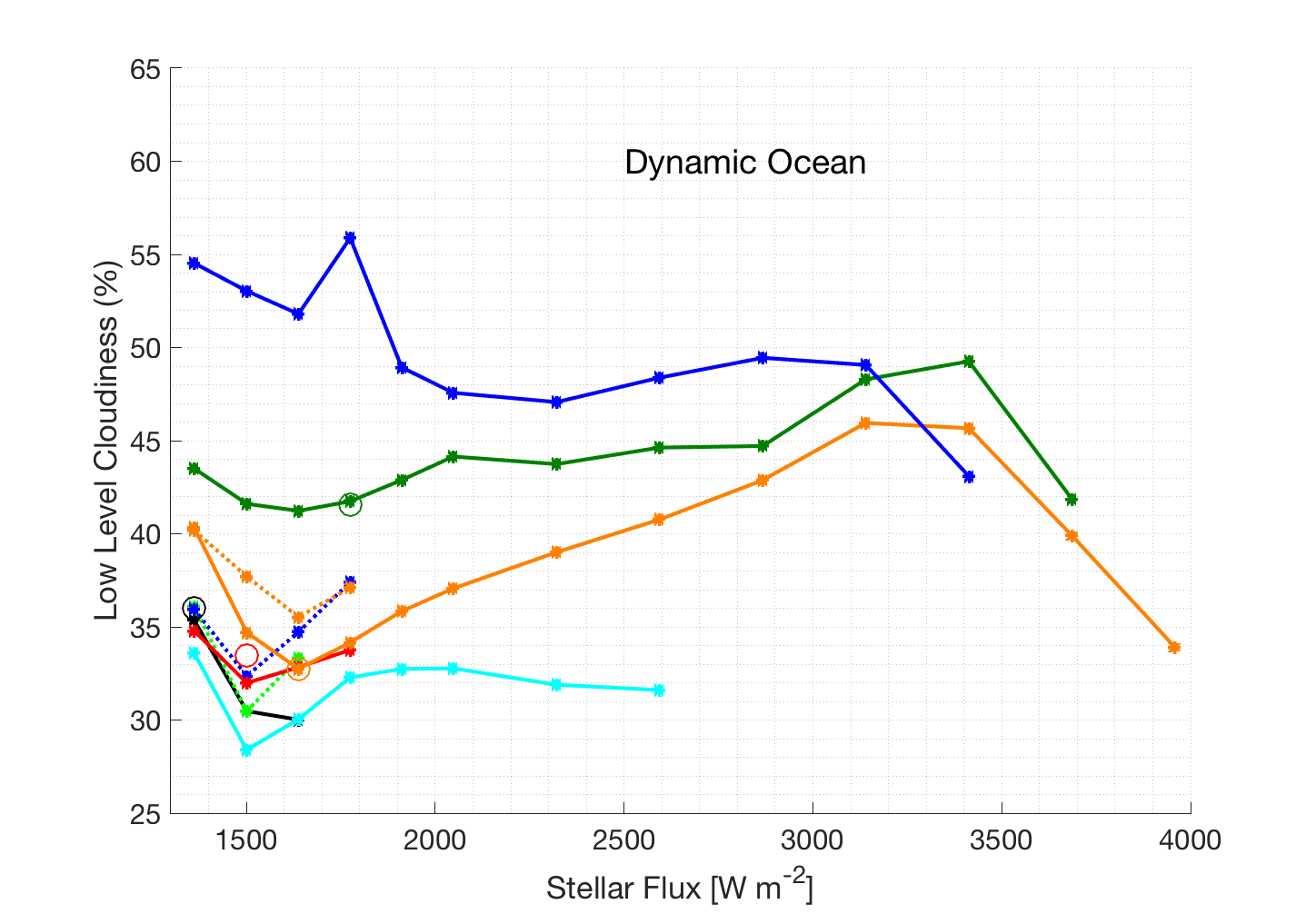}
\caption{\small Cloud Fractions at highest level (top), middle level (middle)
and bottom level (bottom) for Q-flux=0 (left) and Dynamic Ocean runs (right).
Note: the y-axes have different values from top to bottom.}
\label{fig:s0xClouds}
\end{figure}

% Figure 7
\begin{figure}[!htb]
\includegraphics[scale=0.15]{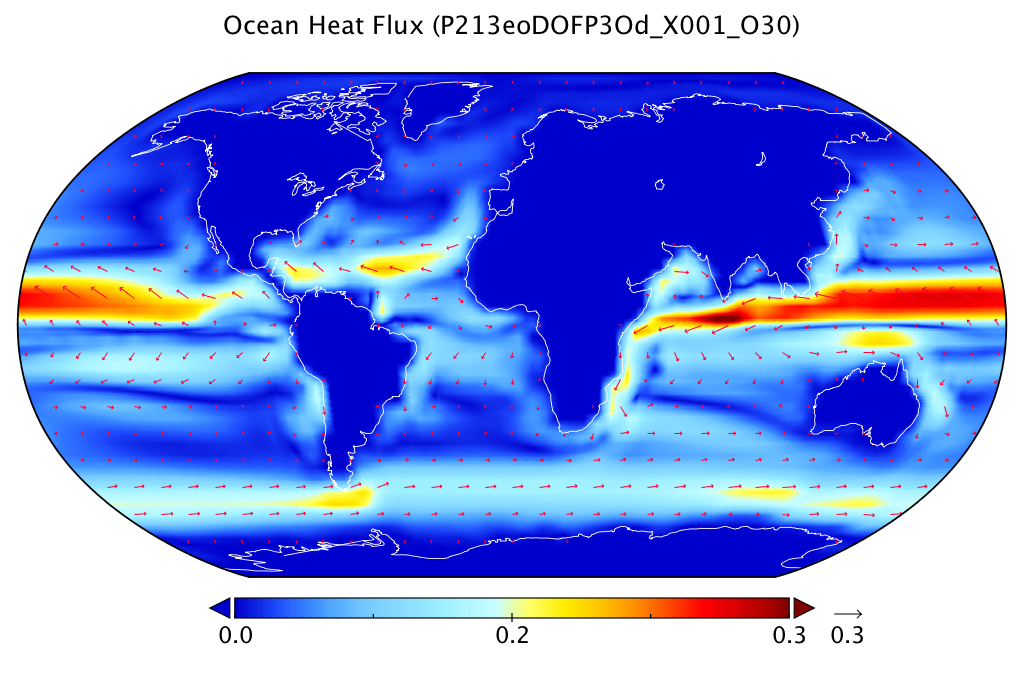}
\includegraphics[scale=0.15]{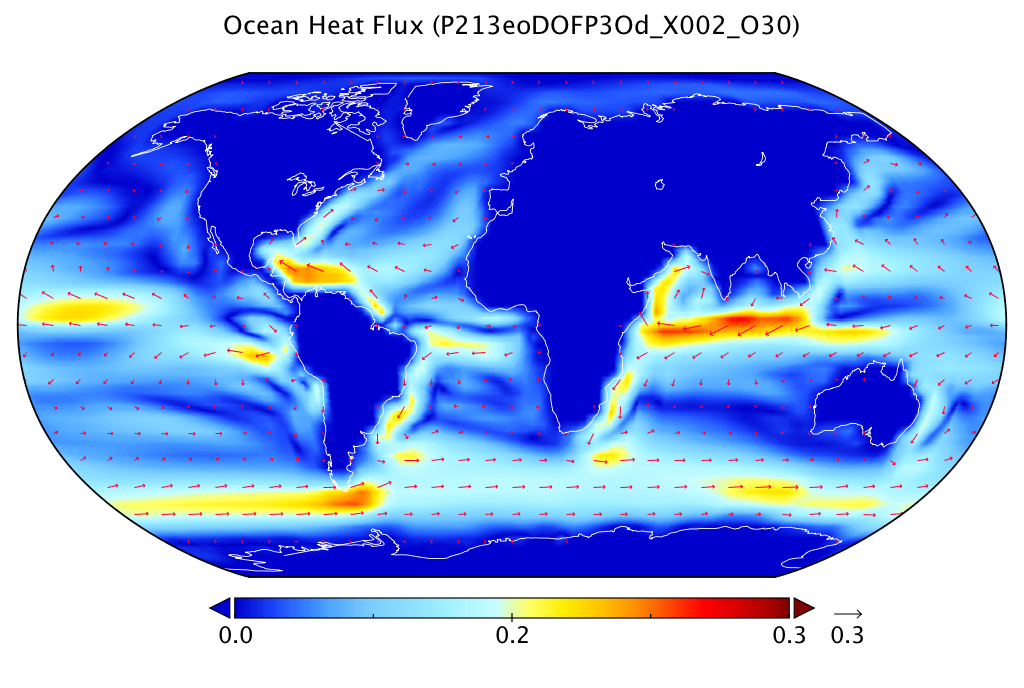}
\includegraphics[scale=0.15]{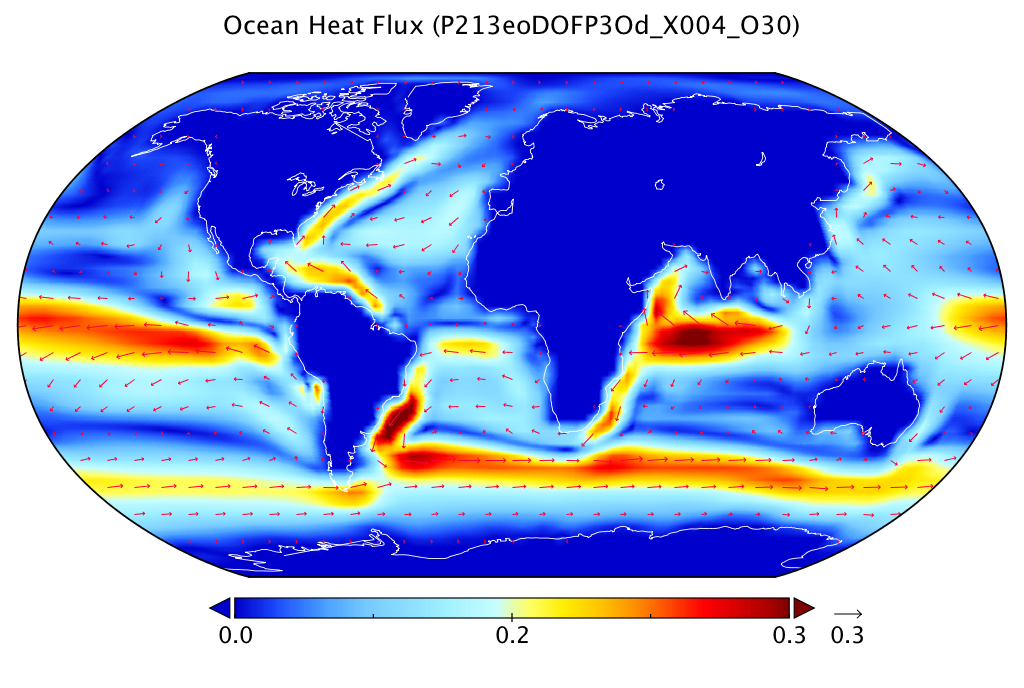}\\
\includegraphics[scale=0.15]{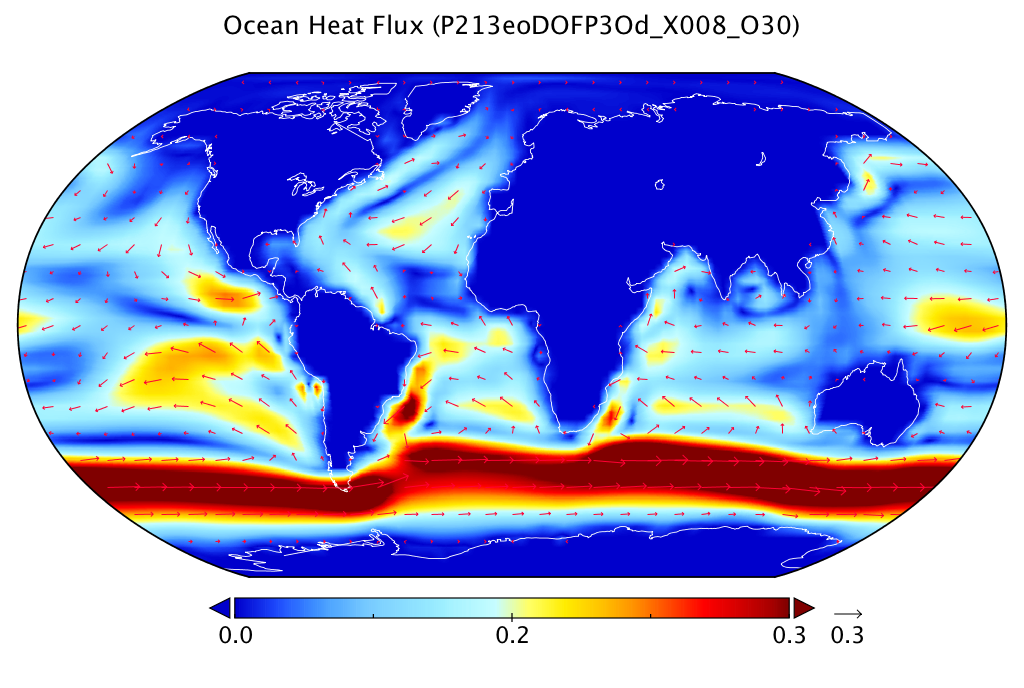}
\includegraphics[scale=0.15]{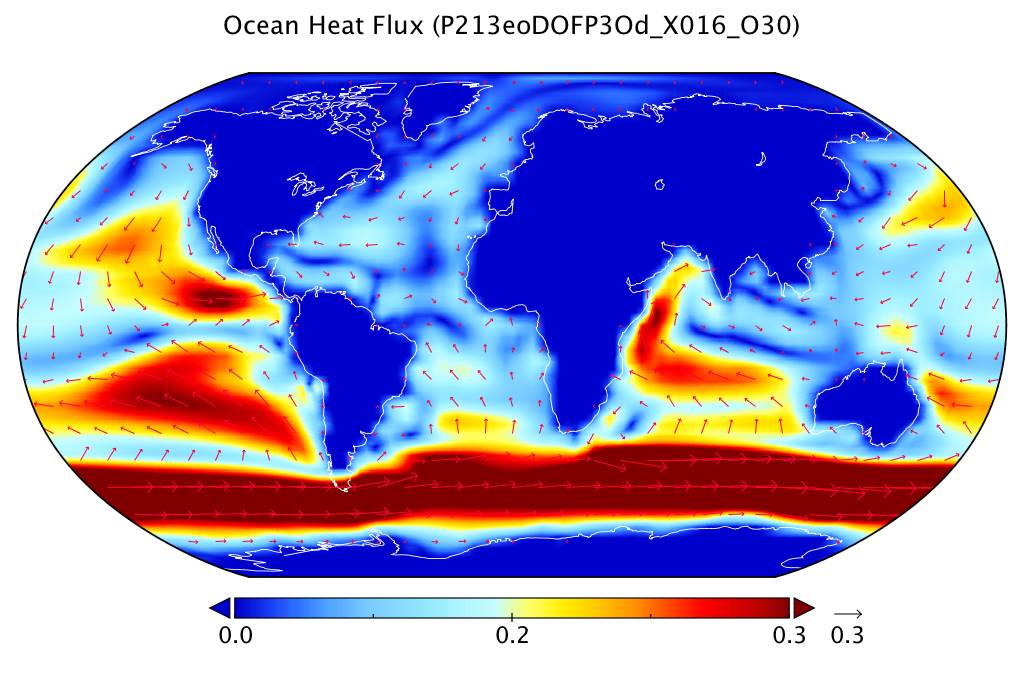}
\includegraphics[scale=0.15]{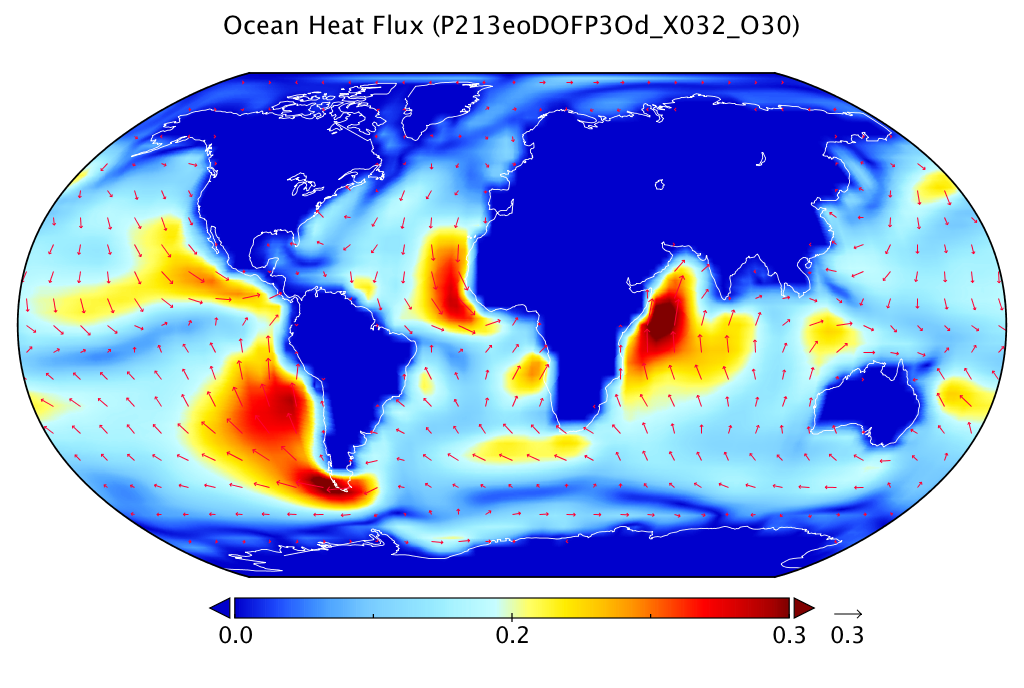}\\
\includegraphics[scale=0.15]{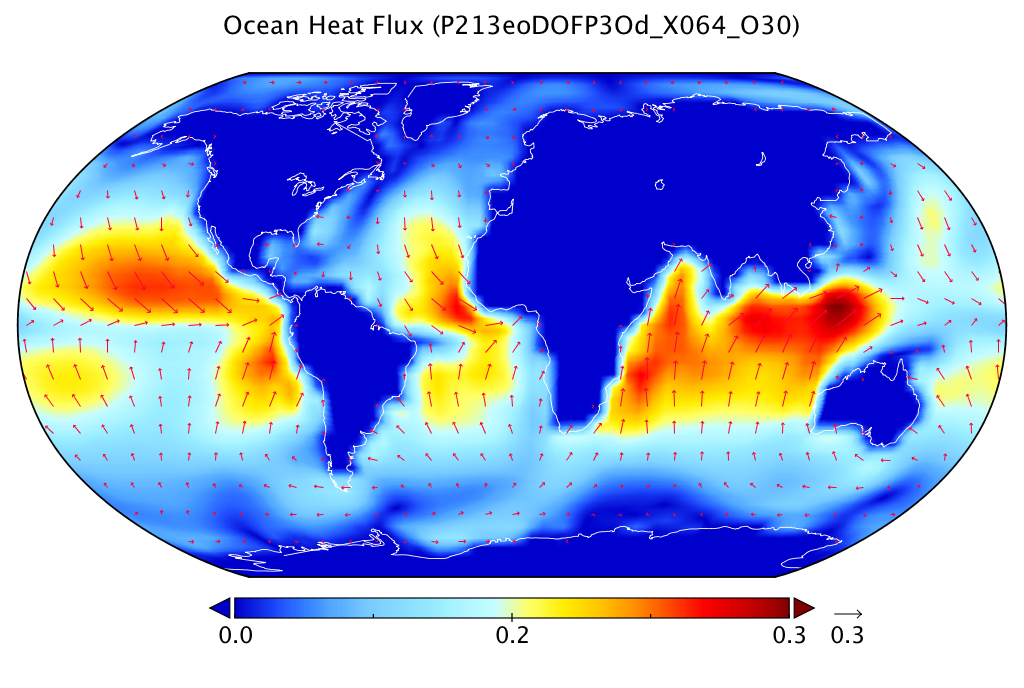}
\includegraphics[scale=0.15]{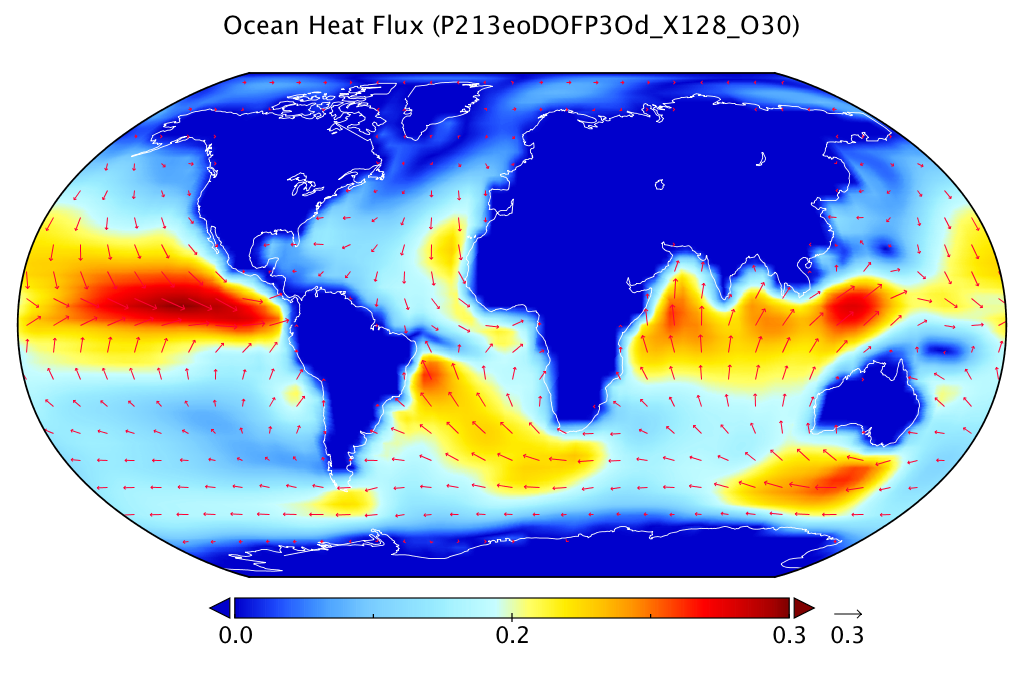}
\includegraphics[scale=0.15]{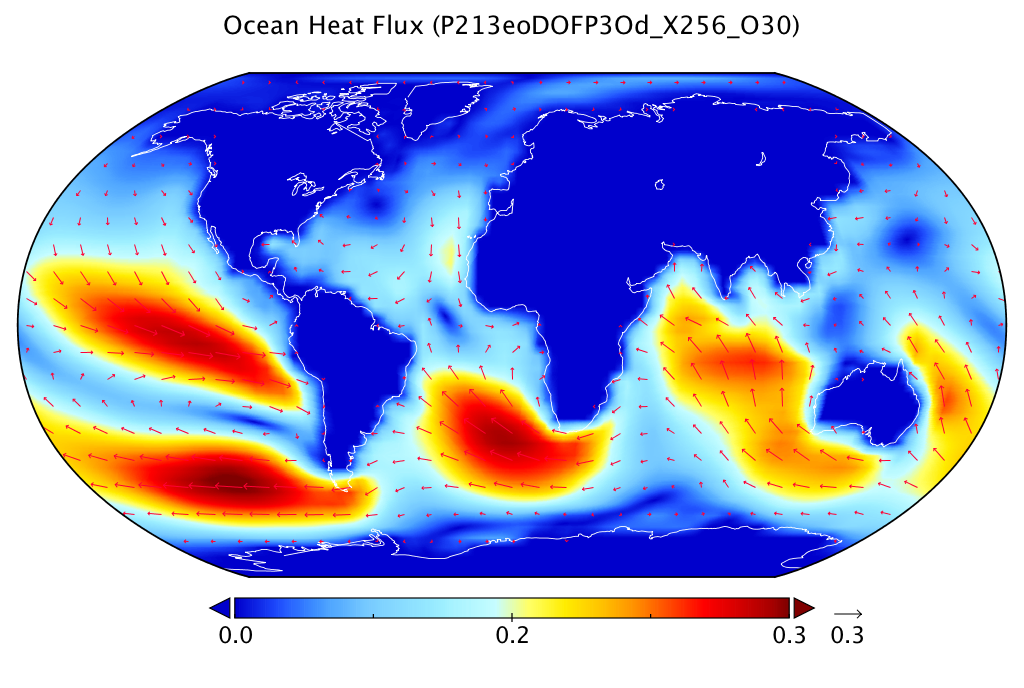}
\caption{\small Meridional plus zonal ocean heat flux
($\sqrt{\rm{Meridional}^2 + \rm{Zonal}^2}$) in units of 10$^{15}$ Watts. These
are 9 runs at fixed insolation of S0X=1.2 from sidereal length of day = X001 (1
x Earth), all the way up to X256 sidereal days in length.}
\label{fig:oceanflux}
\end{figure}

The ocean dynamics are strongly tied to the rotation rate of the planet. This
can most clearly be shown by looking at a range of rotation rates for a fixed
insolation as shown in Figure \ref{fig:oceanflux}.  At the faster rotations
(X001--X004) the ocean current are very Earth-like, with gyres in the northern
hemisphere in both ocean basins and an Antarctic circumpolar current in the
southern mid-latitudes.  The gyres clearly transport heat poleward, as the real
Gulf Stream does on Earth, but because there is more sea ice on this planet due
to the zero obliquity the transport is very important at low S0X and fast
rotation.  The dynamic ocean runs have less sea ice than their Q-flux=0
equivalents and thus the dynamic ocean climates are generally warmer as
described earlier and shown in Figure \ref{fig:s0xdiff}.  The trend goes up
until about X016, by which time the Hadley cell now stretches to the pole and
so the ocean heat transport is directed equator-ward with no gyres.  By this
rotation the sea ice is already gone so the addition of ocean heat transport
doesn't do very much, hence the insensitivity of the results to ocean heat
transport at slower rotations.

\begin{figure}[!htb]
\includegraphics[scale=0.5]{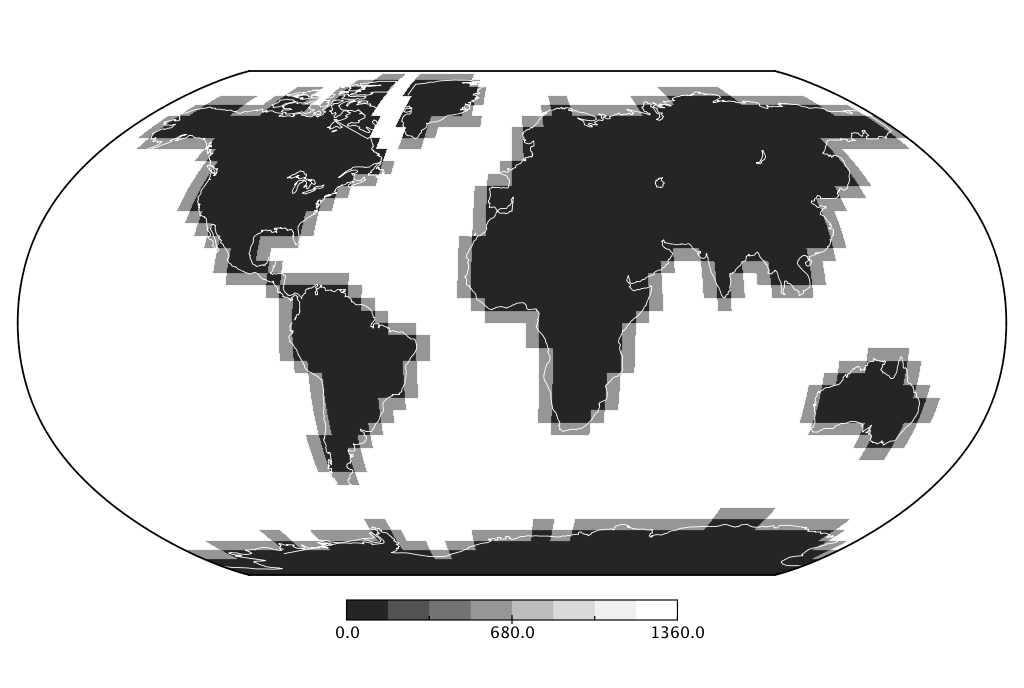}
\caption{\small Ocean bathymetry. Gridboxes at continent boundaries have an
ocean depth of 591m (light gray), while the rest of the ocean is 1360m (white).
Some seas/bays like the Hudson Bay, Baltic Sea, Black Sea, Caspian Sea, and
Mediterranean Sea have been filled in with land (black). We have removed the
island chain from Australia to south east Asia, much of the Malay and Kamchatka
peninsulas, and islands like Indonesia, Japan, Madagascar, Malaysia, New
Zealand and the Carribbean have been replaced with ocean. The channel between
Greenland and the northern Nunavut islands (e.g. Ellesmere) and Drake's Passage
has also been opened slightly.} \label{fig:bathtub}
\end{figure}

Despite the advantages of a dynamic ocean limitations exist as well. The
\rocke{} ocean model utilizes the \cite{gm1990} Earth-specific parameterization
for mesoscale mixing by unresolved eddies, which is appropriate to the
quasi-geostrophic ocean flow on a rapidly rotating planet. Little information
exists about the behavior of mesoscale ocean eddies in the slowly rotating
dynamical regime, although several studies have explored the rotation
dependence of overall ocean heat transport
\citep{FarnetiVallis2009,cullum2014}. However, in a Proxima Centauri b
simulation with \rocke{} it was found that changing the mesoscale diffusivity
had little effect \citep{DelGenio2018}. As mentioned previously, the fully
coupled ocean simulations have only two depths: 591m near continent boundaries
and 1360m elsewhere (see Figure \ref{fig:bathtub}). This allows the model to
come into thermodynamic equilibrium faster than it would if we used all 13
ocean model layers to 5000m, and hence lowers the total computational time per
simulation. The results of \cite{russell2013} and \cite{huyang2014} suggest
that ocean heat transport increases with ocean depth, with concomitant effects
on climate, although this is likely to be partly compensated by increased
atmospheric transport \citep{Stone1978}. Ocean bottom topography, the location
of continents, and the steepness of continental boundaries all influence ocean
circulation and thus heat transports. Likewise, salinity is specified as that
of Earth's ocean, which need not be the case on another planet. This has
implications for the density-driven component of the ocean circulation and
especially the freezing temperature \citep[see][]{DelGenio2018}, which in turn
affects albedo. Finally, the impact of tides, which are strongly correlated
with the size/mass and distance of the Earth's moon, is neglected in \rocke{}.
These are not strictly limitations, since exoplanets can be expected to have a
large variety of ocean depths and either no moons or moons of varying size and
distance, and these will likely be unconstrained by observations for the
foreseeable future.  It may however be a consideration for deep paleo-Earth
studies simulating time periods when the moon was closer to Earth and the tides
were higher.

% Figure 8

\subsection{Cloud parameterization sensitivity}\label{cloudparam}

Clouds are considered the most uncertain aspect of GCMs, accounting for much of
the inter-model spread in estimates of the global climate sensitivity of Earth
to increases in greenhouse gas concentrations \citep{Vial2013}.  A coupled
ocean-atmosphere climate model must be in global top-of-atmosphere radiation
balance at something close to Earth's observed surface temperature in order to
be used for climate change applications.  Given the uncertainties in GCM cloud
parameterizations, free parameters are usually adjusted within reasonable
limits to achieve balance while not overly compromising other aspects of the
simulated climate \citep{Mauritsen2012}. For exoplanets, no effective
observational constraints exist, nor is there any guarantee that a model
developed for modern Earth will reach radiative balance on its own. Thus
adjustments of the free parameters may be necessary. Here we consider two cloud
uncertainties to illustrate the limitations of 3-D studies of exoplanets and
the conclusions that can be drawn despite the limitations.

The \cite{Yang2014} baseline model has a global surface temperature of 287 K
for Earth's rotation period (their Figure 1c), remarkably close to Earth's
observed surface temperature. Given that the \cite{Yang2014} runs are
aquaplanets while those herein are Earth-like it should not be surprising that
ROCKE3D's Q-flux=0 run (see Table \ref{tab:qflux:simulations}) is cooler (265
K) while our dynamic ocean simulation (Table \ref{tab:dyn:simulations}) has a
temperature of 284 K. The reason for the difference between these temperatures
and that of modern Earth is that the planet being simulated is clearly not
Earth (e.g., it has no aerosols or vegetation). Most importantly, the planet
has zero obliquity and eccentricity. Earth's present day obliquity warms the
poles and cools the tropics. We expect the first of these to be more important
because it reduces sea ice and snow cover and decreases the Bond albedo, at
least for planets in the quasi-geostrophic dynamical regime that have large
equator-pole temperature gradients \citep{DelGenio2013,Madeleine2014}.  Thus,
although the global temperature of our idealized planet cannot be constrained,
if such a zero-obliquity planet existed, it would likely be colder than Earth.
The warmer temperature of the dynamic ocean model reflects the role that ocean
thermal inertia and heat transport play in limiting sea ice extent.

To test the sensitivity of the Q-flux=0 model temperature to the choice of
model tuning, we created an alternate version of our baseline planet (See
legend for `t' in Table \ref{tableqflux:rot-sox} and X001B in Figure
\ref{fig:s0xtsurf}a) with free parameters in the cloud parameterization chosen
differently but still within the range of uncertainty for modern Earth to
produce radiation balance at a different climatic state. Specifically, we
changed the values of two parameters that determine the threshold relative
humidity at which stratiform clouds begin to form (one for free troposphere
clouds, one for boundary layer clouds) and thus the cloud fraction. We also
changed a parameter that affects how quickly small cloud ice crystals grow into
large snowflakes by gravitational coalescence and precipitate out of the cloud,
thus altering the cloud's ice water content and optical thickness.

Figure \ref{fig:s0xtsurf}a (X001B, dashed black line) shows the global mean
surface temperature of this alternate model as a function of insolation for
Earth's rotation period, for comparison with the baseline model (X001, solid
black line).  The alternate model has a global mean surface temperature of
259.5 K for Earth's insolation (S0X=1.0) compared to the baseline model's 265.4
K (a difference of $\sim$ 6 K), but its sensitivity to insolation change is
fairly similar to that of the baseline model (see Figure
\ref{fig:s0xclimate}a). The result is that this planet can sustain a 20\%
increase in insolation and remain only slightly warmer (11K) than the baseline
model does for a 10\% increase in insolation. This implies that model-based
estimates of the edges of the habitable zone, whether 1-D or 3-D, are only weak
constraints on where habitable planets may be found, since model clouds can be
tuned to give a variety of climates.  

The more important question is whether any definitive statements can be made at
all about the effects of clouds on exoplanet climates.  The primary finding of
\cite{Yang2014} is that a planet can remain habitable at much higher values of
insolation if it is slowly rotating than if it is rapidly rotating. This occurs
because at rotation rates slow enough (long day lengths) for significant
day-night temperature differences to arise, convergence on the dayside leads to
rising motion and moist convection that produces a shield of high, optically
thick clouds that limit warming there and stabilize the planet's climate. The
fact that \rocke{} produces qualitatively similar behavior (Figure
\ref{fig:s0xtsurf}) is an encouraging sign that this might be a robust feature
of planetary climates. In our own Solar System, this process may be relevant to
the question of whether ancient Venus was habitable \citep{Way2016}.

On the other hand, moist convection is one of the most challenging and
uncertain aspects of terrestrial climate models \citep{delgenio2015}.  The
dayside cloud that stabilizes the climate at slow rotation is the end result of
a series of parameterized cloud and convection processes: A decision to
initiate convection, an assumption about the updraft mass flux and vertical
velocity, and an assumption about how much of the condensate that forms in the
updraft is transported upward and detrained along with saturated updraft vapor.
\rocke{} does this by diagnosing an updraft speed profile, assuming particle
size distributions and size-dependent fall speeds, and interactively
calculating the fraction of the condensate that precipitates vs. being
transported upward to form thick anvil cloud \citep{delgenio2005,delgenio2007}.
Each of these steps introduces uncertainty, e.g., the parameterization in the
baseline model overestimates the ice carried upward by convective events
\citep{Elsaesser2017}. Since $\sim$80 percent of the cloud ice that results
from deep convection in the terrestrial GCM is the result of upward transport
of particles rather than in-situ ice formation from supersaturated air, it is
reasonable to ask whether the rotation dependence of surface temperature is
parameterization-dependent.

To address this, we performed another sensitivity test in which we assume that
all condensate in the convective updraft precipitates. We denote such runs
``convective condensate precipitates" in Tables \ref{tableqflux:rot-sox},
\ref{tabledynocn:rot-sox}, \ref{tab:dyn:simulations} and
\ref{tab:qflux:simulations}. Thus, middle and upper level clouds can only form
from supersaturated vapor detrained by convective updrafts or created by
large-scale resolved upward motion.  We conducted one simulation for four
rotation periods at different insolations (open circles in Figure
\ref{fig:s0xtsurf}). The simulations without upward transport of condensate are
predictably warmer, but only by a couple of degrees, than the simulations with
the full convective physics, independent of rotation period. That is, cloud
that forms in-situ from the vapor detrained by the updraft or the resolved
upward motion is primarily what stabilizes the temperature. Insensitivity of
the results to this type of structural modification, rather than simply a
parameter change, strengthens the case for the conclusions of \cite{Yang2014}.

% Figure 9
\begin{figure}[!htb]
\includegraphics[scale=0.25]{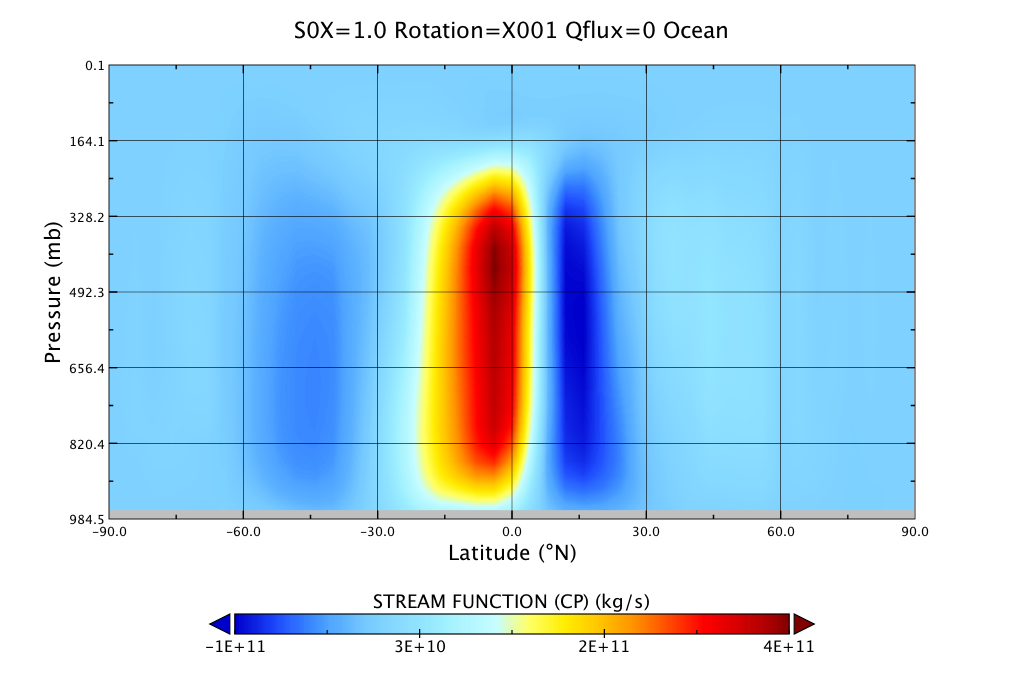}
\includegraphics[scale=0.25]{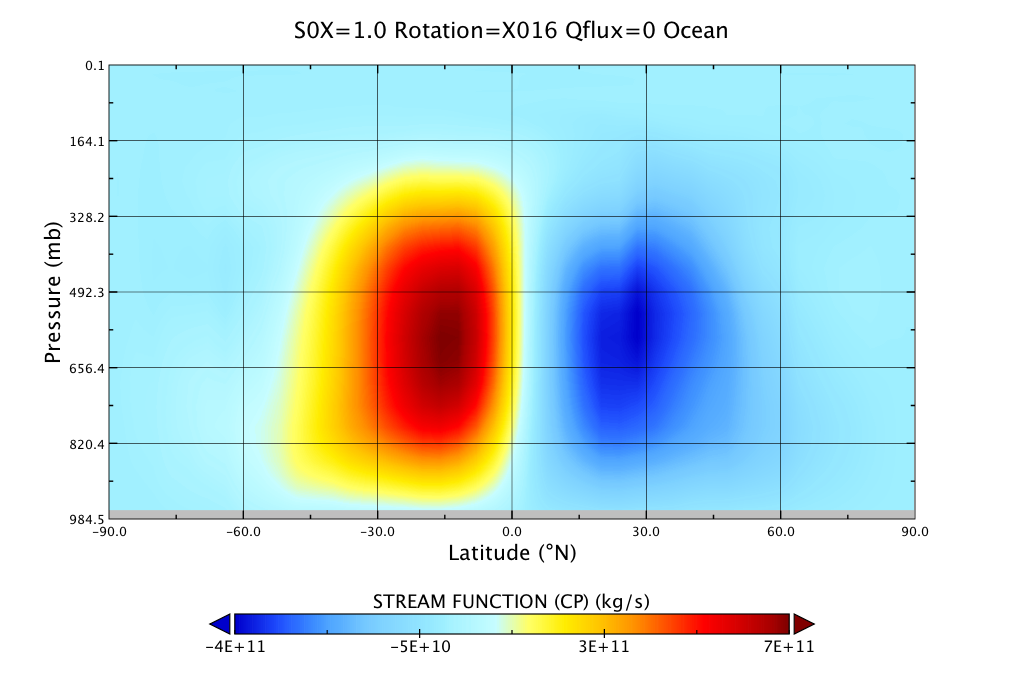}\\
\includegraphics[scale=0.25]{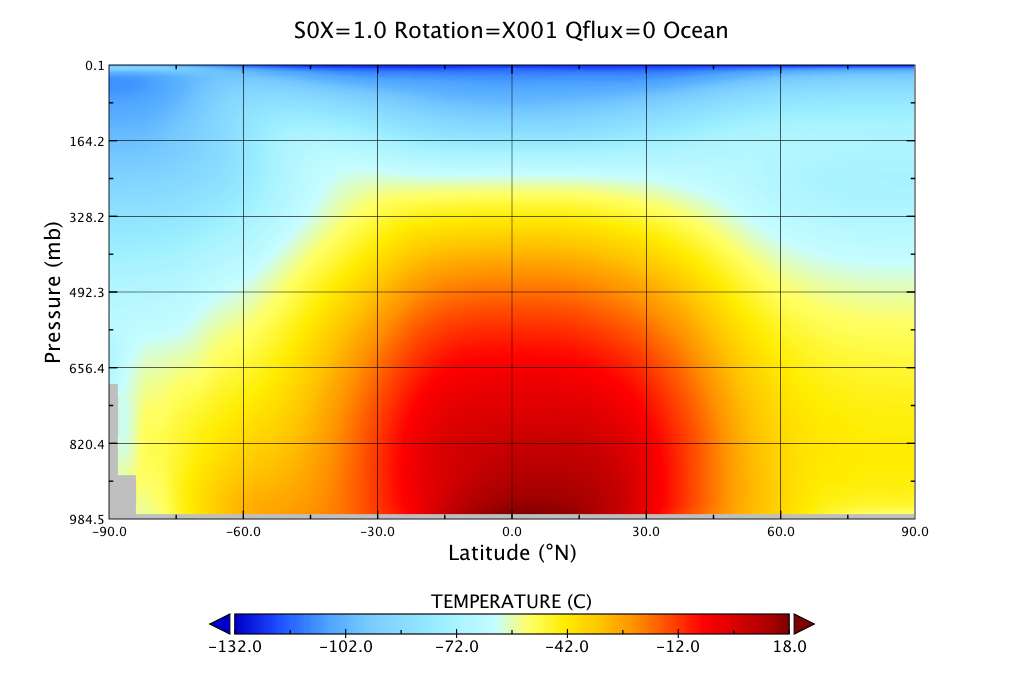}
\includegraphics[scale=0.25]{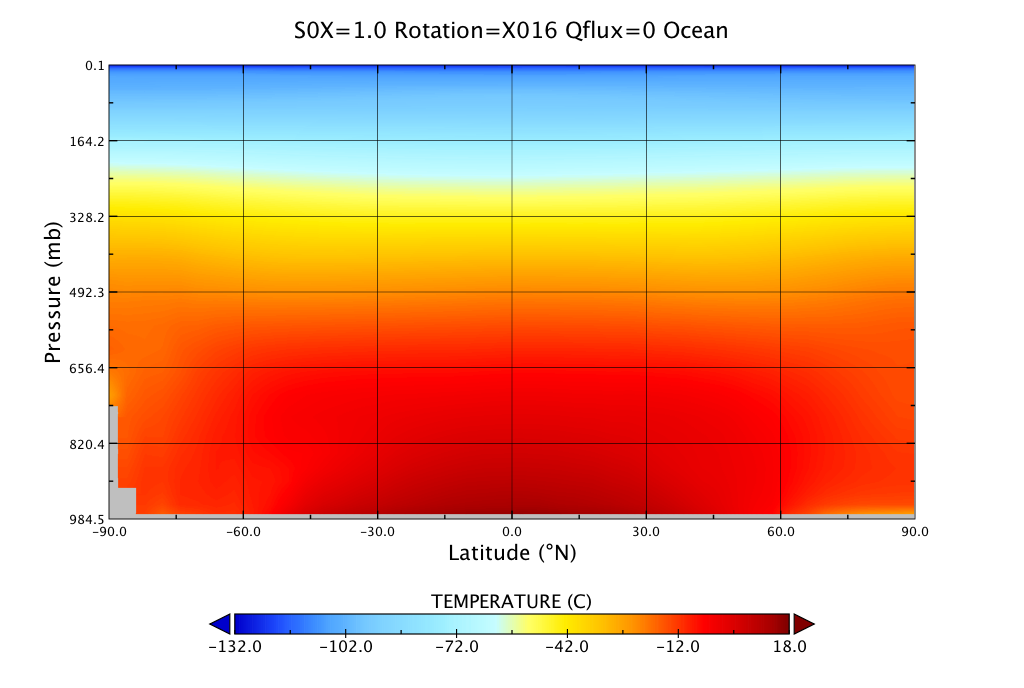}
\caption{\small Top panels: model stream function diagnostic showing increase
in Hadley Cell size and decrease in number from Rotations X001 to X016. Bottom
panels: The increase in the size of the Hadley cell transports more heat to
higher latitudes. This effect is seen in the latitudinal temperature
differences between X001 and X016. This is also seen in the decrease in the
amount of high latitude Sea Ice (Ocean Ice Fraction) in Figure
\ref{fig:s0xAlbOice} in the lower left panel between the X001 and X016
simulations for the same insolation (in this case S0X=1.0).} \label{fig:stream}
\end{figure}

This supports the idea that the fundamental behavior operating in all these
models is not the details of the cloud physics but the interaction between the
dynamics and radiation. The transition from a very sensitive to a weakly
sensitive climate as day length increases depends on two transitions in
dynamical regime that determine how heat is transported.  The first is from the
quasi-geostrophic regime characteristic of rapidly rotating planets to the
quasi-barotropic regime of more slowly rotating planets. This transition has
been well studied for rocky planets
\citep{WilliamsHolloway1982,delgenio1987,delgenio1993,Allison1995,Navarra2002,Showman2013}.
Both regimes are dominated by poleward heat transport. In the quasi-geostrophic
regime this is accomplished by a Hadley circulation at low latitudes, with mean
rising motion near the equator and sinking motion in the subtropics, and by
baroclinically unstable eddies that produce low and high pressure centers and
fronts at higher latitudes. In the quasi-barotropic regime, the Hadley cell
spans most latitudes and dominates the global heat transport.  This transition
occurs approximately when the Rossby radius of deformation (the spatial scale
of rapid baroclinic eddy growth) approaches the size of the planet
\citep{delgenio1993,Edson2011,Showman2013}.  It differentiates the atmospheric
circulations of Earth and Mars from those of Venus and Titan.  Our simulations
with 1 d rotation period are in the quasi-geostrophic regime, while the 16 d
period simulations are in the quasi-barotropic regime. In Figure
\ref{fig:stream} we show how the Hadley cell changes in size when going between
these regimes and how it clearly effects the surface temperatures and ocean ice
fraction at high latitudes (see Figure \ref{fig:s0xAlbOice}).

The second transition is from a circulation that transports heat poleward to a
diurnally-driven circulation that transports heat from a dayside region of
rising motion to a nightside region of sinking motion, with day-night transport
both over the poles and across the terminators.  The day-night circulation has
been explored by \cite{joshi1997,joshi2003,Yang2013} for synchronously rotating
planets and by \citet{Yang2014} for slowly rotating but asynchronous planets.
In our simulations the transition appears to occur between 32 d and 64 d
rotation period. This can be understood by considering the radiative relaxation
time scale t$_{rad}$ = pc$_{p}$T/(gF), where p is pressure, c$_{p}$ the
specific heat at constant pressure, T the temperature, g the acceleration of
gravity, and F the emitted thermal flux to space. For Earth heated at 1 AU by
the Sun t$_{rad}$ $\sim$ 1--2 months, depending on the pressure level one
chooses for the calculation. When the length of the solar day t$_{sol}$ $<<$
t$_{rad}$, day-night temperature differences are small, and equator-pole
temperature gradients drive the circulation. When t$_{sol}$ $\gtrsim$
t$_{rad}$, day-night temperature contrasts become important and the circulation
develops a strong diurnally-driven component. This regime transition occurs in
the 32--64 d rotation interval (for which t$_{sol}$ is only slightly longer
than the rotation period) for the planets we simulate.  \cite{Showman2015}
invoke a similar argument to define the boundary between circulations
characteristic of weakly irradiated jovian planets and strongly irradiated hot
Jupiters.

These changing transport patterns among the three dynamical regimes matter for
habitability because they determine where optically thick clouds form and thus
where sunlight is more strongly reflected. They also matter for the water cycle
because of their effect on precipitation patterns, as we will explore in future
papers in this series.  Despite uncertainties in parameterized cloud physics,
the only requirement of the cloud/convection physics is that thick clouds form
where large scale upward motion is prevalent. This basic behavior should be
model-independent for any mass flux cumulus parameterization, since it requires
only that rising air adiabatically cool and eventually saturate.

% Figure 10
\begin{figure}[!htb]
\includegraphics[scale=0.44]{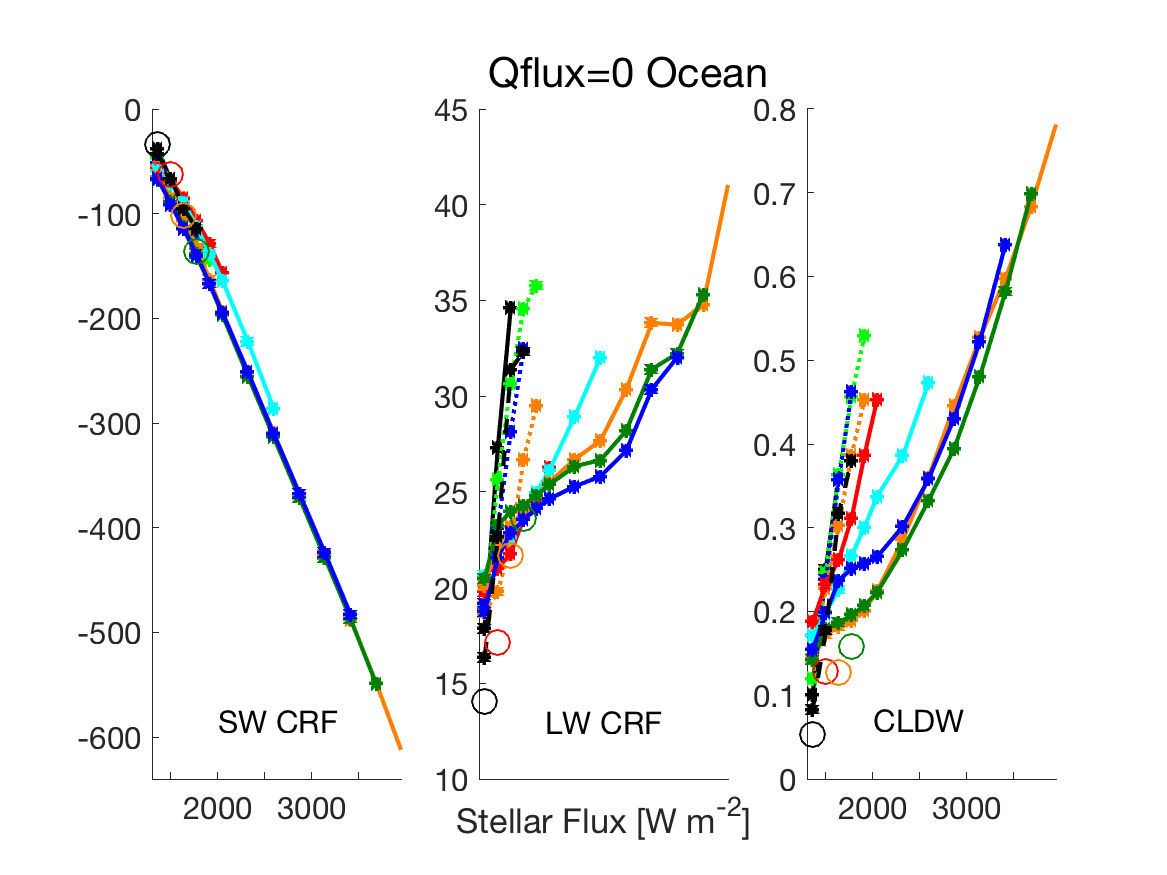}
\includegraphics[scale=0.44]{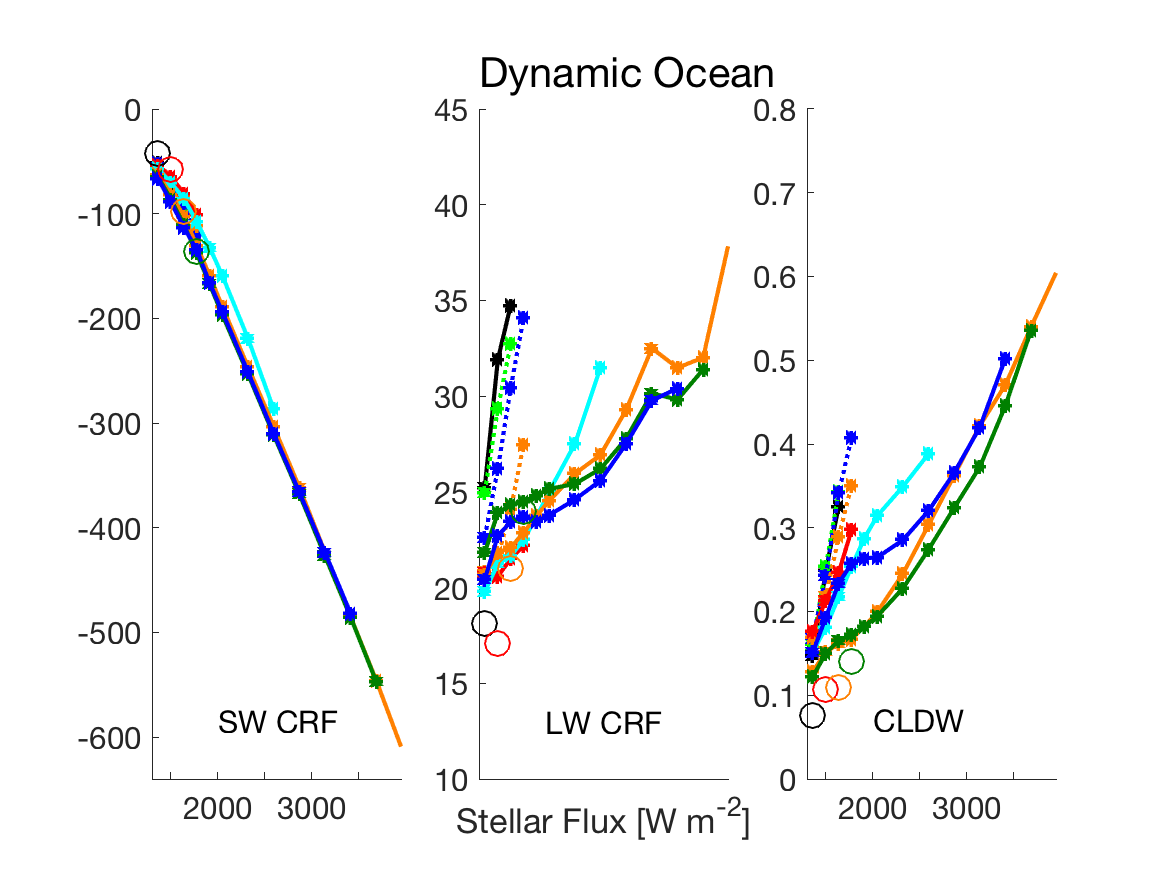}\\
\caption{\small Cloud Radiative Forcing (CRF) for the short wave (SW) and long
wave (LW) in W m$^{-2}$ along with the column-integrated cloud water (CLDW) in
kg m$^{-2}$.} \label{fig:s0xCRF}
\end{figure}

Figure \ref{fig:s0xCRF} shows the global mean shortwave (SW) and longwave (LW)
cloud radiative forcing (CRF), as well as the column-integrated cloud condensed
water (liquid + ice; CLDW) as a function of incident solar flux S$_{o}$ and
rotation period. CRF measures the effect of clouds on a planet's
top-of-atmosphere radiation budget \citep[e.g.,][]{Cess1990}.  We use the
convention that downward/upward fluxes (i.e., heating/cooling of the planet)
are positive/negative and define the absorbed SW flux:

\begin{equation}
\rm{Q} = S_{o}(1-A)/4
\end{equation}

where A is the planetary (Bond) albedo, and the emitted LW flux:

\begin{equation}
\rm{F} = \sigma T_{e}^{4}
\end{equation}

where T$_{e}$ is the equilibrium temperature.

Letting subscript c indicate the fluxes that would exist in clear skies
(calculated by a separate offline call to the radiation parameterization that
ignores the clouds),  and subscript o the fluxes that exist in overcast skies,
we define:
\begin{equation}
\rm{SWCRF}\footnote{Short Wave Cloud Radiative Forcing} = Q - Q_{c} = [fQ_{o} + (1-f)Q_{c}] - Q_{c} = fS_{o}(A_{c}-A_{o})/4
\end{equation}

\begin{equation}
\rm{LWCRF}\footnote{Long Wave Cloud Radiative Forcing} = F_{c}-F = f(F_{c}-F_{o}) = \sigma f(T_{ec}^{4}-T_{eo}^{4})
\end{equation}

where f is the cloud fractional area.

Positive/negative values of CRF indicate heating/warming by clouds,
respectively.  SWCRF $<$ 0 because clouds are more reflective than the clear
sky and most surfaces and thus make a planet cooler than it would be without
clouds, while LWCRF $>$ 0 because clouds absorb upwelling thermal radiation and
re-radiate it at a lower temperature, producing greenhouse warming. Note that
CRF depends not only on the coverage and properties of clouds but also on the
properties of the clear-sky atmosphere and planet surface through the terms
Q$_{c}$ and F$_{c}$.  Thus changes in CRF in response to a climate forcing such
as an increase in incident solar flux capture the cloud feedback only in the
context of the accompanying clear sky changes rather than isolating the cloud
feedback itself \citep{Soden2006}.

Figure \ref{fig:s0xCRF} shows that SWCRF increases monotonically in magnitude
with incident solar flux and is somewhat stronger for the slowly rotating
planets, as expected. Much of the insolation dependence reflects the increase
in S$_o$ itself, but for a doubling of S$_o$, SWCRF increases by about a factor
of 3, indicating that cloud fraction and/or the albedo difference between
overcast and clear skies increases with insolation as well. Cloud vertically
integrated water (CLDW) actually increases more slowly with insolation for the
slowly rotating planets, but this is misleading because the clouds on the slow
rotators are much denser on the dayside, where they can contribute to SWCRF.
LWCRF is smaller than SWCRF and increases slowly as incident solar flux
increases This does not appear to be due to increasing cloud height, given that
high cloud cover decreases and middle level cloud cover increases with
insolation, at least for the lower insolation values (Figure
\ref{fig:s0xClouds}). Furthermore, the high cloud decrease exceeds the middle
cloud increase, so cloud fraction does not appear to be the cause. Instead, the
similarity of the LWCRF and cloud water dependences on temperature and
insolation suggest that cloud opacity, and its implications for the radiating
temperature contrast between cloud and clear skies, controls the LWCRF
behavior.   The net result of the SWCRF and LWCRF behavior is that the clouds
exert a net cooling effect on the climate that becomes stronger as the incident
flux increases, i.e., by this simple measure the net cloud feedback is
negative.

% Figure 11
\begin{figure}[!htb]
\includegraphics[scale=0.8]{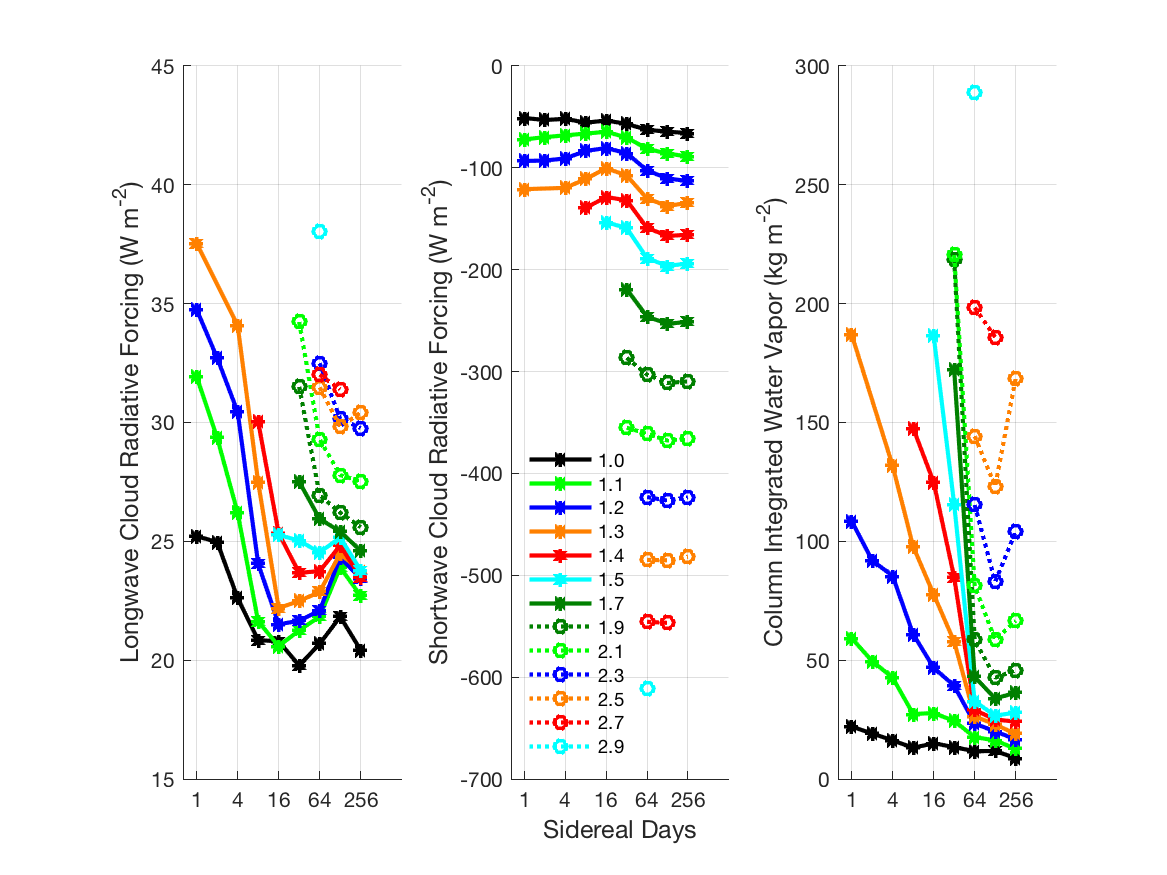}
\caption{\small
(Left) Longwave Cloud Radiative Forcing (CRF), (center) Shortwave CRF, and
(right) column integrated water vapor as a function of rotation period for the
dynamic ocean simulations for different S0X values.} \label{fig:lwswqatm}
\end{figure}

Rotation period does not directly affect temperature, so the cooling of the
planet as rotation slows in Figure \ref{fig:s0xtsurf} must be the result of
climate feedbacks that occur as the dynamical regime changes in response to the
changing rotation. Figure \ref{fig:lwswqatm} shows the LWCRF, SWCRF, and column
integrated water vapor as a function of rotation period for different values of
S0X.  From 1 d to 16 d period, SW cloud cooling generally weakens, though this
is partly offset by an increase in LW cloud warming.  Furthermore, sea ice
decreases with increasing rotation period as the Hadley cell expands (Figure
\ref{fig:stream}). Thus, neither cloud nor sea ice feedbacks can explain the
generally cooler temperatures and lower climate sensitivity as rotation slows.
Instead, the dramatic decrease in water vapor with increasing rotation period
($>$ 50\% for most values of S0X) reduces the greenhouse effect of the clear
sky atmosphere.  Water vapor continues to decrease with longer rotation period
for periods up to 128 d, but SWCRF becomes increasingly negative and LWCRF
slightly less positive.  All three feedbacks are thus negative, consistent with
the large separation in temperature in Figure \ref{fig:s0xtsurf} between the
more rapidly and more slowly rotating planets.

\section{Discussion}

Our results support the conclusions of previous work \citep[e.g.][]{Yang2014}
that rapidly rotating planets like Earth are much more susceptible to a runaway
greenhouse than are slowly rotating planets at high insolations. As mentioned
above, some of the differences between the work herein and previous results may
be related to the different GCMs being used, the types of oceans (dynamic ocean
versus thermodynamic ocean), and an Earth-like topography and land/sea mask
versus aquaplanet configurations.

As mentioned in Section \ref{sec:Experimental_Setup} \rocke{} does not yet use
high temperature line lists, nor does it account for the effect of water vapor
mass on the dynamics, so beyond the fact that \rocke{} lies within the range of
previous GCM studies that concluded that the inner edge of the habitable zone
must be significantly closer to the Sun than 1-D models estimate, we cannot
estimate a precise inner edge location.

\begin{figure}[!htb]
\includegraphics[scale=0.35]{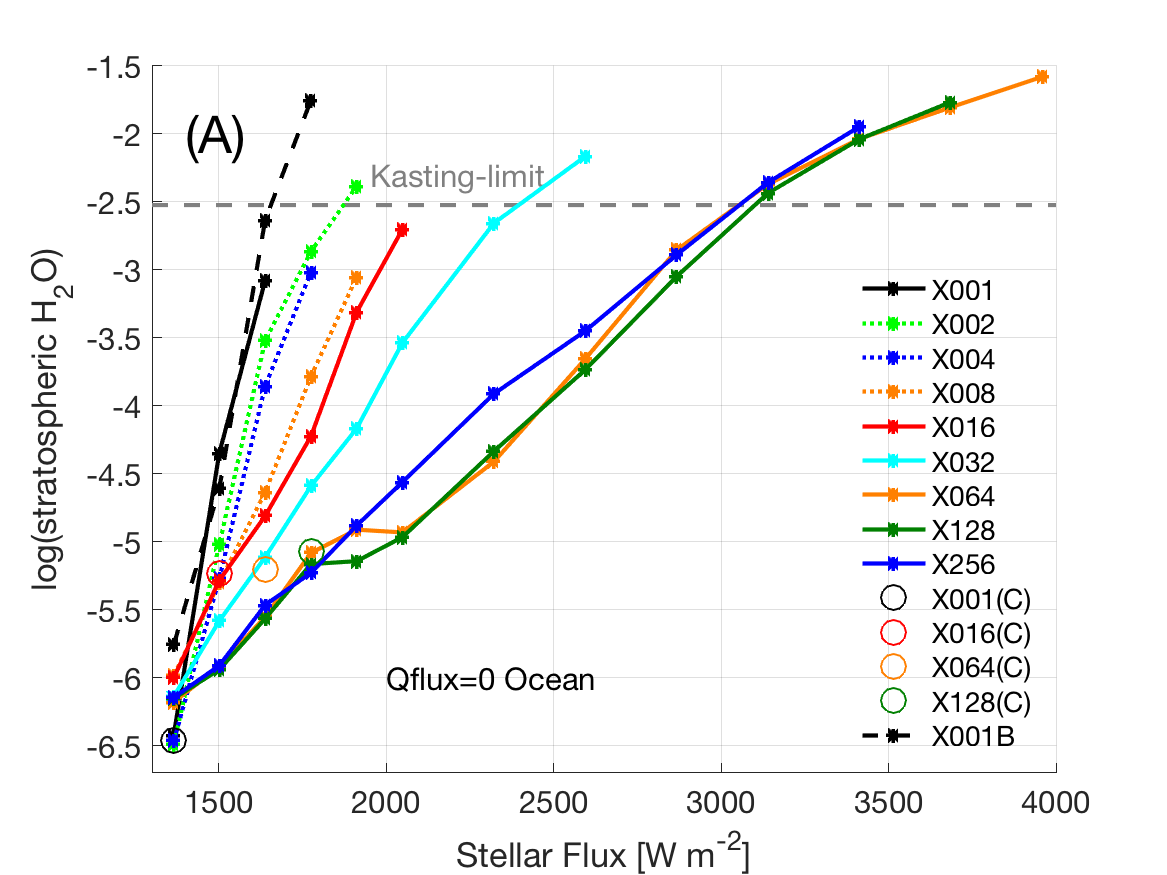}
\includegraphics[scale=0.35]{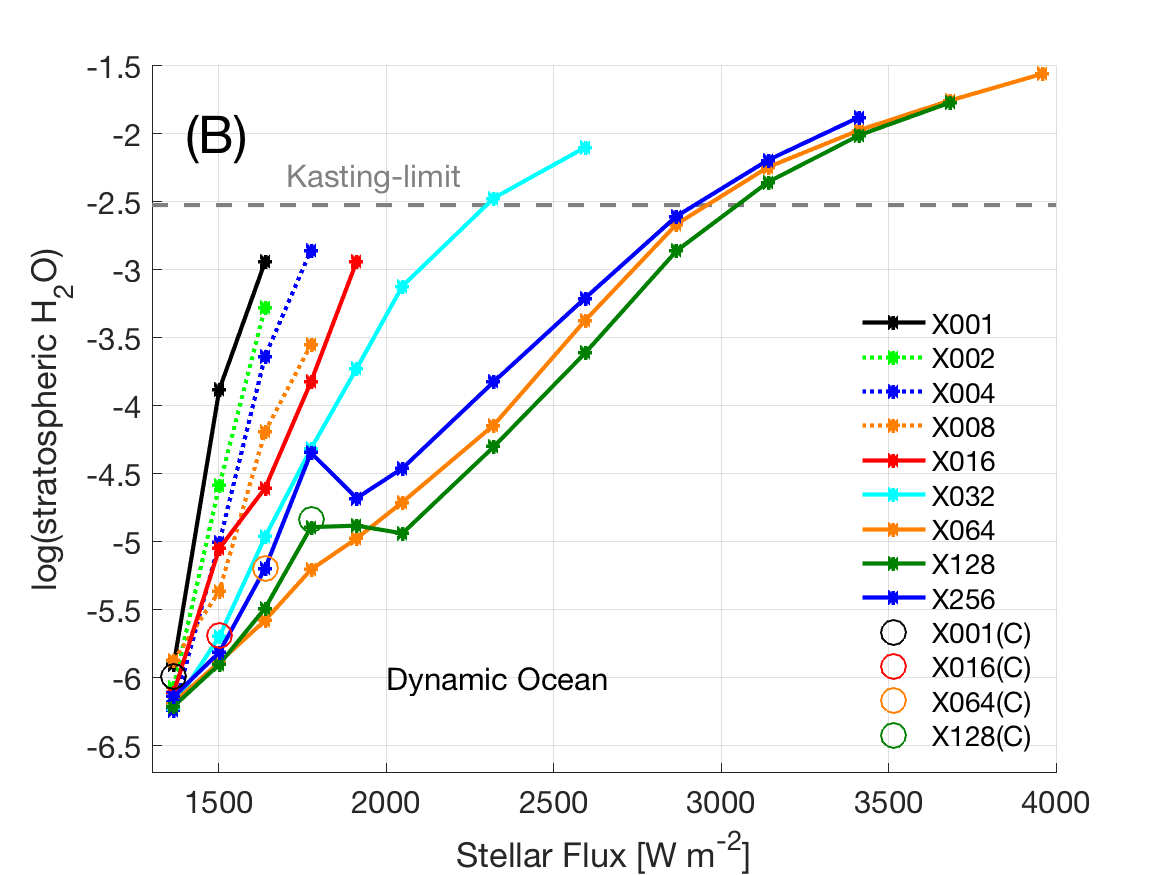}\\
\includegraphics[scale=0.35]{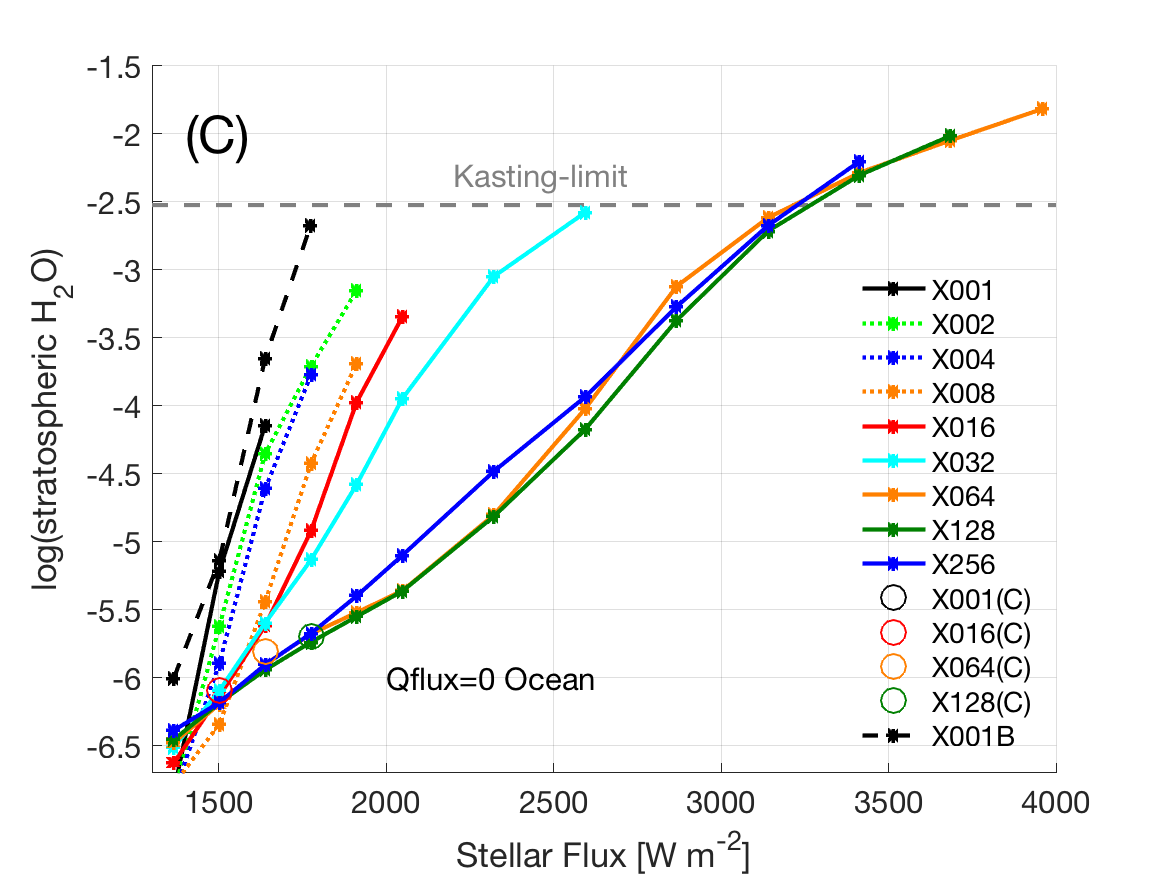}
\includegraphics[scale=0.35]{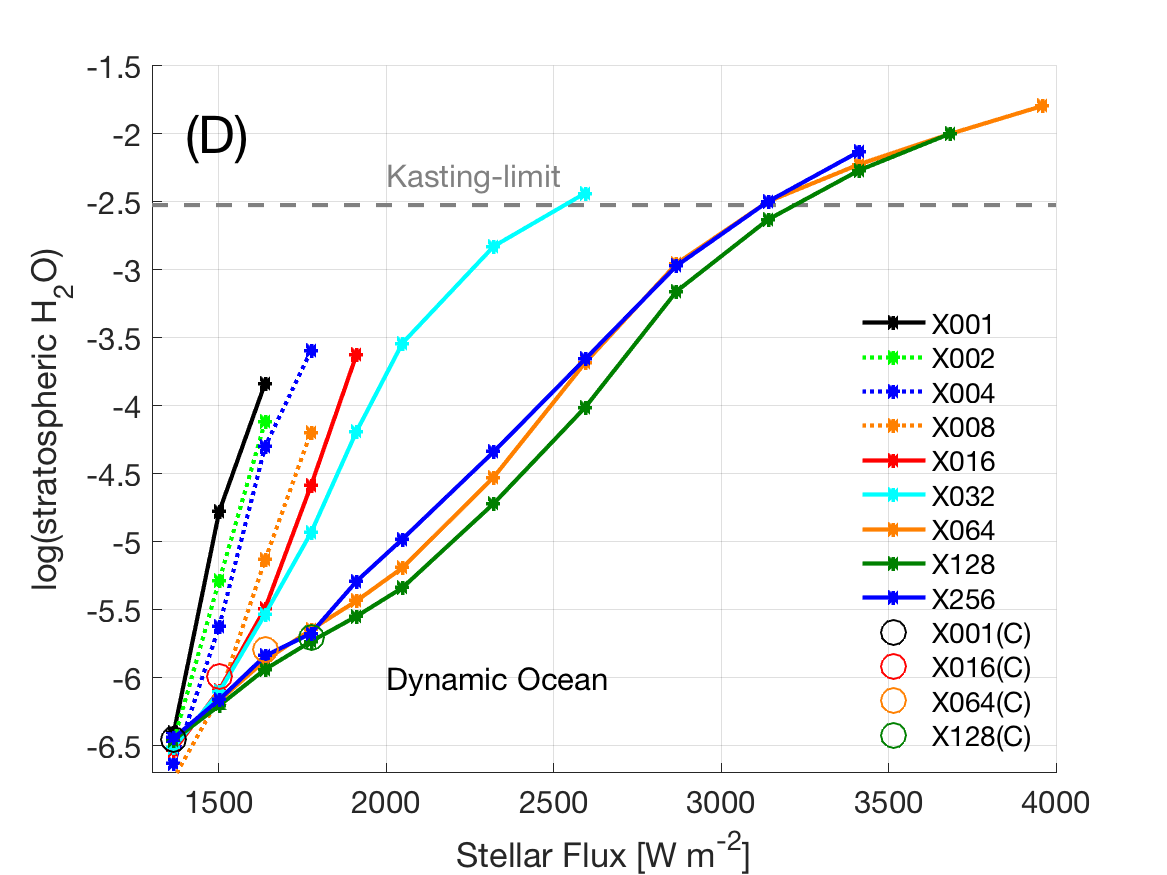}
\caption{Stratospheric water vapor content for Q-flux=0 ocean runs (A/C, left)
and Dynamic Ocean runs (B/D, right). Y-axis is the log$_{10}$ of the specific
humidity at 0.1 hPa for H$_{2}$O/Air (kg/kg). A \& B are values for the grid
cell with the largest value at the highest layer in each simulation whereas C
\& D are the mean at the highest layer.} \label{fig:h2oclimate}
\end{figure}

On the other hand one can look at diagnostics of water vapor transported into
the stratosphere, where it can potentially be photodissociated, leading to
hydrogen escape and onset of the moist greenhouse, a more conservative
definition of the inner edge of the habitable zone. To that end in Figure
\ref{fig:h2oclimate} we show water vapor molar concentration in the highest
model layer for the grid cell with the largest value (A \& B) and the mean in
the highest layer (C \& D) to investigate whether any of our simulations begin
to approach the moist greenhouse limit of \cite{kasting1993}, i.e.
f(H$_{2}$O)$=$3$\times$10$^{-3}$. Figure \ref{fig:h2oclimate}A shows that for
the highest insolation runs of most rotations with a Q-flux=0 ocean the
stratosphere exceeds the 3$\times$10$^{-3}$ limit in a given grid cell whereas
the mean Figure \ref{fig:h2oclimate}C does not in general. With the dynamic
ocean in Figure \ref{fig:h2oclimate}B most of the high insolation rapidly
rotating planets (rotations $<$X032) almost reach the moist greenhouse state,
whereas for the slower rotating models where S0X$>$2.9 they exceed the
3$\times$10$^{-3}$ limit for the maximum grid cell value. However, if one takes
the mean of the highest layer Figure \ref{fig:h2oclimate}D then one finds they
are roughly similar to the Q-flux=0 runs in Figure \ref{fig:h2oclimate}B.

% Figure 12

We have shown that the representation of ocean dynamics is just as important a
consideration for extreme exoplanet climates as is the representation of
radiation, particularly so for rapidly rotating planets as shown in the 1 d --
32 d cases herein. It should be noted that \rocke{} cannot actually reach the
runaway greenhouse definition of the inner edge of the habitable zone for
terrestrial worlds, unlike in many of the works cited above. The radiative
transfer in \rocke{} cannot accurately calculate heating rates for temperatures
over 400K. In addition, the dynamics begins to lose accuracy when water vapor
becomes more than 10\% of the mass of the atmosphere and begins to
significantly affect pressure gradients and the assumed ideal gas behavior of
the atmosphere. When water vapor starts to becomes a non-negligible fraction of
the atmospheric mass it also becomes important to be able to adjust the
atmospheric constants such as the atmospheric mass, and heat capacity at
constant pressure and volume, which \rocke{} is not presently capable of doing.

Future papers in this series will address different habitability metrics such
as that of \cite{Spiegel2008} and others related to water availability that are
crucial to life on present day Earth.

The data from this paper is open source and at publication time will be made
available on the \rocke{} NCCS\footnote{NASA Center for Climate Simulation}
data portal website: https://portal.nccs.nasa.gov/GISS\_modelE/ROCKE-3D An
additional copy of the files will also be made available at
https://archive.org/details/Climates\_of\_Warm\_Earth\_like\_Planets

\newpage
\appendix
\section{Calendar}\label{appendix:calendar}

As detailed in \cite{Way2017} \rocke{} uses a 120 day calendar system when the
numbers of days per year is less than 120 days. There is an option for
overriding this feature, but it was not utilized for the simulations herein.
For this reason we are publishing details from the calendar in use for these
simulations in Tables \ref{tablecal1test}, \ref{tablecal1b} and
\ref{tablecal3a}. The 120 day default calendar goes into effect for any
simulations with sidereal day lengths longer than 2 Earth sidereal days. 

\rocke{} places the calendar output into a runtime text file with additional
details of the run as it moves forward in time. For all runs (regardless of the
calendar in use) the Longitude at Periapsis is fixed to 282.9$\degr$.

\begin{deluxetable}{|l|r|r|}
\tablecaption{Calendar day lengths 1 \& 2 \label{tablecal1test}}
\tablehead{\multicolumn{1}{|l|}{Name} & \multicolumn{1}{r|}{1 day \tablenotemark{1}} & \multicolumn{1}{|r|}{2 days \tablenotemark{2}}}
\startdata
\hline
Mean solar day           &   1.000000     &   2.005329\\
Sidereal Rotation Period &   0.997268     &   1.994379\\
Sidereal Orbital Period  & 365.000000     & 365.256369\\
Solar Days per year      & 365.000000     & 182.142857\\
\hline
\enddata
\tablenotetext{1}{Modern Earth Sidereal Day (ESD) length.}
\tablenotetext{2}{Twice the length of modern ESD length.}
\end{deluxetable}

\begin{deluxetable}{|l|c|r|r||c|r|r||c|r|r|}
\tablecaption{Calendars for day lengths 1 \& 2 \label{tablecal1b}}
\tablehead{& \multicolumn{3}{c||}{Modern Earth} & \multicolumn{3}{c|}{2 x Modern Earth\tablenotemark{1}} & \multicolumn{3}{c|}{120 day calendar\tablenotemark{2}}}
\startdata
\hline
Month & day length & first day & last day&length &first&last&length &first&last\\
\hline
January   & 32 &   1   &  32 & 16 &   1  &  16 & 11 &   1 &  11\\
February  & 29 &  33   &  61 & 14 &  17  &  30 & 09 &  12 &  20\\
March     & 31 &  62   &  92 & 16 &  31  &  46 & 10 &  21 &  30\\
April     & 30 &  93   & 122 & 15 &  47  &  61 & 10 &  31 &  40\\
May       & 30 & 123   & 152 & 15 &  62  &  76 & 10 &  41 &  50\\
June      & 29 & 153   & 181 & 14 &  77  &  90 & 10 &  51 &  60\\
July      & 30 & 182   & 211 & 15 &  91  & 105 & 09 &  61 &  69\\
August    & 31 & 212   & 242 & 15 & 106  & 120 & 10 &  70 &  79\\
September & 29 & 243   & 271 & 15 & 121  & 135 & 10 &  80 &  89\\
October   & 31 & 272   & 302 & 16 & 136  & 151 & 10 &  90 &  99\\
November  & 31 & 303   & 333 & 15 & 152  & 166 & 10 & 100 & 109\\
December  & 32 & 334   & 365 & 16 & 167  & 182 & 11 & 110 & 120\\
\hline
\enddata
\tablenotetext{1}{Calendar for when sidereal day length is twice modern Earth's.}
\tablenotetext{2}{Calendar for all sidereal day lengths greater than two Earth sidereal days.}
\end{deluxetable}

\begin{deluxetable}{|l|r|r|r|r|r|r|r|}
\tabletypesize{\footnotesize}
\tablecaption{Additional calendar day lengths in multiples of Earth sidereal days\label{tablecal3a}}
\tablehead{\multicolumn{1}{|l|}{Name} & \multicolumn{1}{r|}{4 days} & \multicolumn{1}{r|}{8 days} & \multicolumn{1}{r|}{16 days} & \multicolumn{1}{r|}{32 days} & \multicolumn{1}{r|}{64 days} & \multicolumn{1}{r|}{128 days} & \multicolumn{1}{r|}{256 days}}
\startdata
\hline
Mean solar day           & 4.032799   & 8.156210 & 16.684998& 34.967671& 77.339399 & 196.228155 & 848.066511\\
Sidereal Rotation Period & 3.988759   & 7.978059 & 15.956119& 31.912537&  63.825075& 127.650149&255.300299\\
Sidereal Orbital Period  & 365.256369 & - & - & - & - & - & -\\
Solar Days per year      & 90.571429  & 44.782609& 21.891304& 10.445545&   4.722772&  1.861386&   0.430693\\
\enddata
\end{deluxetable}

The movement of the substellar point month-by-month is an intuitive way for the
reader to understand the implications of the 120 day calendar. In Table
\ref{tab:substellarpoint} we show an example of the movement of the substellar
point in each monthly output for each of the runs. We do not show results for
rotation rates faster than X032 as the substellar point is too smeared out over
monthly output to be visually meaningful.

\startlongtable
\begin{deluxetable}{|m{.7in}|m{1.2in}m{1.2in}m{1.2in}m{1.3in}|}
\tablecaption{Substellar point in monthly output\label{tab:substellarpoint}}
\tablehead{
\multicolumn{1}{|l|}{Month} & 
\multicolumn{1}{c}{X032}   & 
\multicolumn{1}{c}{X064}   &
\multicolumn{1}{c}{X128}   &
\multicolumn{1}{c|}{X256} 
}
\startdata
January &
\includegraphics[width=1.3in,trim=0 40 0 65,clip]{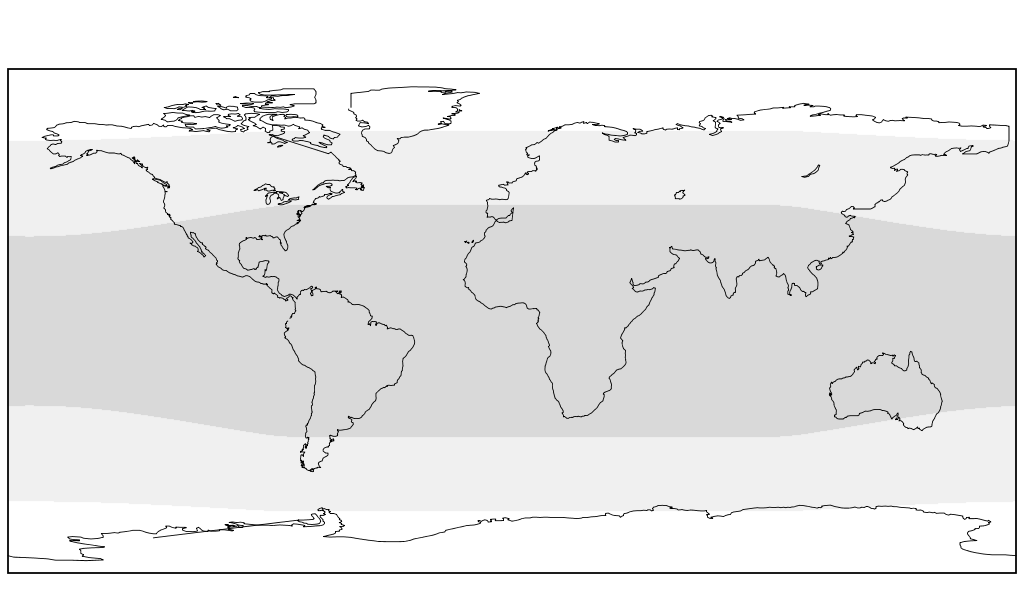} &
\includegraphics[width=1.3in,trim=0 40 0 65,clip]{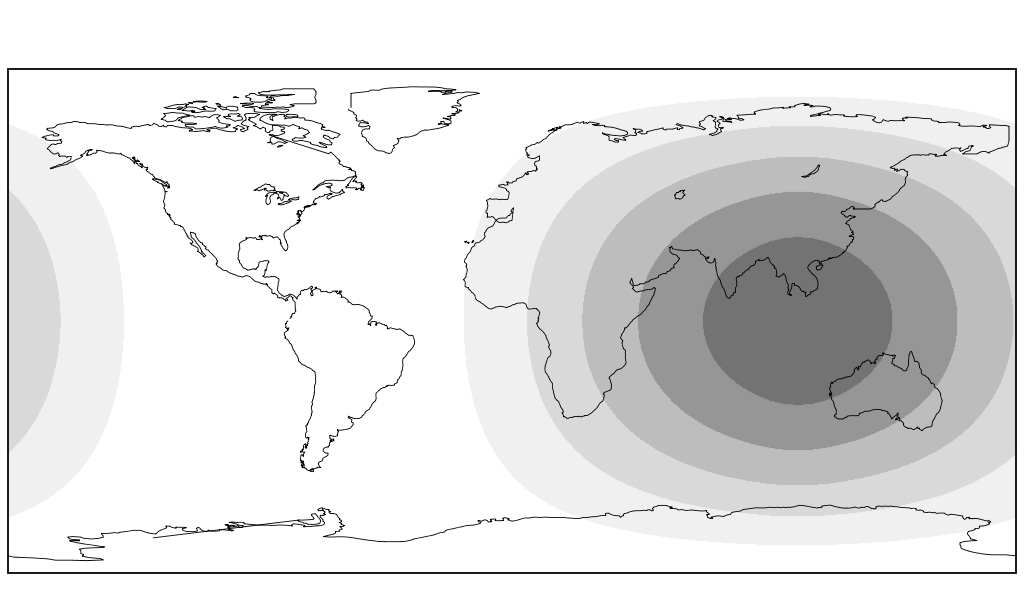} &
\includegraphics[width=1.3in,trim=0 40 0 65,clip]{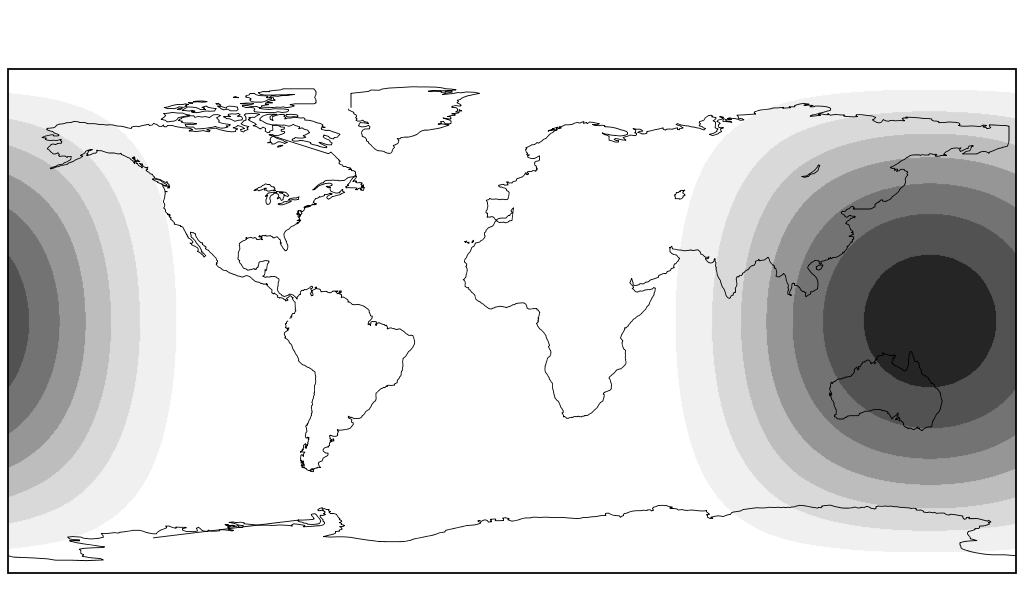} &
\includegraphics[width=1.3in,trim=0 40 0 65,clip]{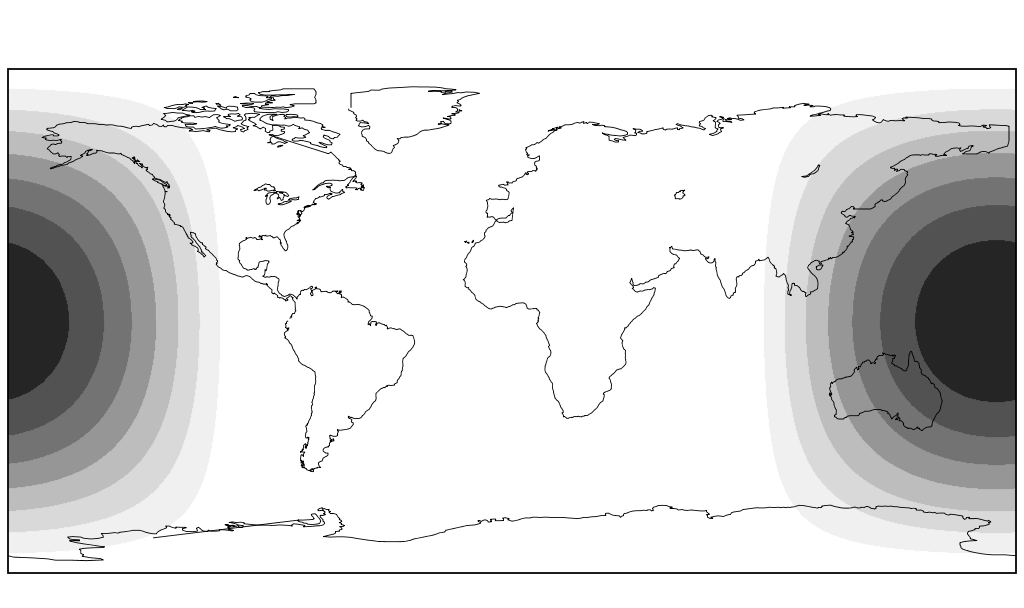} \\
February &
\includegraphics[width=1.3in,trim=0 40 0 75,clip]{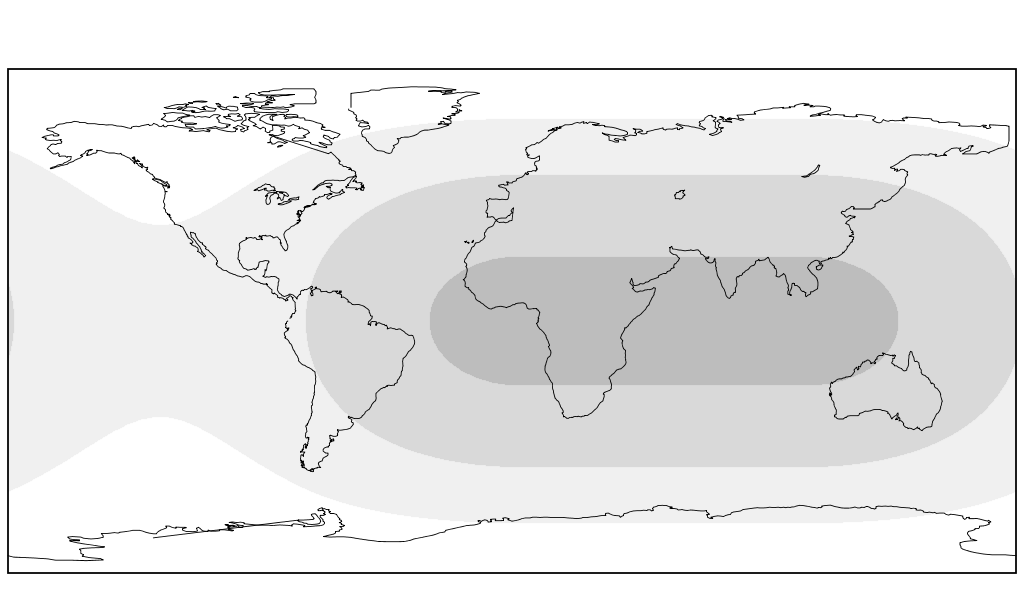} &
\includegraphics[width=1.3in,trim=0 40 0 75,clip]{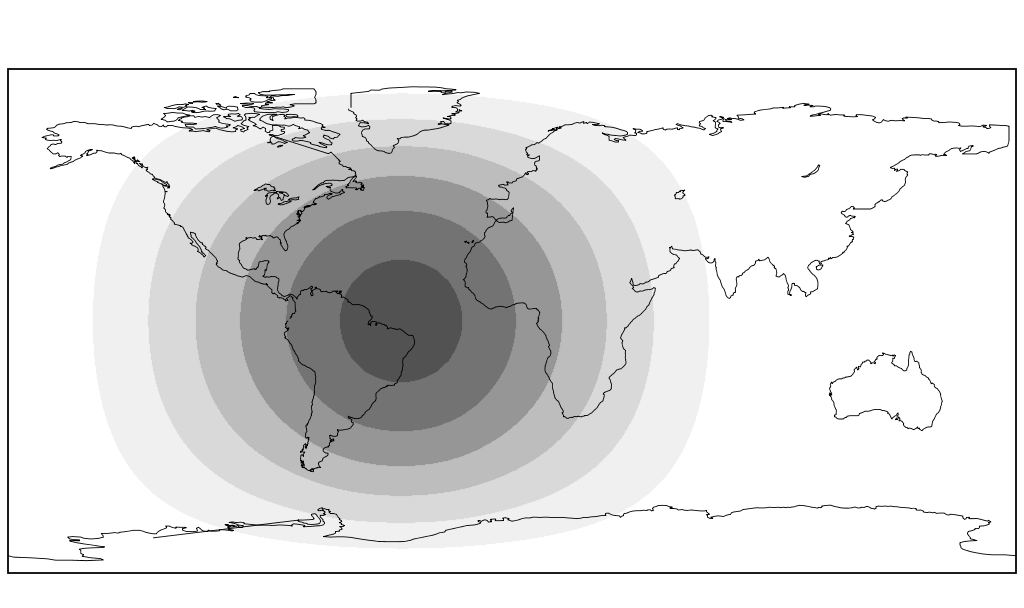} &
\includegraphics[width=1.3in,trim=0 40 0 75,clip]{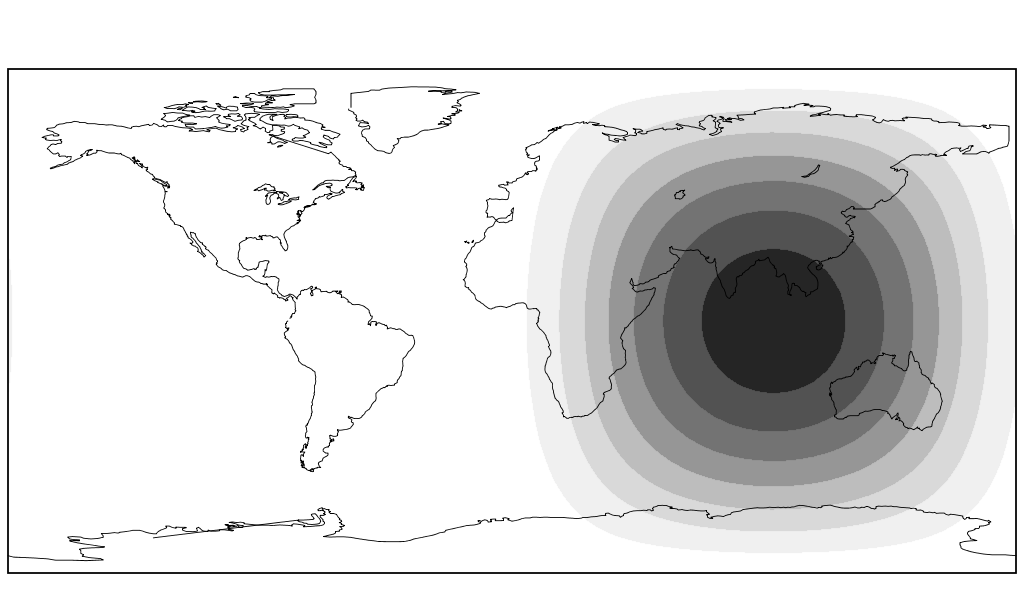} &
\includegraphics[width=1.3in,trim=0 40 0 75,clip]{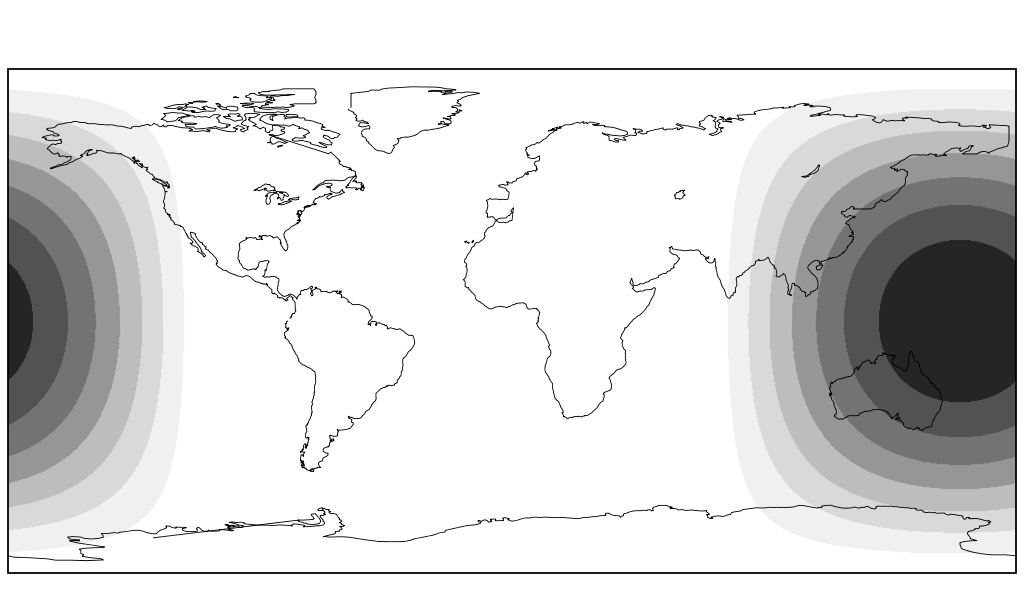} \\
March &
\includegraphics[width=1.3in,trim=0 40 0 65,clip]{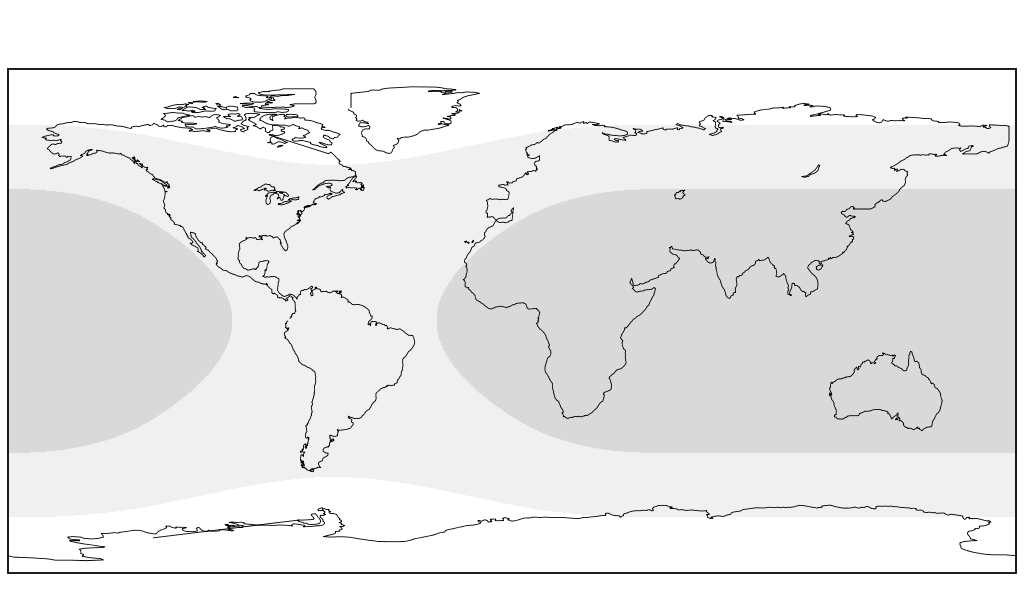} &
\includegraphics[width=1.3in,trim=0 40 0 65,clip]{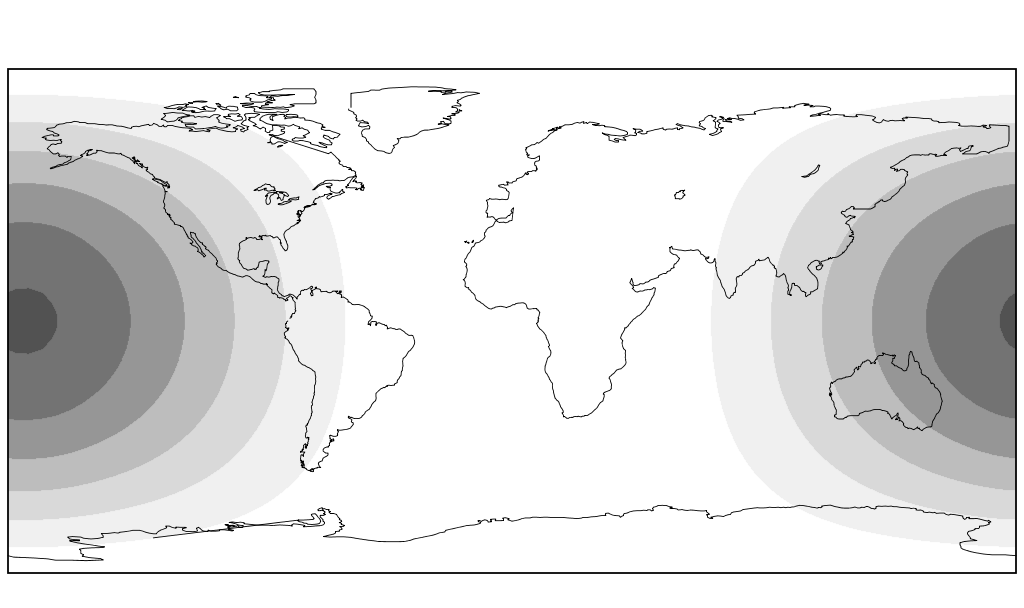} &
\includegraphics[width=1.3in,trim=0 40 0 65,clip]{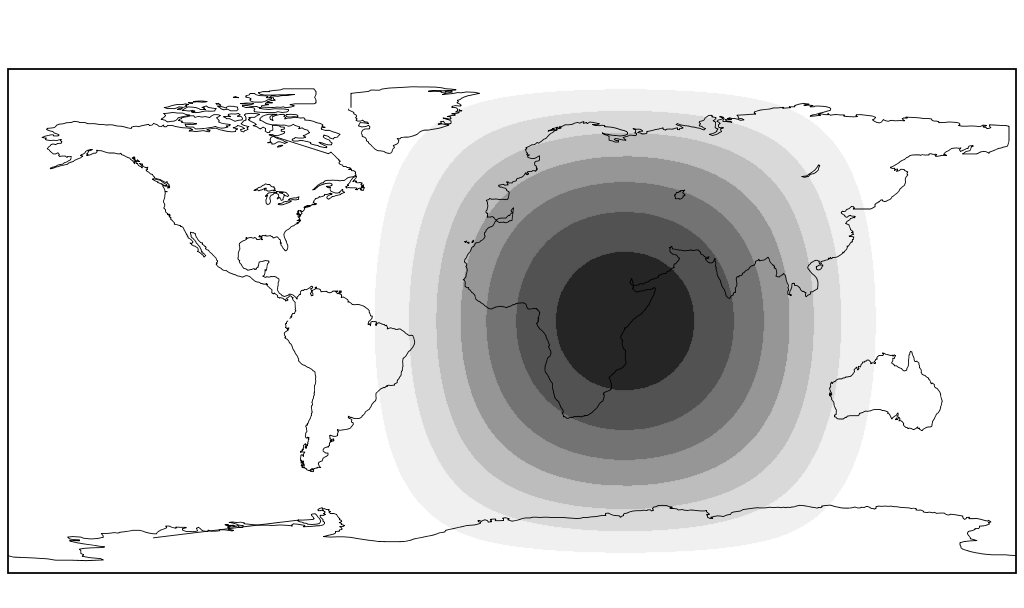} &
\includegraphics[width=1.3in,trim=0 40 0 65,clip]{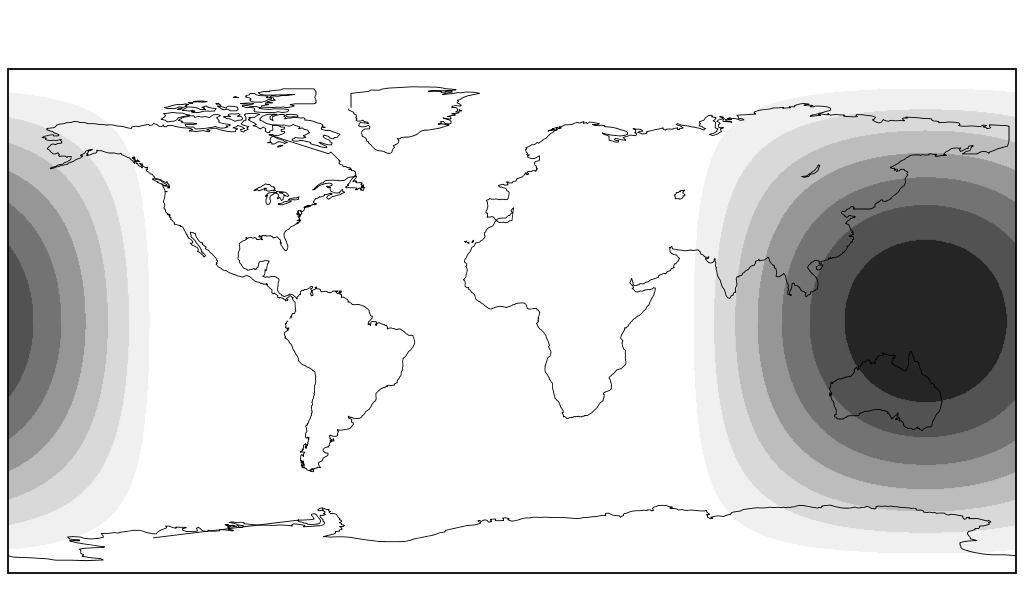} \\
April &
\includegraphics[width=1.3in,trim=0 40 0 65,clip]{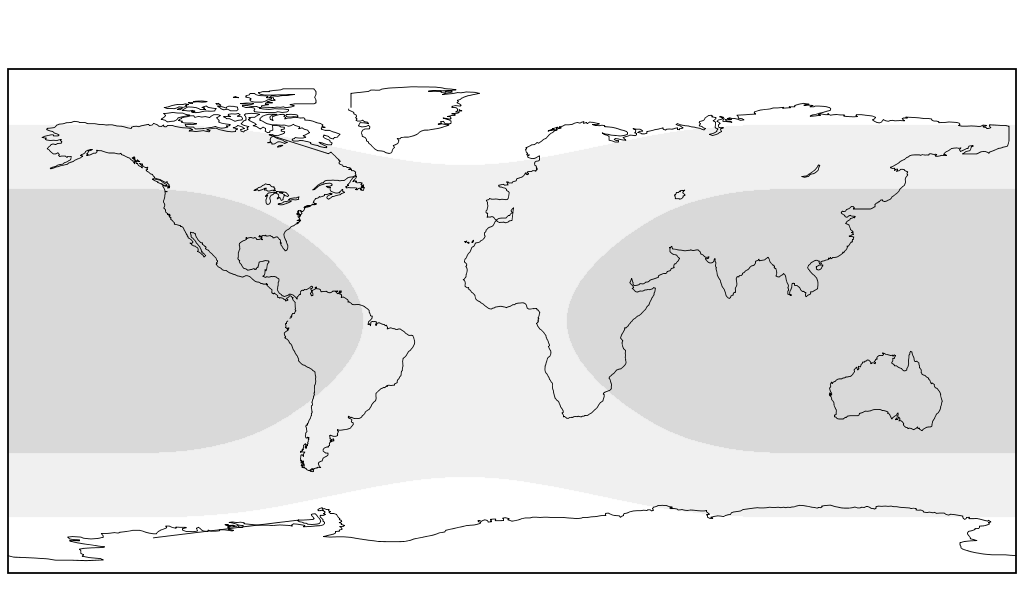} &
\includegraphics[width=1.3in,trim=0 40 0 65,clip]{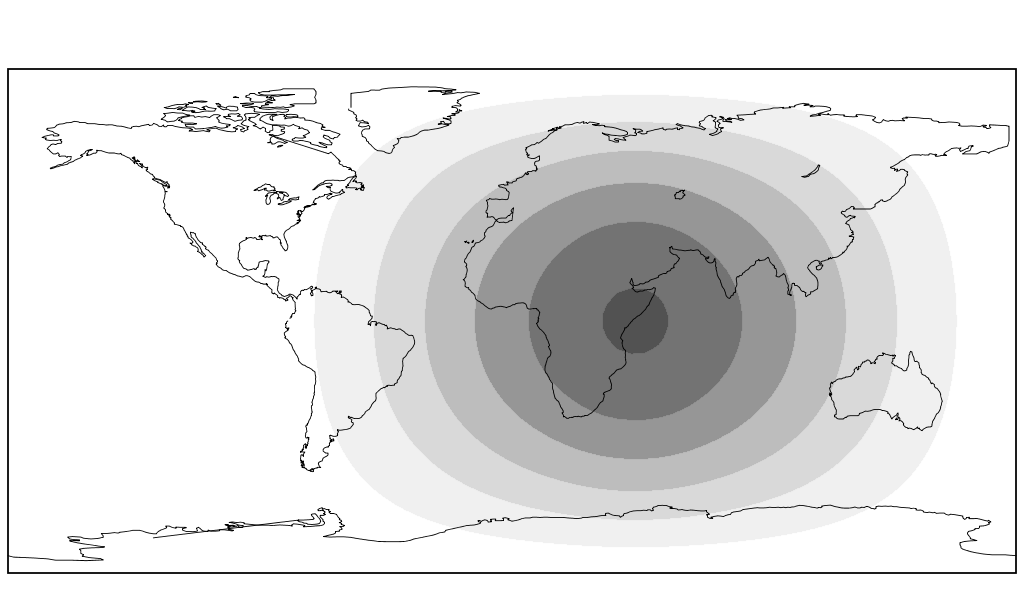} &
\includegraphics[width=1.3in,trim=0 40 0 65,clip]{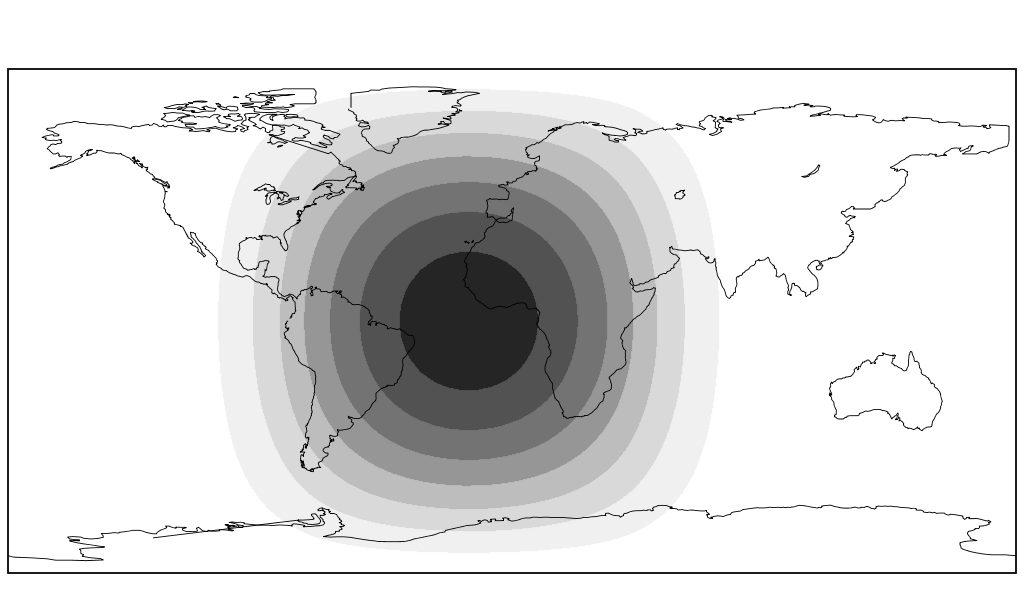} &
\includegraphics[width=1.3in,trim=0 40 0 65,clip]{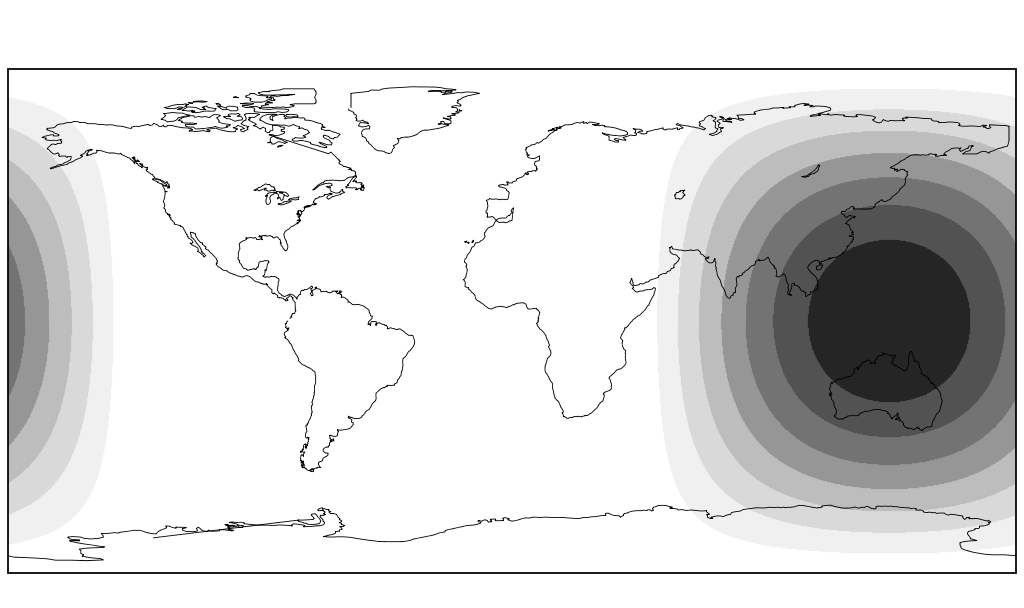} \\
May &
\includegraphics[width=1.3in,trim=0 40 0 65,clip]{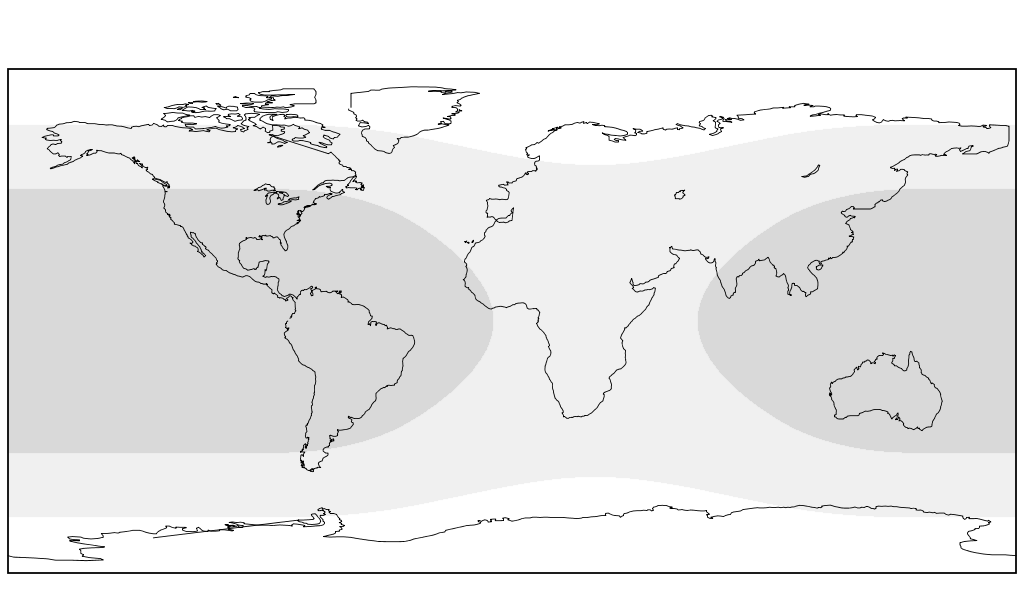} &
\includegraphics[width=1.3in,trim=0 40 0 65,clip]{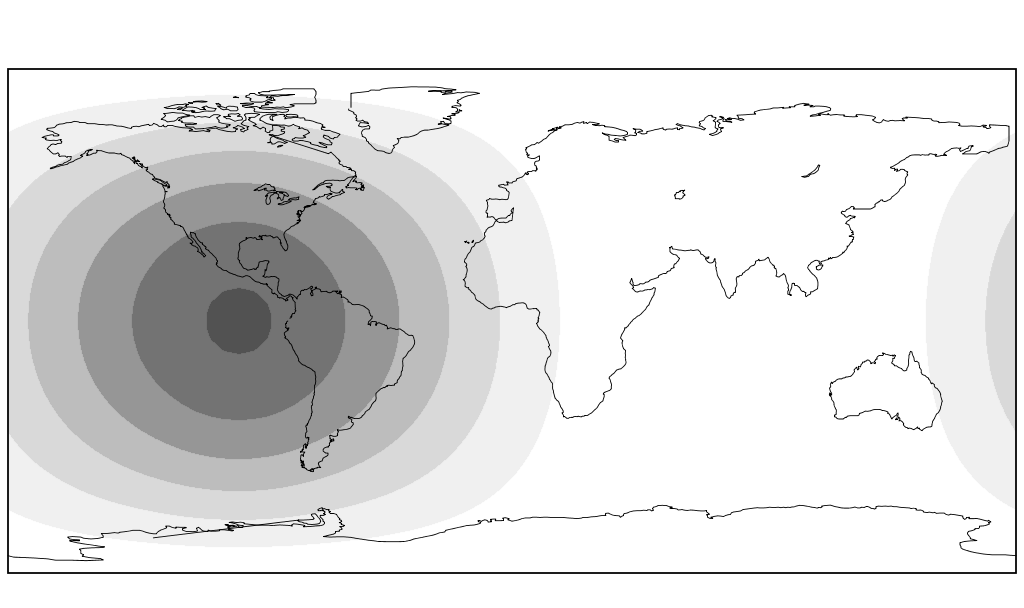} &
\includegraphics[width=1.3in,trim=0 40 0 65,clip]{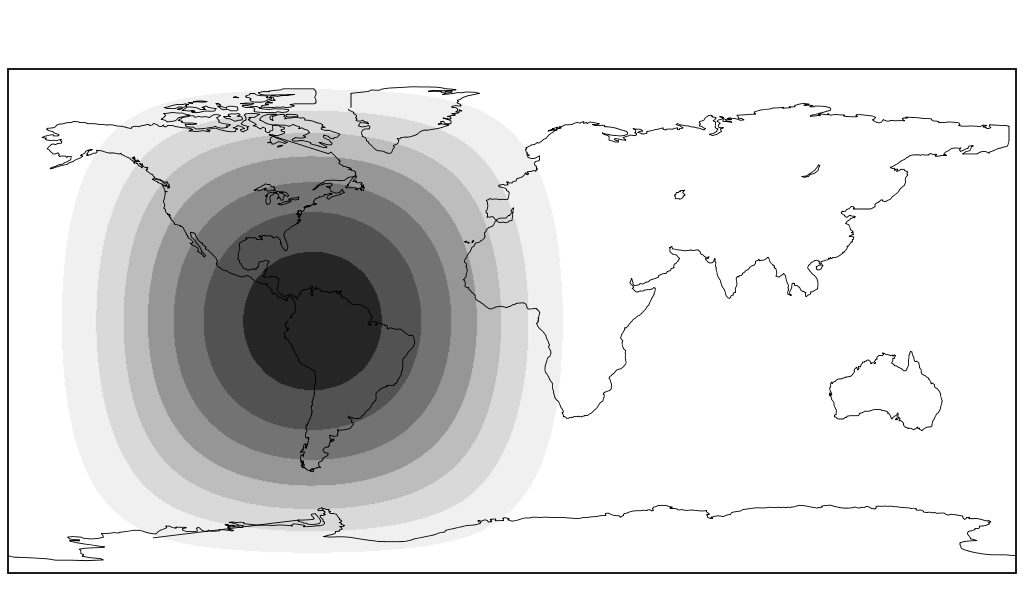} &
\includegraphics[width=1.3in,trim=0 40 0 65,clip]{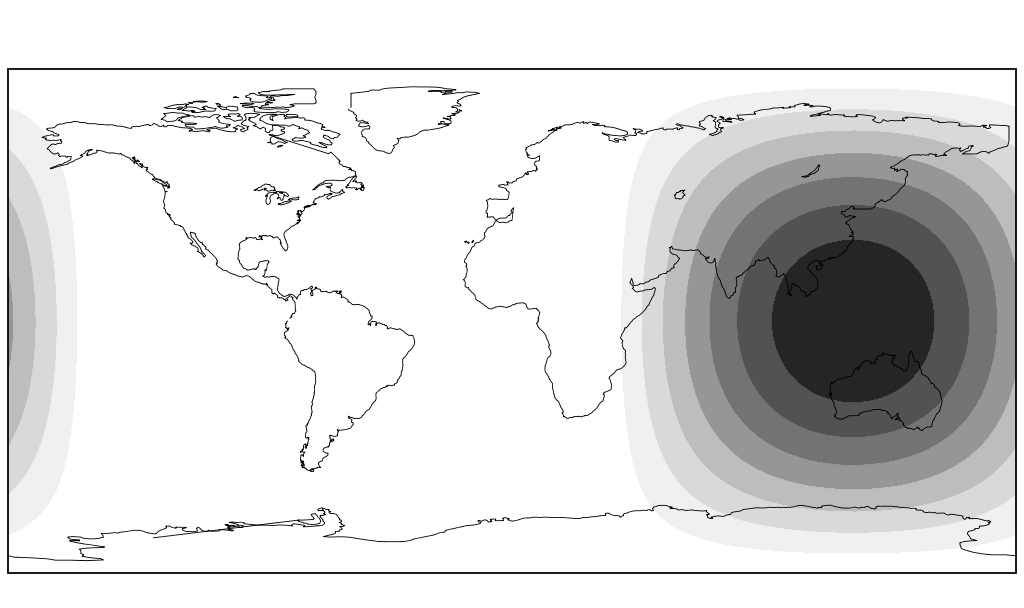} \\
June &
\includegraphics[width=1.3in,trim=0 40 0 65,clip]{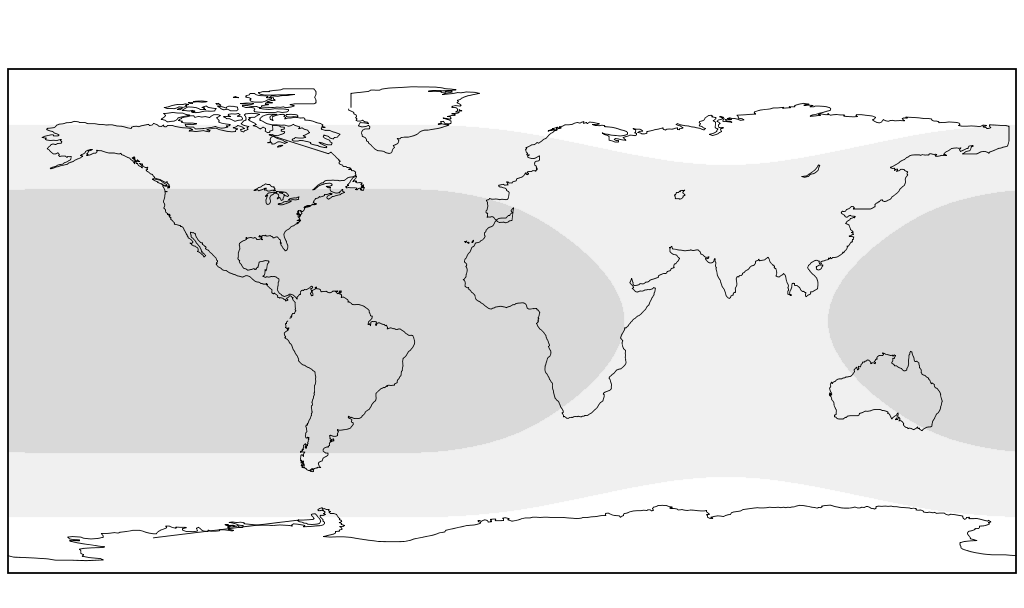} &
\includegraphics[width=1.3in,trim=0 40 0 65,clip]{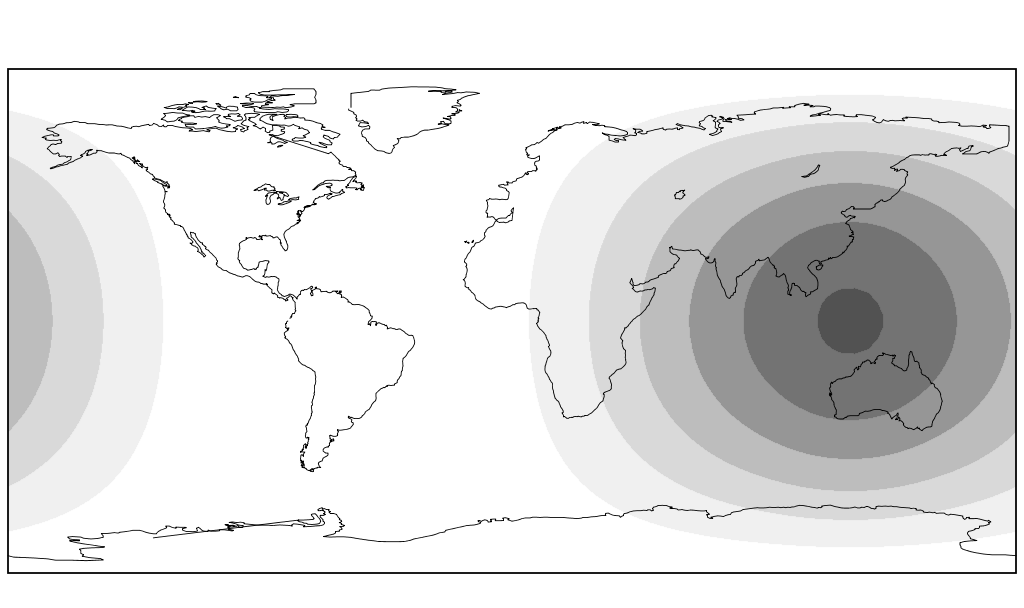} &
\includegraphics[width=1.3in,trim=0 40 0 65,clip]{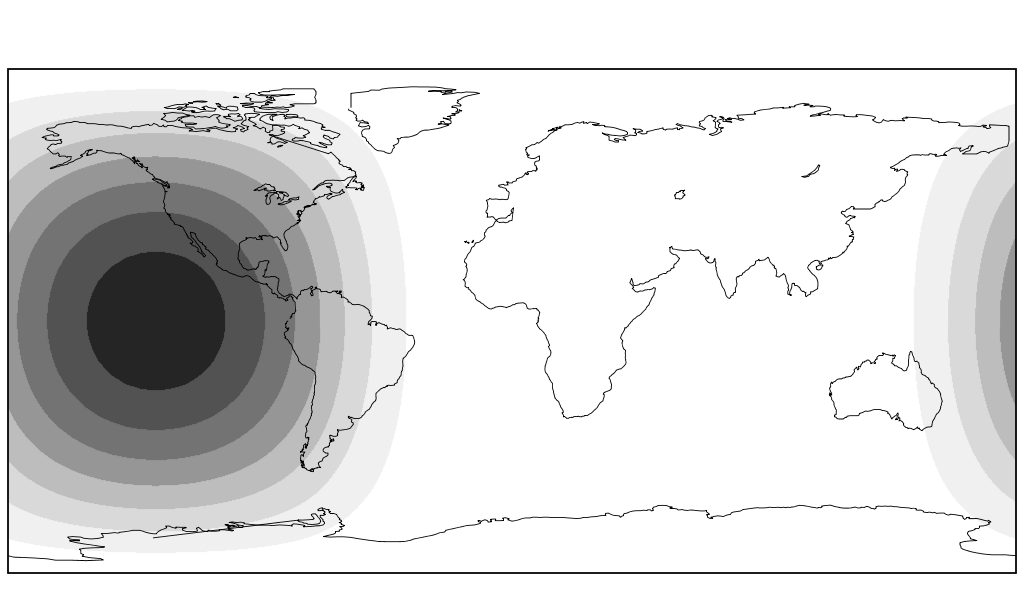} &
\includegraphics[width=1.3in,trim=0 40 0 65,clip]{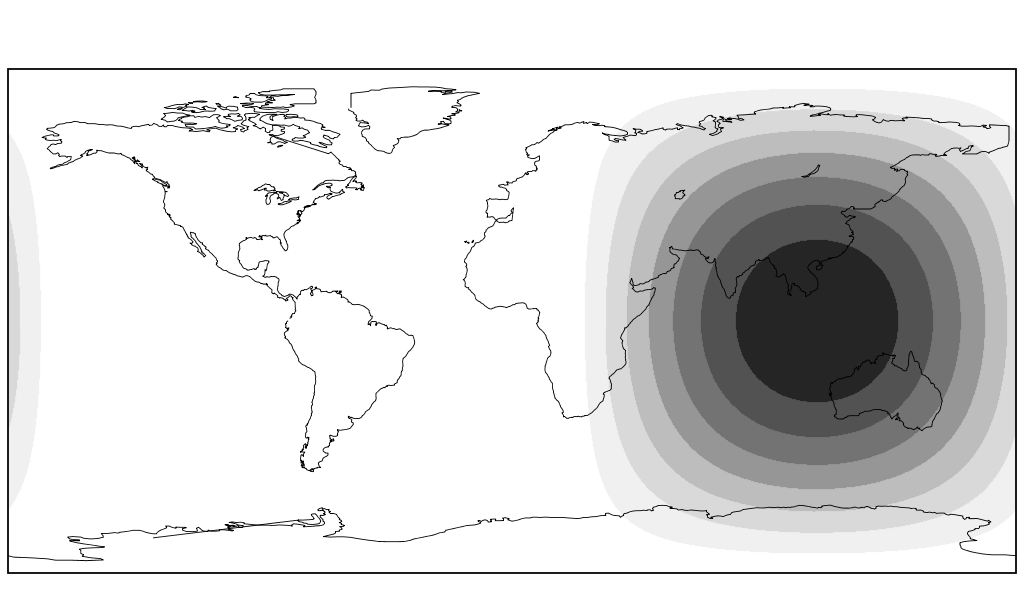} \\
July &
\includegraphics[width=1.3in,trim=0 40 0 65,clip]{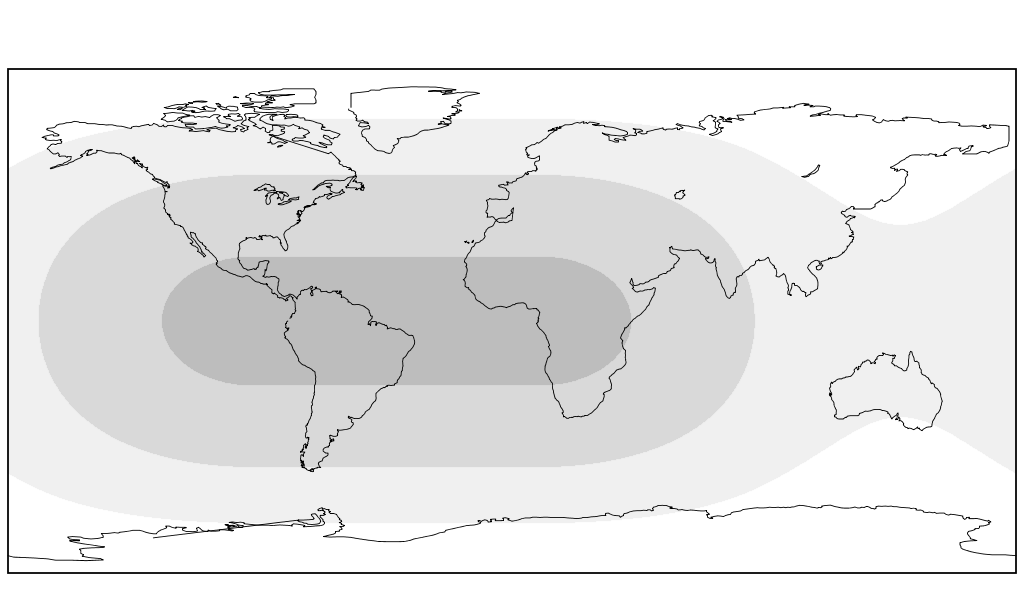} &
\includegraphics[width=1.3in,trim=0 40 0 65,clip]{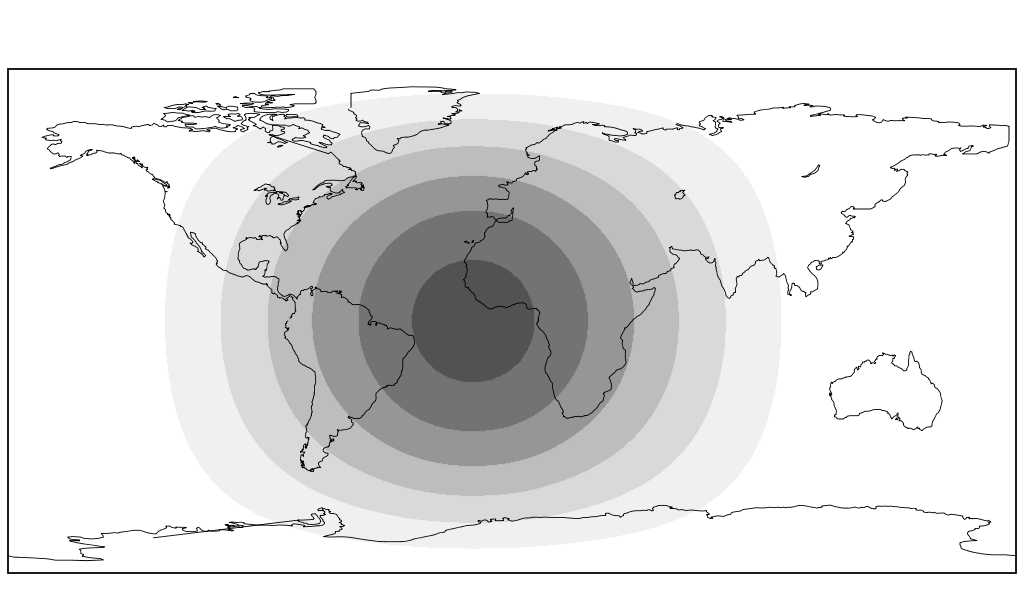} &
\includegraphics[width=1.3in,trim=0 40 0 65,clip]{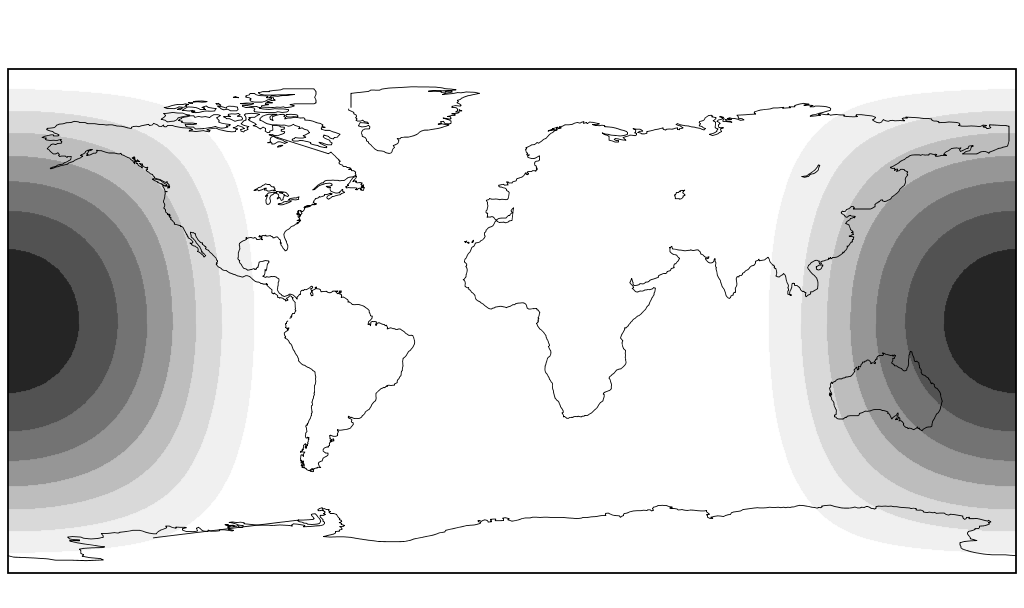} &
\includegraphics[width=1.3in,trim=0 40 0 65,clip]{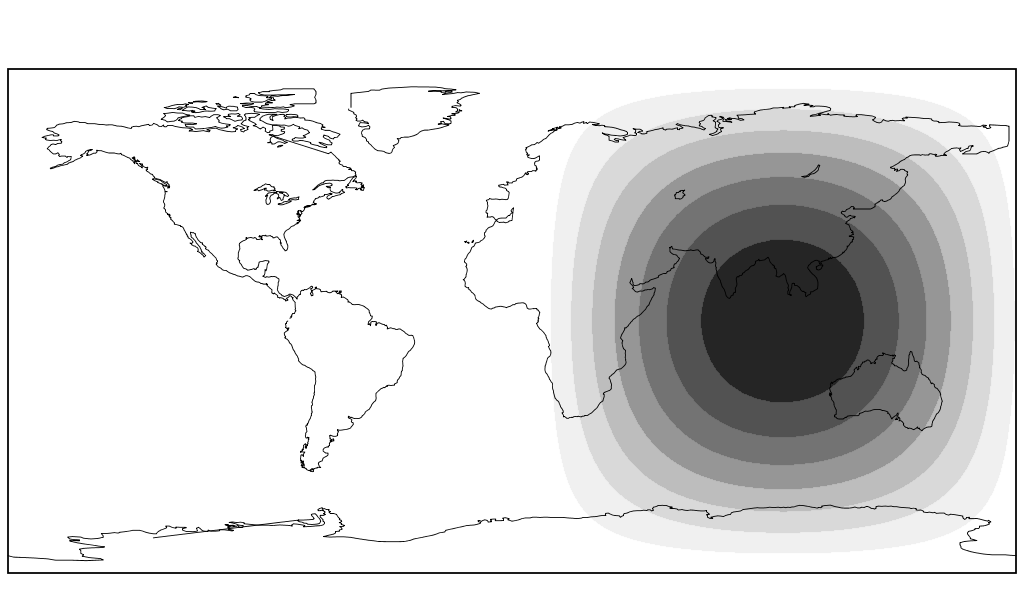} \\
August &
\includegraphics[width=1.3in,trim=0 40 0 65,clip]{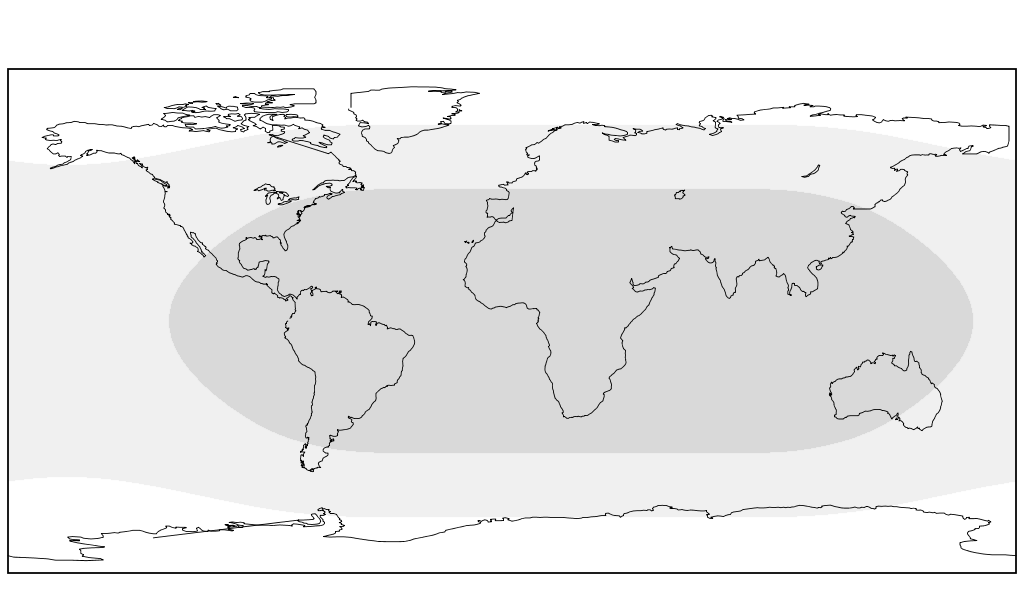} &
\includegraphics[width=1.3in,trim=0 40 0 65,clip]{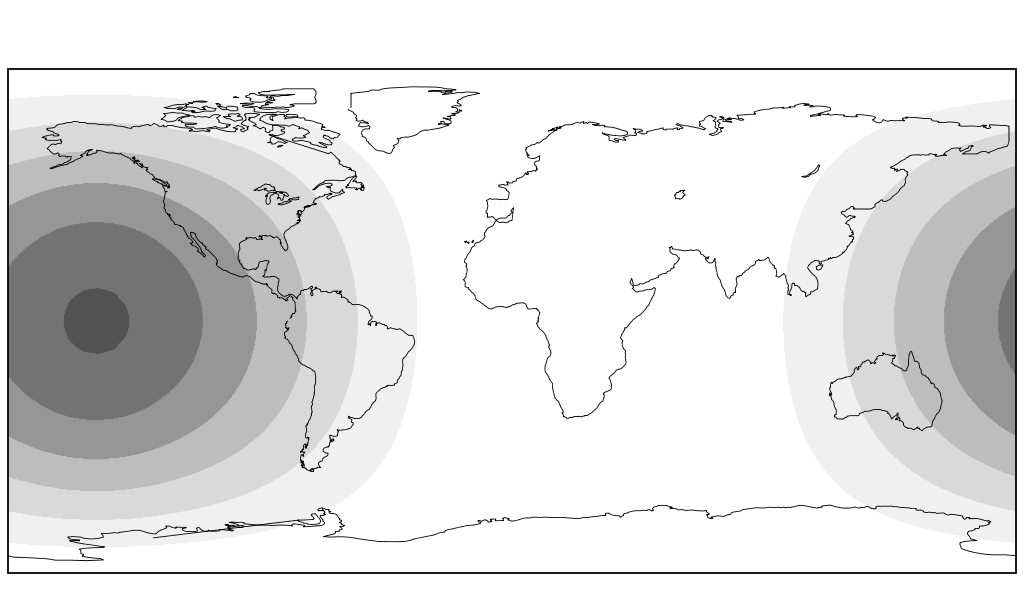} &
\includegraphics[width=1.3in,trim=0 40 0 65,clip]{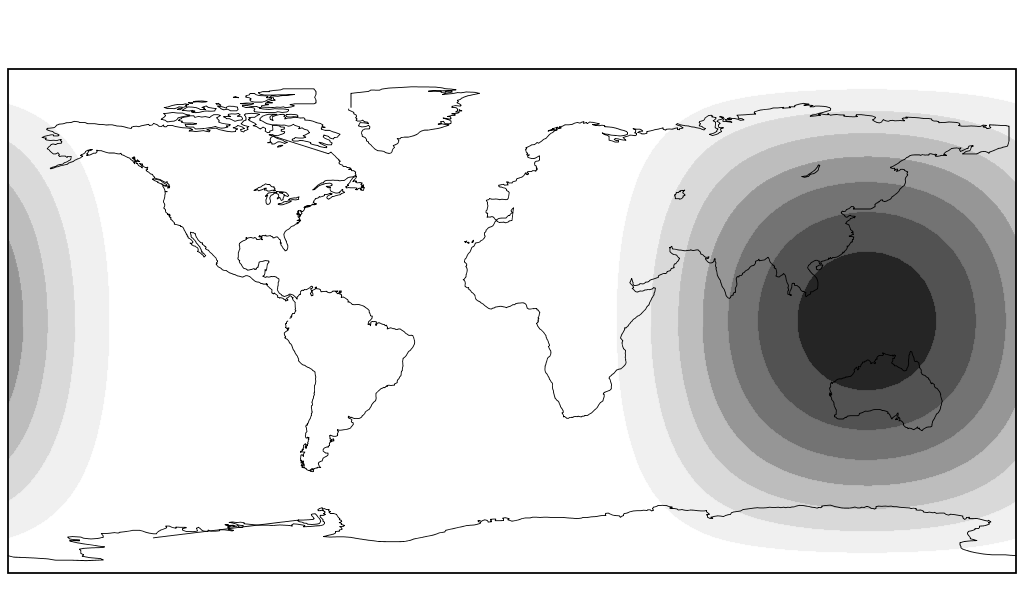} &
\includegraphics[width=1.3in,trim=0 40 0 65,clip]{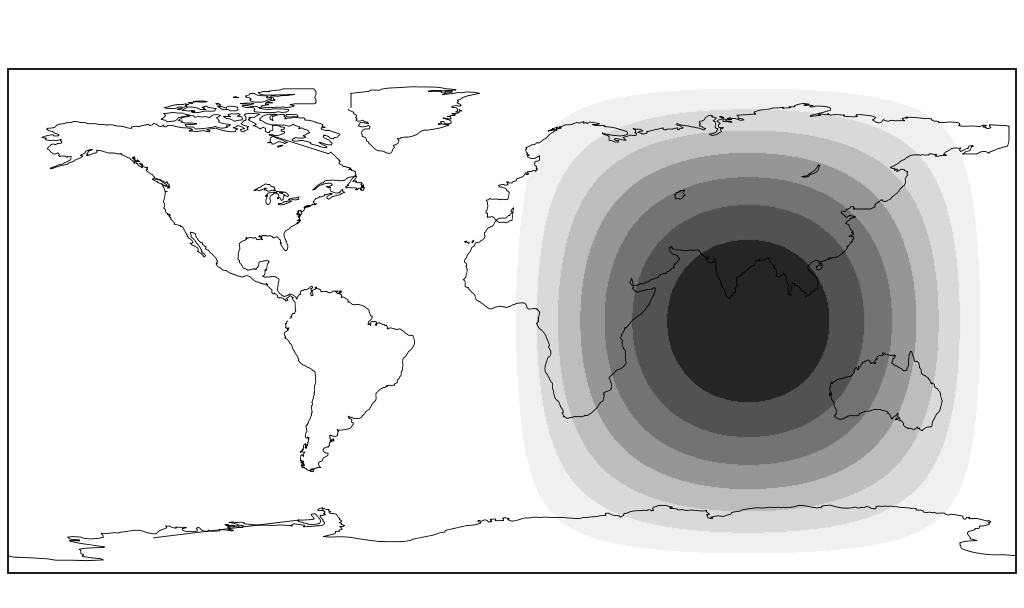} \\
September &
\includegraphics[width=1.3in,trim=0 40 0 65,clip]{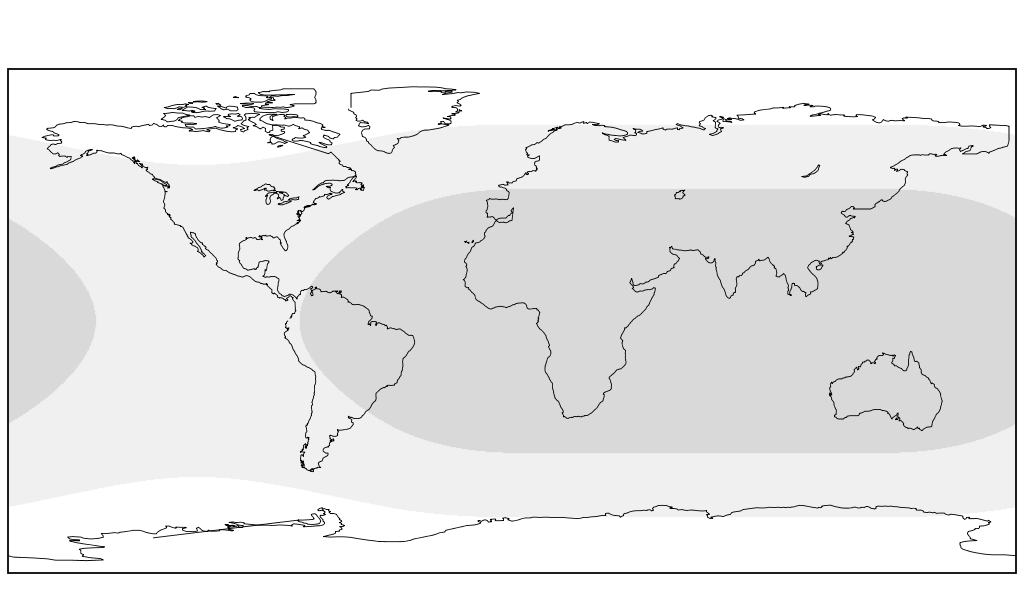} &
\includegraphics[width=1.3in,trim=0 40 0 65,clip]{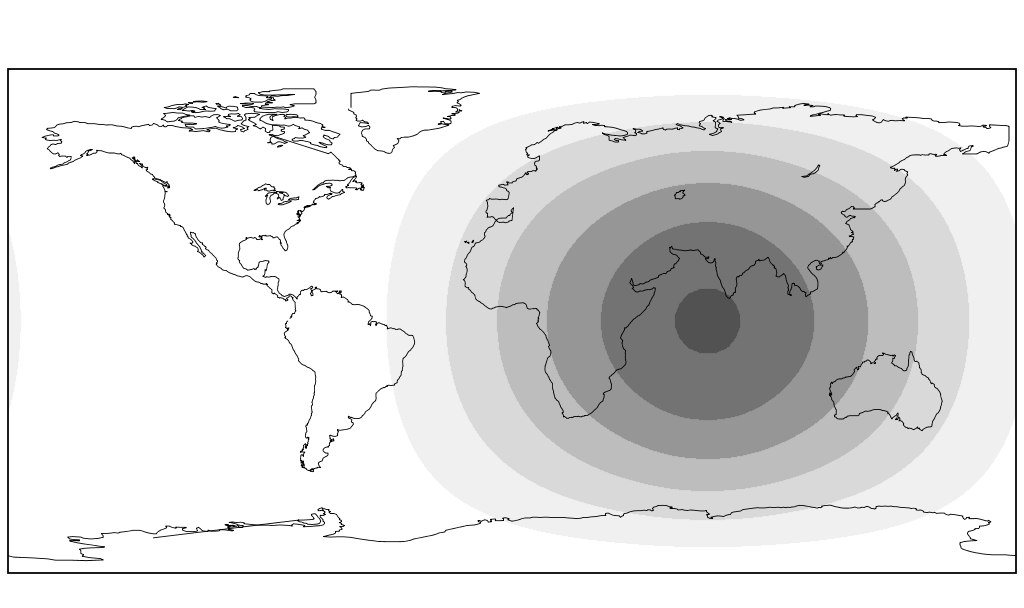} &
\includegraphics[width=1.3in,trim=0 40 0 65,clip]{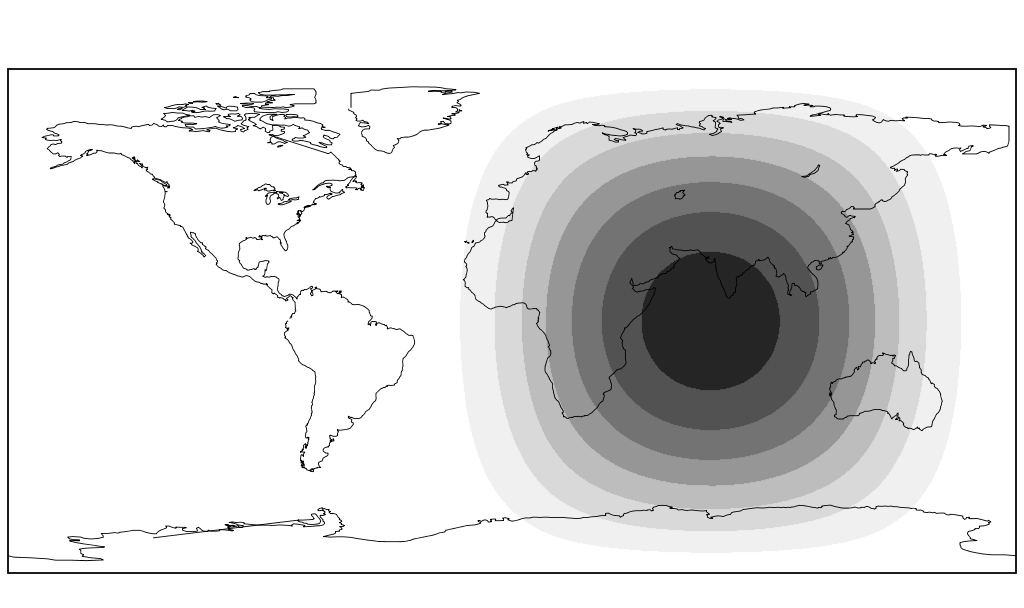} &
\includegraphics[width=1.3in,trim=0 40 0 65,clip]{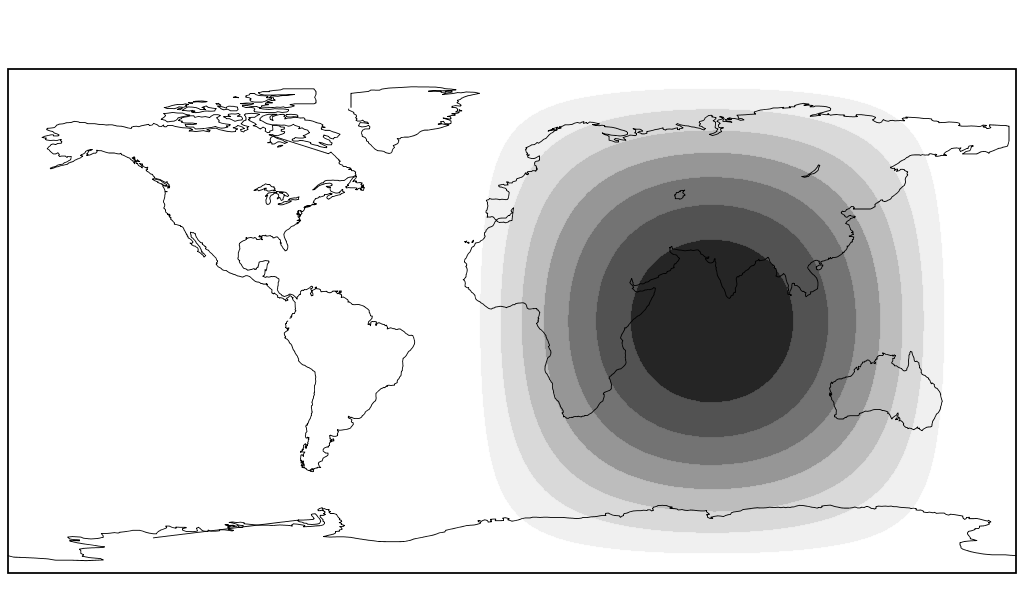} \\
October &
\includegraphics[width=1.3in,trim=0 40 0 65,clip]{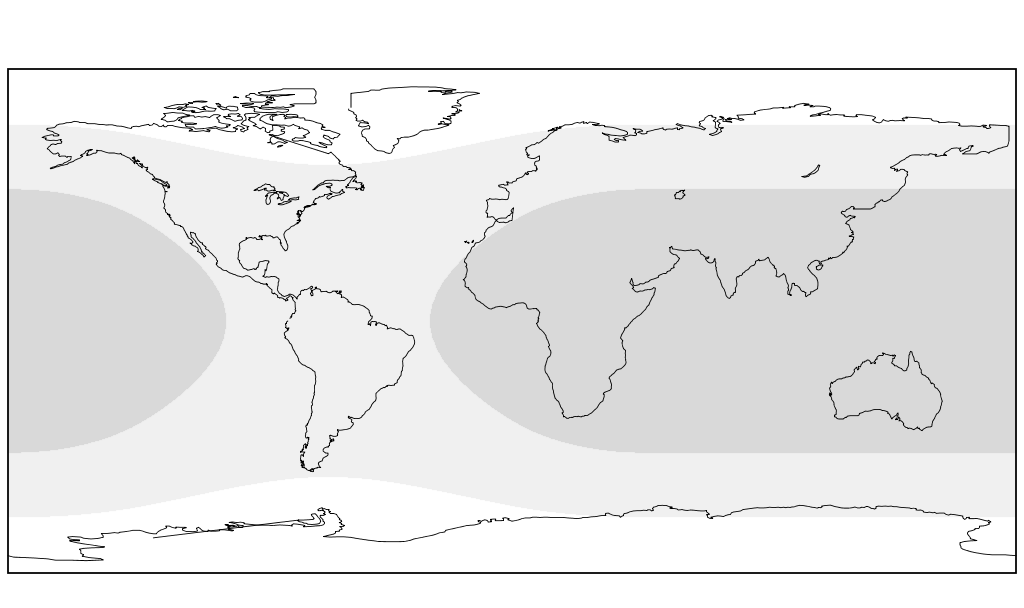} &
\includegraphics[width=1.3in,trim=0 40 0 65,clip]{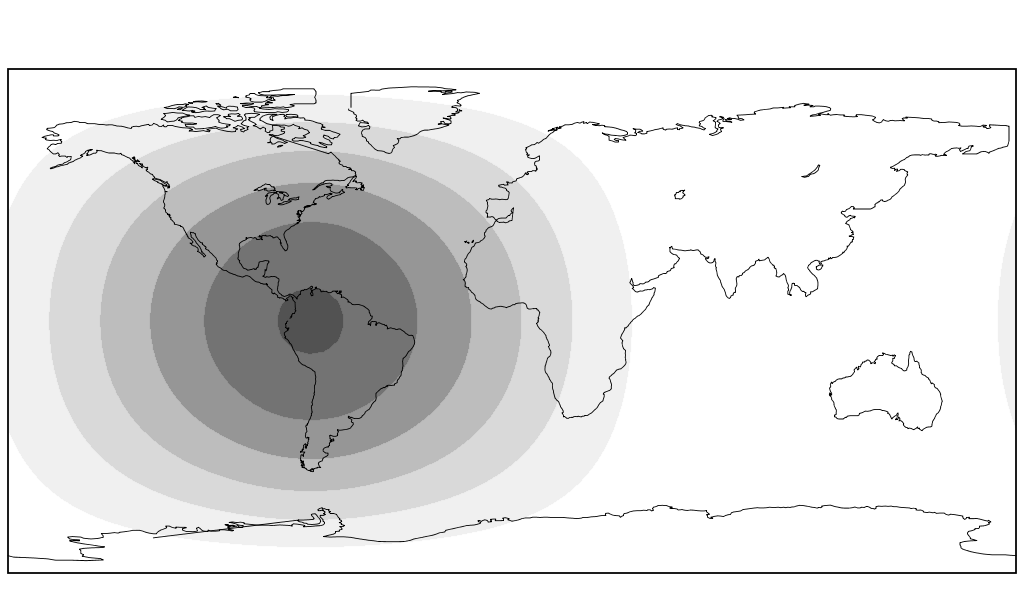} &
\includegraphics[width=1.3in,trim=0 40 0 65,clip]{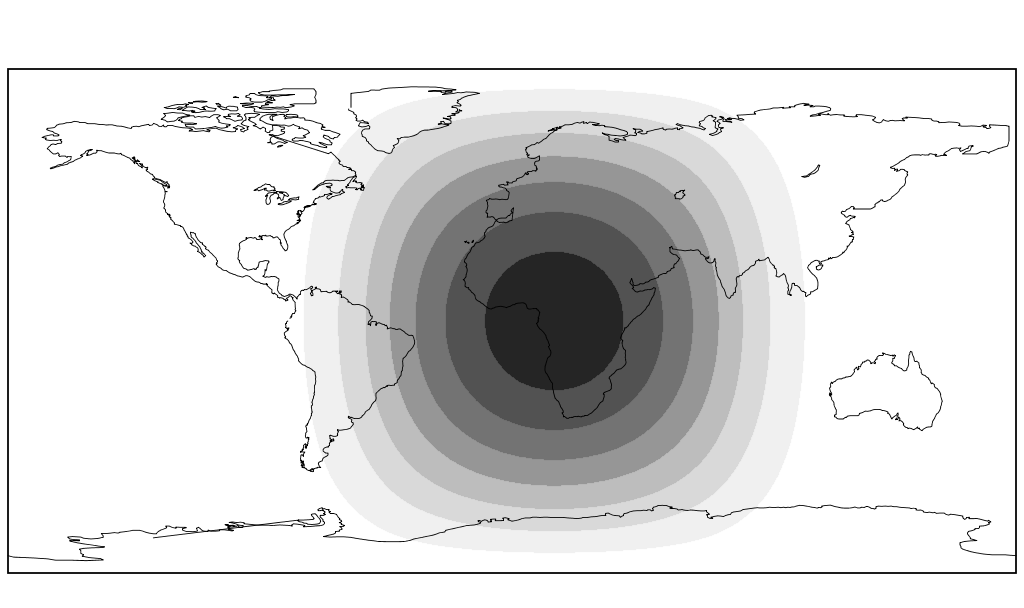} &
\includegraphics[width=1.3in,trim=0 40 0 65,clip]{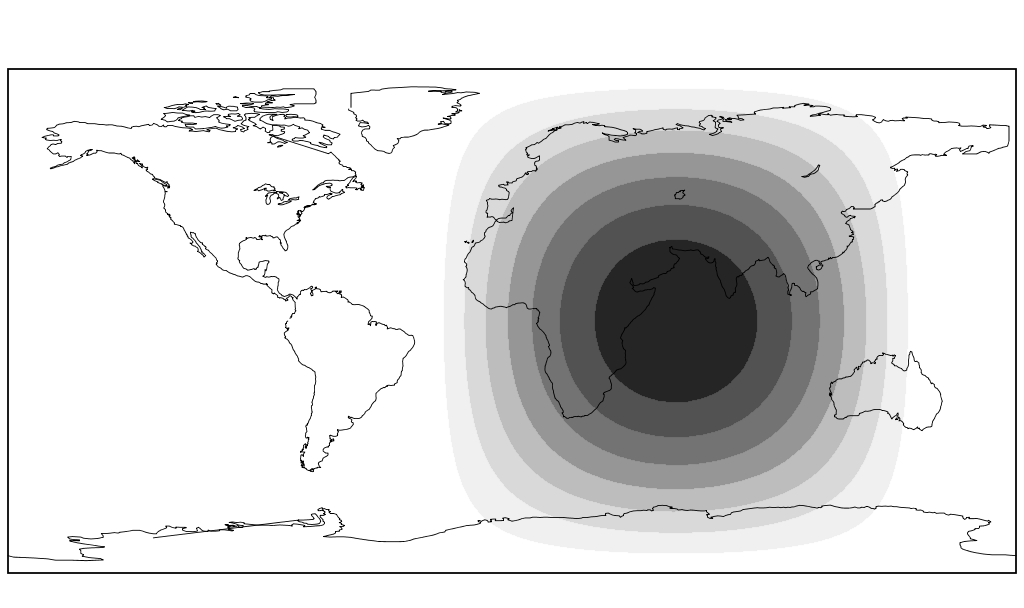} \\
November &
\includegraphics[width=1.3in,trim=0 40 0 65,clip]{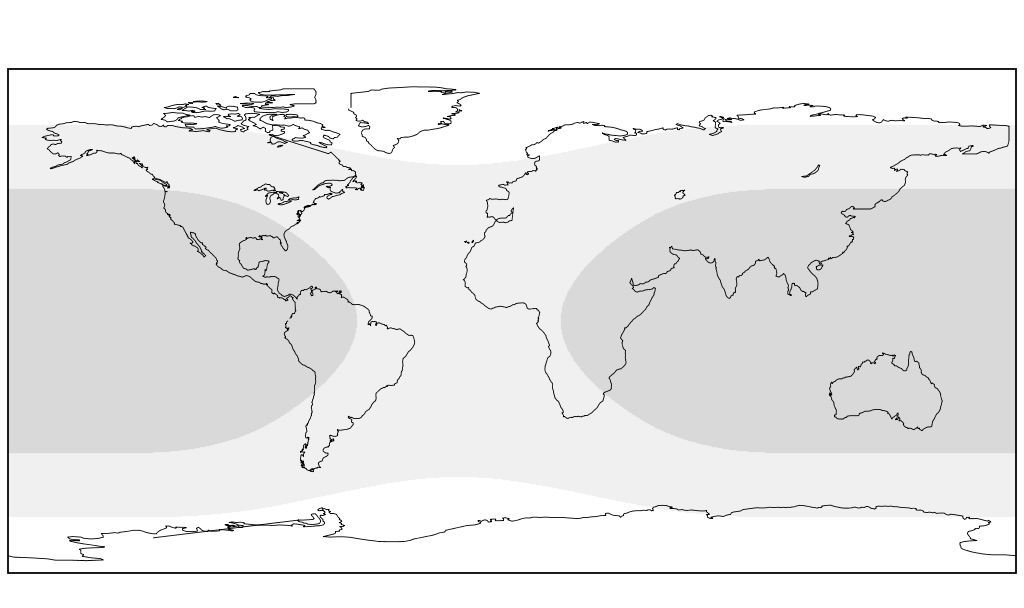} &
\includegraphics[width=1.3in,trim=0 40 0 65,clip]{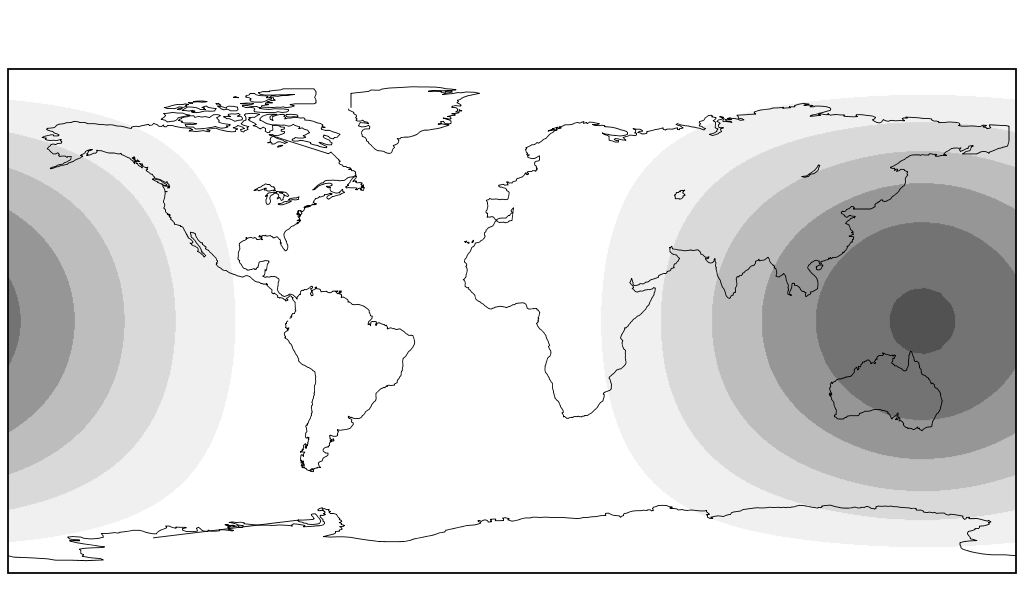} &
\includegraphics[width=1.3in,trim=0 40 0 65,clip]{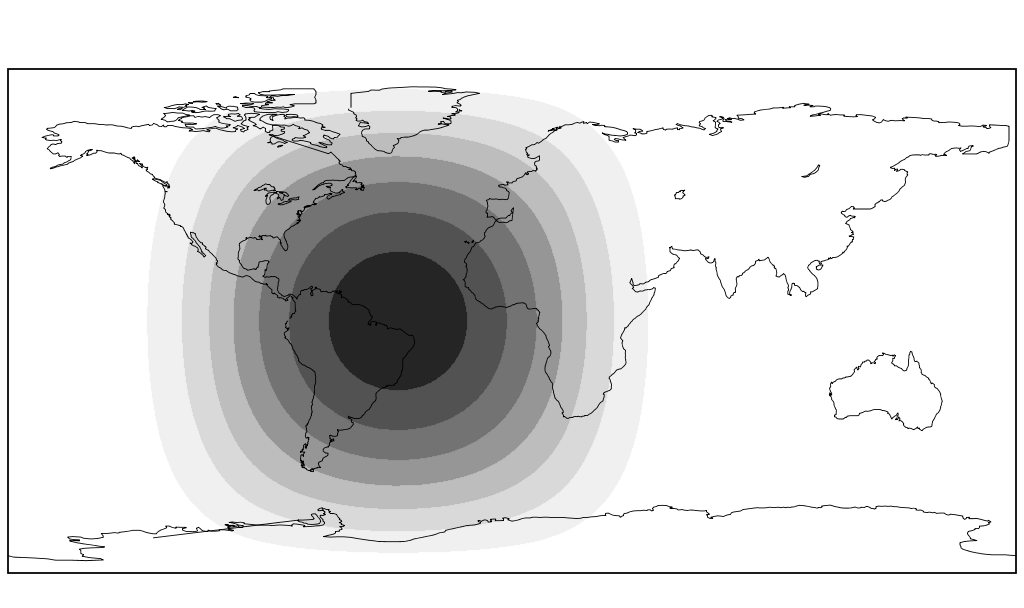} &
\includegraphics[width=1.3in,trim=0 40 0 65,clip]{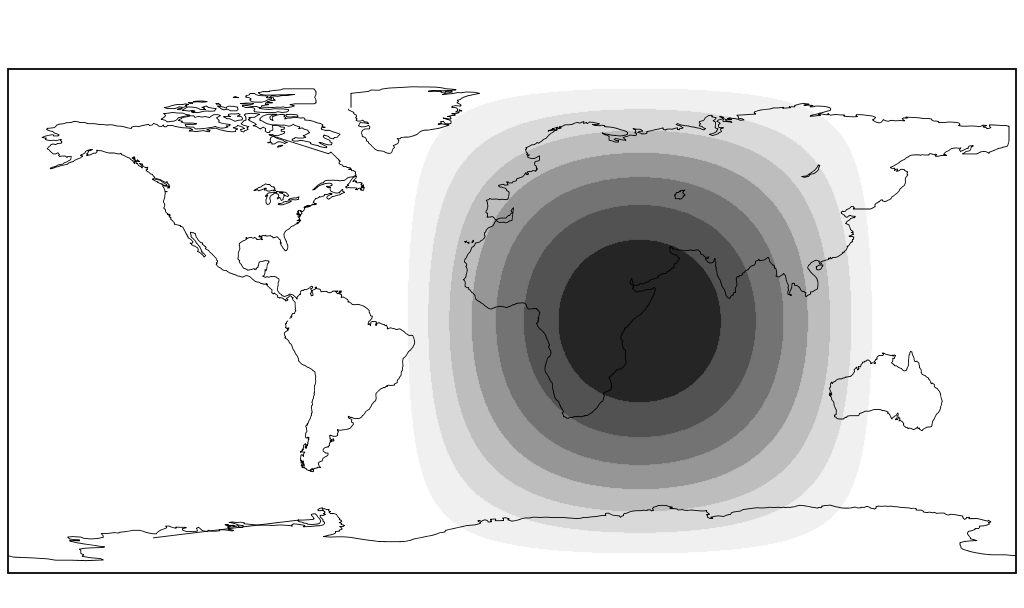} \\
December &
\includegraphics[width=1.3in,trim=0 40 0 65,clip]{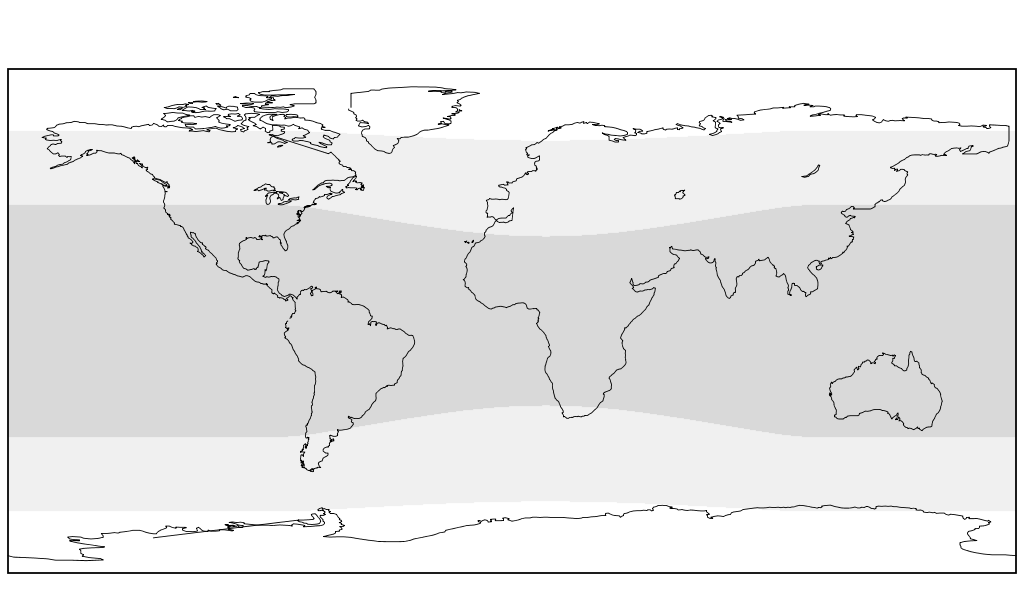} &
\includegraphics[width=1.3in,trim=0 40 0 65,clip]{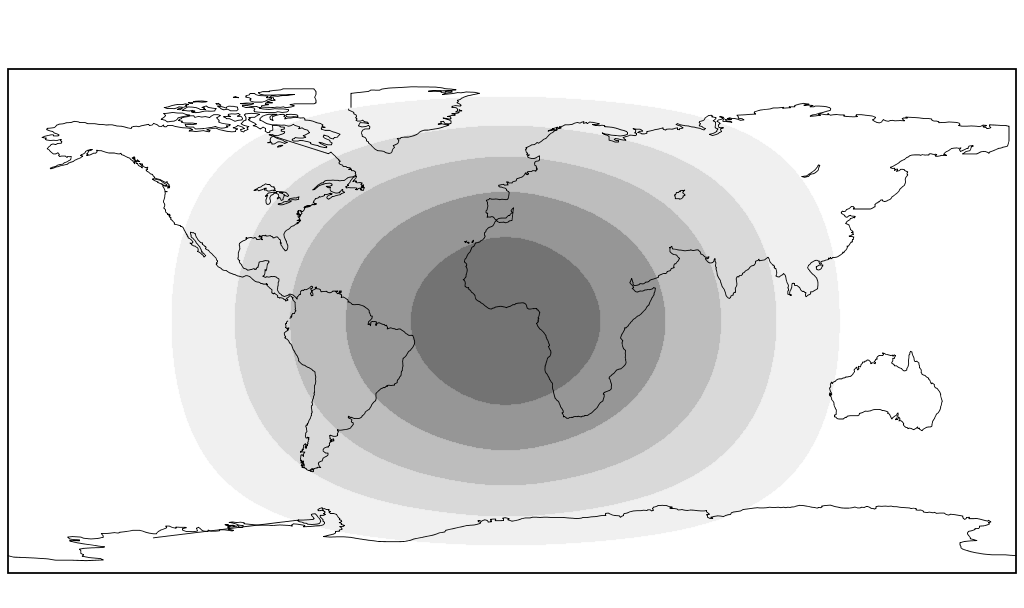} &
\includegraphics[width=1.3in,trim=0 40 0 65,clip]{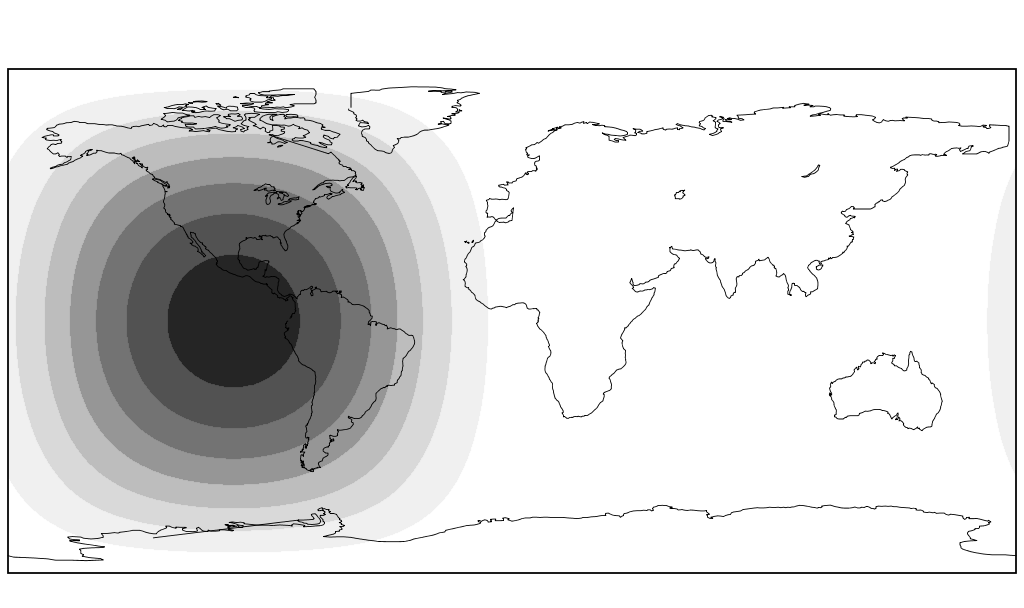} &
\includegraphics[width=1.3in,trim=0 40 0 65,clip]{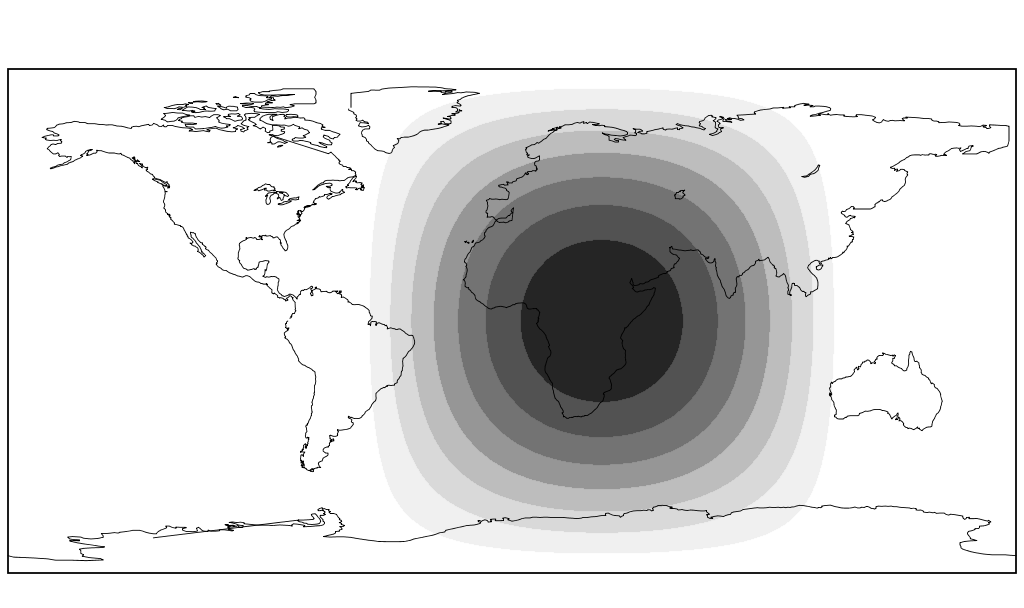} \\
\hline
\enddata
\end{deluxetable}

\section{Equilibrium}\label{appendix:equilibrium}

\begin{figure}[!htb]
\includegraphics[scale=0.35]{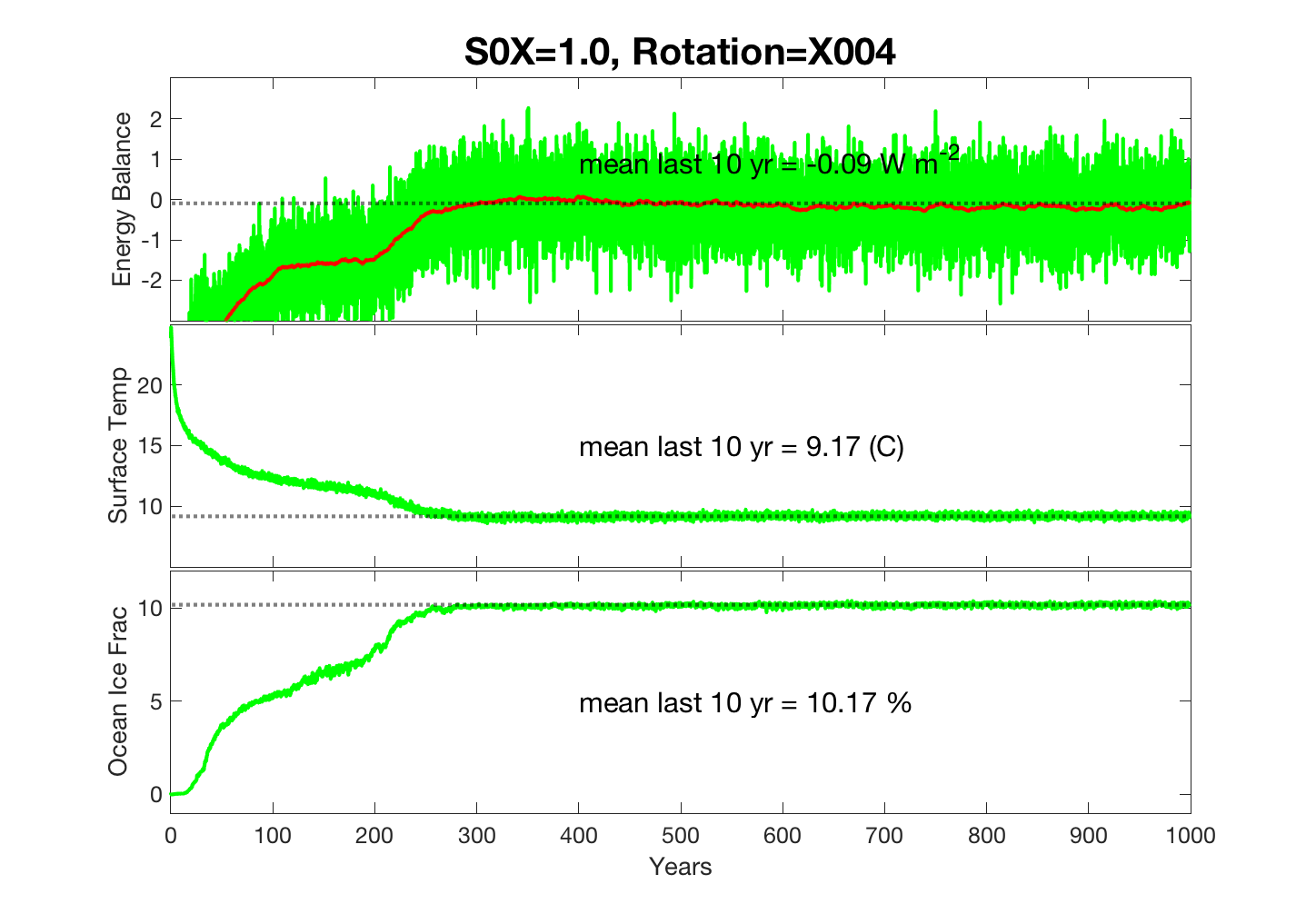}
\includegraphics[scale=0.35]{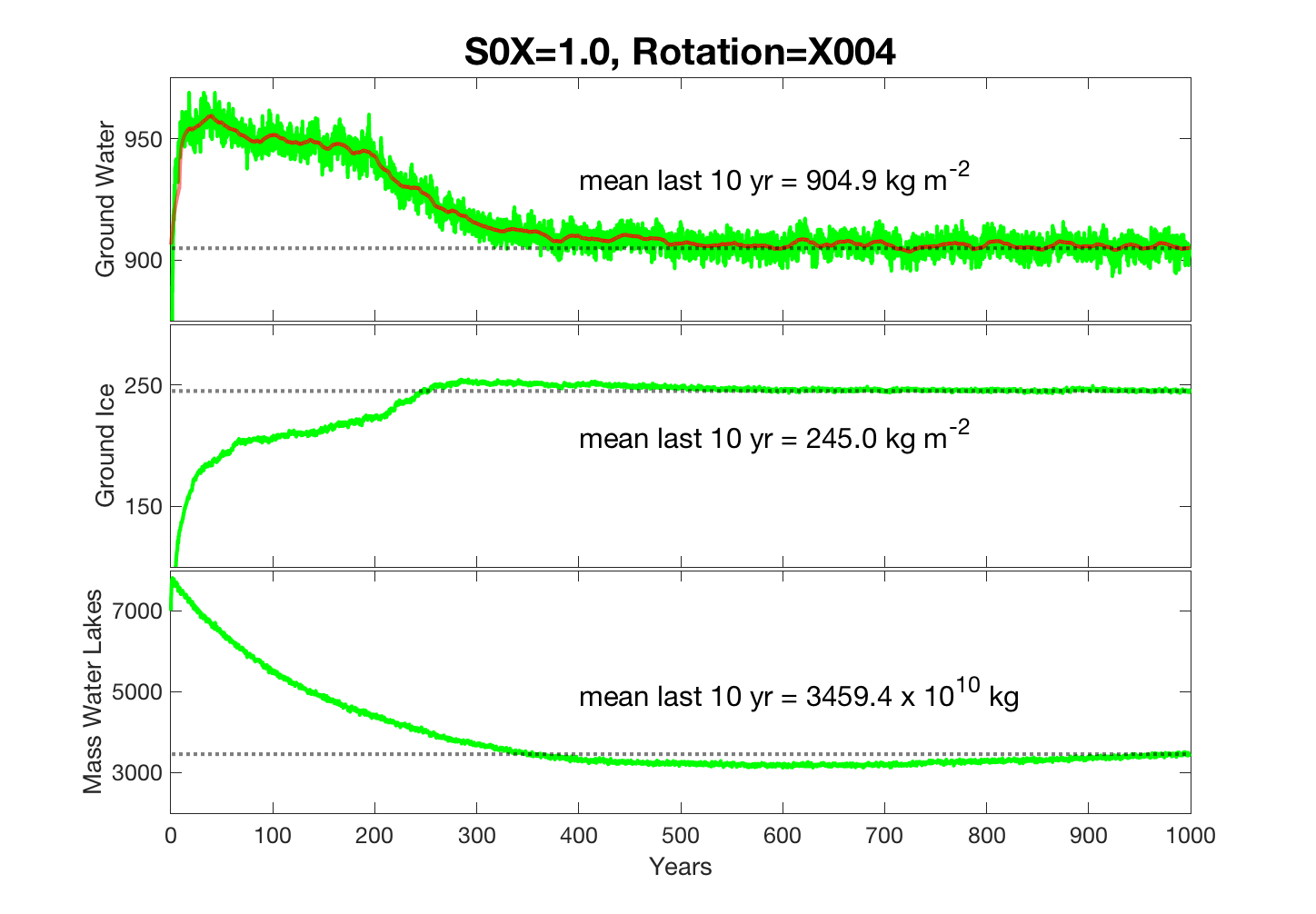}
\caption{\small S0X=1.0 with sidereal day length set to 4 days (X004). Left (a)
from top to bottom y-axes are net radiative balance in W m$^{-2}$, surface
temperature in Celsius, and ocean ice fraction in total percentage of ocean
area. Right (b) from top to bottom y-axes are total global ground water amounts
in kg m$^{-2}$, total ground ice in kg m$^{-2}$ and the total mass of water in
lakes and rivers in kg.} \label{fig:bal1}
\end{figure}

\begin{figure}[!htb]
\includegraphics[scale=0.35]{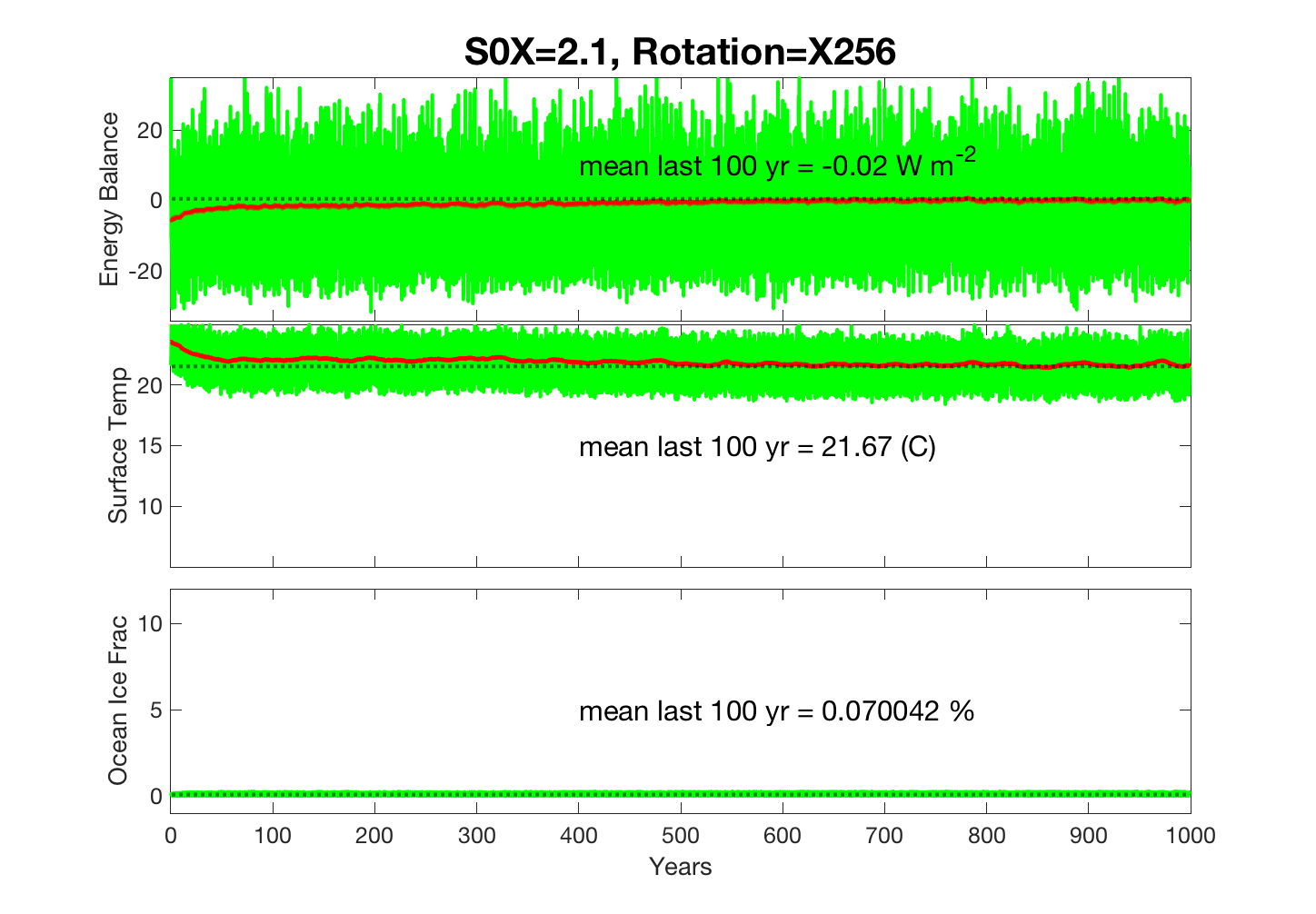}
\includegraphics[scale=0.35]{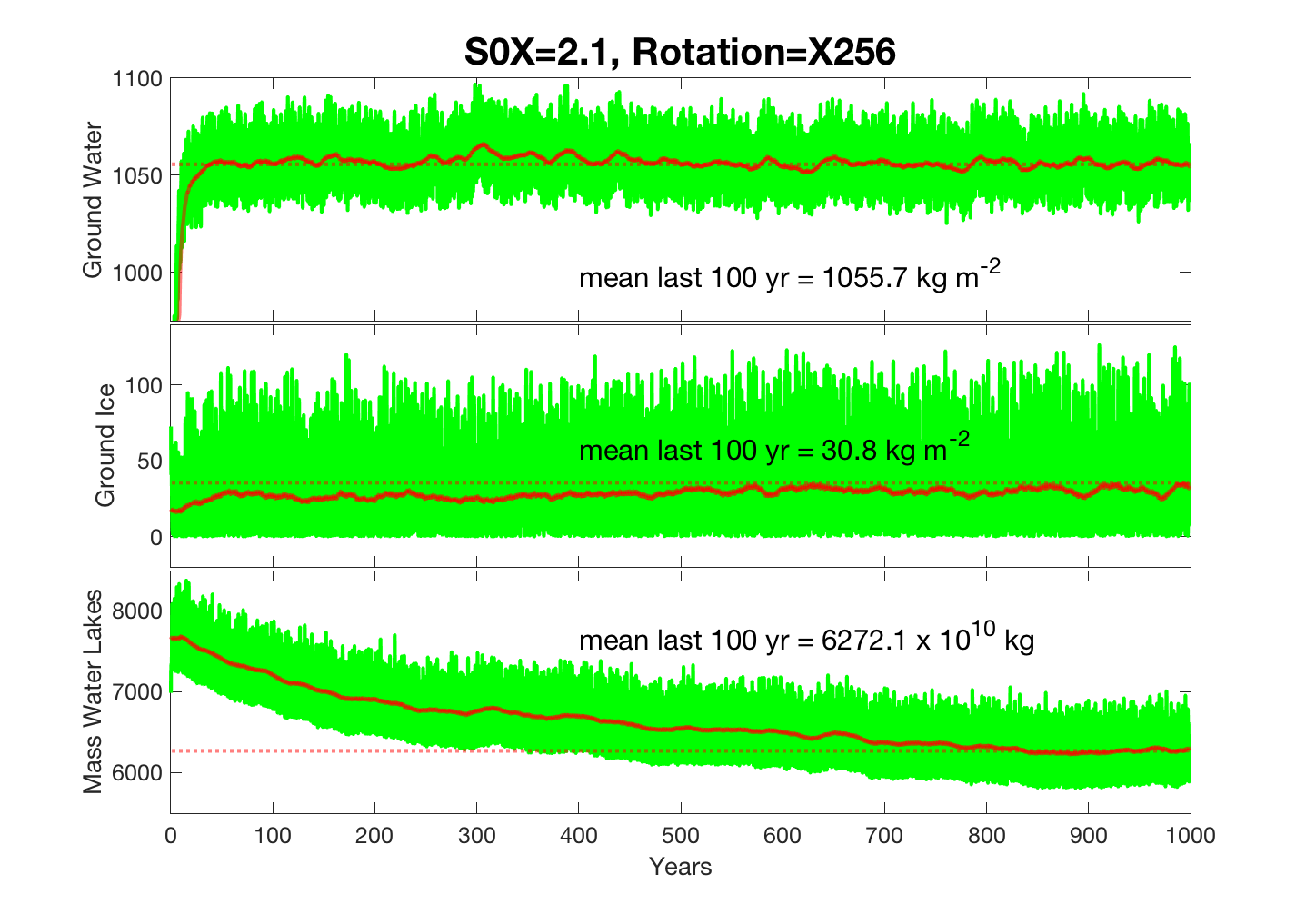}
\caption{\small S0X=2.1 with sidereal day length set to 256 days (X256). Axes are the same as in Figure \ref{fig:bal1}.}
\label{fig:bal2}
\end{figure}

To determine whether a GCM under a particular forcing (be that  via differences
in concentrations of radiatively active gases or via changes in insolation) has
reached equilibrium, one typically looks at the energy balance at the top of
the atmosphere. For \rocke{} we prefer the net radiative balance to be within
$\pm$ 0.2 W m$^{-2}$ of zero, which is as good as or better than most current
Earth climate models \citep{Forster2013}. For the simulations of water metrics
to be discussed in Paper II we need to also look at surface and sub-surface
water diagnostics over land to ensure that they are in steady state, which can
take longer than is required to achieve energy balance. Figures \ref{fig:bal1}
\& \ref{fig:bal2} show two examples from different insolation and day length
regimes.

% Figure 13

% Figure 14

In Figure \ref{fig:bal1}a it is apparent that     it takes the model
approximately 300 years to come into energy balance. As expected both the
global mean surface temperature and ocean ice fraction also reach a stable
state about the same time. The latter can have a large effect given the high
albedo of ocean ice versus ocean liquid water. In Figure \ref{fig:bal1}b one
can see that the ground water (liquid and ice) takes a bit longer to stabilize
at around 500 years. The ground ice also stabilizes around 500 years. The last
diagnostic in Figure \ref{fig:bal1}b is the mass of water in lakes and rivers.
One can see that at model start it is far from stability, and it takes nearly
500 model years for it to begin to stabilize. This is partly because of the
different non-modern-Earth topography used given that we start with modern
Earth topographic ground initial conditions. There is also the fact that the
obliquity and eccentricity is zero. This means that water sources and sinks
need to adjust to the new boundary conditions in this run.

Figure \ref{fig:bal2}a is similar to Figure \ref{fig:bal1}a except that it is a
world with much higher insolation (2.1 versus 1.0) and a much longer sidereal
day length (X256 versus X001). The energy balance varies over a much broader
range than in Figure \ref{fig:bal1}a, likely due to the slower rotation rate.
The surface temperature and ocean ice fraction stabilize much more quickly with
the higher insolation. In Figure \ref{fig:bal2}b we see a similar trend, fast
stabilization for ground water, but less so for ground ice hydrologies.
However, the mass of water in lakes and rivers (MWL) take nearly all 1000 years
before it begins to stablize. We find that runs with  longer day lengths at
lower insolation take longer for MWL to come into equilibrium. 

\section{Simulations}\label{appendix:simulations}

Tables \ref{tab:dyn:simulations} and \ref{tab:qflux:simulations} contain the
list of simulations and rundeck model configuration files available on-line. A
number of boundary condition files are  included in Table
\ref{tab:boundarycondfiles} and are also available on-line.

\startlongtable
\begin{deluxetable}{|l|l|l|c|c|l|l|}
\tabletypesize{\scriptsize}
\tablecaption{Dynamic Ocean Simulations \label{tab:dyn:simulations}}
\tablehead{
\multicolumn{1}{|l|}{ID} & \multicolumn{1}{l|}{S0X\tablenotemark{1}} & \multicolumn{1}{l|}{SDL\tablenotemark{2}} &
\multicolumn{1}{c|}{Mean\tablenotemark{3}} & \multicolumn{1}{c|}{Years\tablenotemark{4}} & \multicolumn{1}{l|}{Mean model output name\tablenotemark{5}} & \multicolumn{1}{l|}{Rundeck Name\tablenotemark{6}}
}
\startdata
01D & 1.0 & X001 & 10 & 1000 & ANN0990-0999.accP211eoDOFP3Od\_X001\_O30.nc & P211eoDOFP3Od\_X001\_O30.R\\
02D & 1.0 & X001 & 10 & 500  & ANN0490-0499.accP211eoDOFP3OdCOND\_X001\_O30.nc\tablenotemark{7} & P211eoDOFP3OdCOND\_X001\_O30.R\\
03D & 1.1 & X001 & 10 & 1000 & ANN0990-0999.accP212eoDOFP3Od\_X001\_O30.nc & P212eoDOFP3Od\_X001\_O30.R\\
04D & 1.2 & X001 & 10 & 1300 & ANN1290-1299.accP213eoDOFP3Od\_X001\_O30.nc & P213eoDOFP3Od\_X001\_O30.R\\
\hline
05D & 1.0 & X002 & 10 & 1000 & ANN0990-0999.accP211eoDOFP3Od\_X002\_O30.nc & P211eoDOFP3Od\_X002\_O30.R\\
06D & 1.1 & X002 & 10 & 1000 & ANN0990-0999.accP212eoDOFP3Od\_X002\_O30.nc & P212eoDOFP3Od\_X002\_O30.R\\
07D & 1.2 & X002 & 10 & 1000 & ANN0990-0999.accP213eoDOFP3Od\_X002\_O30.nc & P213eoDOFP3Od\_X002\_O30.R\\
\hline
08D & 1.0 & X004 & 10 & 1000 & ANN0990-0999.accP211eoDOFP3Od\_X004\_O30.nc & P211eoDOFP3Od\_X004\_O30.R\\
09D & 1.1 & X004 & 10 & 1000 & ANN0990-0999.accP212eoDOFP3Od\_X004\_O30.nc & P212eoDOFP3Od\_X004\_O30.R\\
10D & 1.2 & X004 & 10 & 1000 & ANN0990-0999.accP213eoDOFP3Od\_X004\_O30.nc & P213eoDOFP3Od\_X004\_O30.R\\
11D & 1.3 & X004 & 10 & 1000 & ANN0990-0999.accP214eoDOFP3Od\_X004\_O30.nc & P214eoDOFP3Od\_X004\_O30.R\\
\hline
12D & 1.0 & X008 & 10 & 1000 & ANN0990-0999.accP211eoDOFP3Od\_X008\_O30.nc & P211eoDOFP3Od\_X008\_O30.R\\
13D & 1.1 & X008 & 10 & 1000 & ANN0990-0999.accP212eoDOFP3Od\_X008\_O30.nc & P212eoDOFP3Od\_X008\_O30.R\\
14D & 1.2 & X008 & 10 & 1000 & ANN0990-0999.accP213eoDOFP3Od\_X008\_O30.nc & P213eoDOFP3Od\_X008\_O30.R\\
15D & 1.3 & X008 & 10 & 1300 & ANN1290-1299.accP214eoDOFP3Od\_X008\_O30.nc & P214eoDOFP3Od\_X008\_O30.R\\
16D & 1.4 & X008 & 10 & 1500 & ANN1490-1499.accP215eoDOFP3Od\_X008\_O30.nc & P215eoDOFP3Od\_X008\_O30.R\\
\hline
17D & 1.0 & X016 & 10 & 1000 & ANN0990-0999.accP211eoDOFP3Od\_X016\_O30.nc & P211eoDOFP3Od\_X016\_O30.R\\
18D & 1.1 & X016 & 10 & 1000 & ANN0990-0999.accP212eoDOFP3Od\_X016\_O30.nc & P212eoDOFP3Od\_X016\_O30.R\\
19D & 1.1 & X016 & 10 & 1000 & ANN0990-0999.accP212eoDOFP3OdCOND\_X016\_O30.nc & P212eoDOFP3OdCOND\_X016\_O30.R\\
20D & 1.2 & X016 & 10 & 1000 & ANN0990-0999.accP213eoDOFP3Od\_X016\_O30.nc & P213eoDOFP3Od\_X016\_O30.R\\
21D & 1.3 & X016 & 10 & 1000 & ANN0990-0999.accP214eoDOFP3Od\_X016\_O30.nc & P214eoDOFP3Od\_X016\_O30.R\\
22D & 1.3 & X016 & 10 & 1000 & ANN0990-0999.accP214eoDOFP3Od\_X016\_O30S.nc\tablenotemark{8} & P214eoDOFP3Od\_X016\_O30S.R\\
23D & 1.4 & X016 & 10 & 2200 & ANN2190-2199.accP215eoDOFP3Od\_X016\_O30.nc & P215eoDOFP3Od\_X016\_O30.R\\
\hline
24D & 1.0 & X032 & 10 & 1000 & ANN0990-0999.accP211eoDOFP3Od\_X032\_O30.nc & P211eoDOFP3Od\_X032\_O30.R\\
25D & 1.1 & X032 & 10 & 1000 & ANN0990-0999.accP212eoDOFP3Od\_X032\_O30.nc & P212eoDOFP3Od\_X032\_O30.R\\
26D & 1.2 & X032 & 10 & 1000 & ANN0990-0999.accP213eoDOFP3Od\_X032\_O30.nc & P213eoDOFP3Od\_X032\_O30.R\\
27D & 1.3 & X032 & 10 & 1300 & ANN1290-1299.accP214eoDOFP3Od\_X032\_O30.nc & P214eoDOFP3Od\_X032\_O30.R\\
28D & 1.4 & X032 & 10 & 1500 & ANN1490-1499.accP215eoDOFP3Od\_X032\_O30.nc & P215eoDOFP3Od\_X032\_O30.R\\
29D & 1.5 & X032 & 10 & 1500 & ANN1290-1299.accP216eoDOFP3Od\_X032\_O30.nc & P216eoDOFP3Od\_X032\_O30.R\\
30D & 1.7 & X032 & 10 & 2500 & ANN2490-2499.accP217eoDOFP3Od\_X032\_O30.nc & P217eoDOFP3Od\_X032\_O30.R\\
31D & 1.9 & X032 & 10 & 2500 & ANN2490-2499.accP219eoDOFP3Od\_X032\_O30.nc & P219eoDOFP3Od\_X032\_O30.R\\
32D & 2.1 & X032 & 50 & 3000 & ANN2950-2999.accP221eoDOFP3Od\_X032\_O30.nc & P221eoDOFP3Od\_X032\_O30.R\\
\hline
33D & 1.0 & X064 & 10 & 1000 & ANN0990-0999.accP211eoDOFP3Od\_X064\_O30.nc & P211eoDOFP3Od\_X064\_O30.R\\
34D & 1.1 & X064 & 10 & 1000 & ANN0990-0999.accP212eoDOFP3Od\_X064\_O30.nc & P212eoDOFP3Od\_X064\_O30.R\\
35D & 1.2 & X064 & 10 & 1000 & ANN0990-0999.accP213eoDOFP3Od\_X064\_O30.nc & P213eoDOFP3Od\_X064\_O30.R\\
36D & 1.2 & X064 & 10 & 1000 & ANN0990-0999.accP213eoDOFP3OdCOND\_X064\_O30.nc & P213eoDOFP3OdCOND\_X064\_O30.R\\
37D & 1.3 & X064 & 10 & 1000 & ANN0990-0999.accP214eoDOFP3Od\_X064\_O30.nc & P214eoDOFP3Od\_X064\_O30.R\\
38D & 1.4 & X064 & 10 & 1000 & ANN0990-0999.accP215eoDOFP3Od\_X064\_O30.nc & P215eoDOFP3Od\_X064\_O30.R\\
39D & 1.5 & X064 & 10 & 1000 & ANN0990-0999.accP216eoDOFP3Od\_X064\_O30.nc & P216eoDOFP3Od\_X064\_O30.R\\
40D & 1.7 & X064 & 10 & 1000 & ANN0990-0999.accP217eoDOFP3Od\_X064\_O30.nc & P217eoDOFP3Od\_X064\_O30.R\\
41D & 1.9 & X064 & 10 & 1000 & ANN0990-0999.accP219eoDOFP3Od\_X064\_O30.nc & P219eoDOFP3Od\_X064\_O30.R\\
42D & 2.1 & X064 & 10 & 1000 & ANN0990-0999.accP221eoDOFP3Od\_X064\_O30.nc & P221eoDOFP3Od\_X064\_O30.R\\
43D & 2.3 & X064 & 10 & 1500 & ANN1490-1499.accP223eoDOFP3Od\_X064\_O30.nc & P223eoDOFP3Od\_X064\_O30.R\\
44D & 2.5 & X064 & 10 & 1700 & ANN1690-1699.accP225eoDOFP3Od\_X064\_O30.nc & P225eoDOFP3Od\_X064\_O30.R\\
45D & 2.7 & X064 & 10 & 1500 & ANN1490-1499.accP227eoDOFP3Od\_X064\_O30.nc & P227eoDOFP3Od\_X064\_O30.R\\
46D & 2.9 & X064 & 10 & 2000 & ANN1990-1999.accP229eoDOFP3Od\_X064\_O30.nc & P229eoDOFP3Od\_X064\_O30.R\\
\hline
47D & 1.0 & X128 & 50 & 1000 & ANN0950-0999.accP211eoDOFP3Od\_X128\_O30.nc & P211eoDOFP3Od\_X128\_O30.R\\
48D & 1.1 & X128 & 50 & 1000 & ANN0950-0999.accP212eoDOFP3Od\_X128\_O30.nc & P212eoDOFP3Od\_X128\_O30.R\\
49D & 1.2 & X128 & 50 & 1000 & ANN0950-0999.accP213eoDOFP3Od\_X128\_O30.nc & P213eoDOFP3Od\_X128\_O30.R\\
50D & 1.3 & X128 & 50 & 1000 & ANN0950-0999.accP214eoDOFP3Od\_X128\_O30.nc & P214eoDOFP3Od\_X128\_O30.R\\
51D & 1.3 & X128 & 50 & 1000 & ANN0950-0999.accP214eoDOFP3OdCOND\_X128\_O30.nc & P214eoDOFP3OdCOND\_X128\_O30.R\\
52D & 1.4 & X128 & 50 & 1000 & ANN0950-0999.accP216eoDOFP3Od\_X128\_O30.nc & P215eoDOFP3Od\_X128\_O30.R\\
53D & 1.5 & X128 & 50 & 1000 & ANN0950-0999.accP216eoDOFP3Od\_X128\_O30.nc & P216eoDOFP3Od\_X128\_O30.R\\
54D & 1.7 & X128 & 50 & 1000 & ANN0950-0999.accP217eoDOFP3Od\_X128\_O30.nc & P217eoDOFP3Od\_X128\_O30.R\\
55D & 1.9 & X128 & 50 & 1000 & ANN0950-0999.accP219eoDOFP3Od\_X128\_O30.nc & P219eoDOFP3Od\_X128\_O30.R\\
56D & 2.1 & X128 & 50 & 1000 & ANN0950-0999.accP221eoDOFP3Od\_X128\_O30.nc & P221eoDOFP3Od\_X128\_O30.R\\
57D & 2.3 & X128 & 50 & 1000 & ANN0950-0999.accP223eoDOFP3Od\_X128\_O30.nc & P223eoDOFP3Od\_X128\_O30.R\\
58D & 2.5 & X128 & 50 & 1000 & ANN1250-1299.accP225eoDOFP3Od\_X128\_O30.nc & P225eoDOFP3Od\_X128\_O30.R\\
59D & 2.7 & X128 & 50 & 1000 & ANN1250-1299.accP227eoDOFP3Od\_X128\_O30.nc & P227eoDOFP3Od\_X128\_O30.R\\
\hline
60D & 1.0 & X256 & 100 & 1000 & ANN0900-0999.accP211eoDOFP3Od\_X256\_O30.nc & P211eoDOFP3Od\_X256\_O30.R\\
61D & 1.0 & X256 & 100 & 1000 & ANN0900-0999.accP211eoDOFP3Od\_X256\_O30S.nc\tablenotemark{9} & P211eoDOFP3Od\_X256\_O30S.R\\
62D & 1.1 & X256 & 100 & 1000 & ANN0900-0999.accP212eoDOFP3Od\_X256\_O30.nc & P212eoDOFP3Od\_X256\_O30.R\\
63D & 1.2 & X256 & 100 & 1000 & ANN0900-0999.accP213eoDOFP3Od\_X256\_O30.nc & P213eoDOFP3Od\_X256\_O30.R\\
64D & 1.3 & X256 & 100 & 1000 & ANN0900-0999.accP214eoDOFP3Od\_X256\_O30.nc & P214eoDOFP3Od\_X256\_O30.R\\
65D & 1.4 & X256 & 100 & 1000 & ANN1400-1499.accP215eoDOFP3Od\_X256\_O30.nc & P215eoDOFP3Od\_X256\_O30.R\\
66D & 1.5 & X256 & 100 & 1000 & ANN0900-0999.accP216eoDOFP3Od\_X256\_O30.nc & P216eoDOFP3Od\_X256\_O30.R\\
67D & 1.7 & X256 & 100 & 1000 & ANN0900-0999.accP217eoDOFP3Od\_X256\_O30.nc & P217eoDOFP3Od\_X256\_O30.R\\
68D & 1.9 & X256 & 100 & 1000 & ANN0900-0999.accP219eoDOFP3Od\_X256\_O30.nc & P219eoDOFP3Od\_X256\_O30.R\\
69D & 2.1 & X256 & 100 & 1000 & ANN0900-0999.accP221eoDOFP3Od\_X256\_O30.nc & P221eoDOFP3Od\_X256\_O30.R\\
70D & 2.3 & X256 & 100 & 1000 & ANN0900-0999.accP223eoDOFP3Od\_X256\_O30.nc & P223eoDOFP3Od\_X256\_O30.R\\
71D & 2.5 & X256 & 100 & 1000 & ANN0900-0999.accP225eoDOFP3Od\_X256\_O30.nc & P225eoDOFP3Od\_X256\_O30.R\\
72D & 1.7 & X256 & 100 & 1000 & ANN0900-0999.accP225eoDOFP3Od\_X256\_O30S.nc\tablenotemark{8} &P225eoDOFP3Od\_X256\_O30S.R\\
\enddata
\tablenotetext{1}{S0X: S0X1.0=1360.67, all other X's are multiples of 1365.3 W
m$^{-2}$, e.g. S0X=1.1=1365.3 $\times$ 1.1 = 1501.8 W m$^{-2}$.}
\tablenotetext{2}{SDL: Sidereal Day Length in multipules of modern Earth days
(24 hours). e.g. X004 = 4 x modern Earth's sidereal day length.}
\tablenotetext{3}{Mean: Number of years used for mean in column 6.}
\tablenotetext{4}{Years: Length in years/orbits of model run.}
\tablenotetext{5}{Filename of run on public archive. The files are averaged
over the number of orbits listed in the columnn labeled ``mean." To generate
viewable diagnostics these netCDF files must be converted to aij, aj etc. using
the ModelE2/ROCKE-3D scaleacc command. See
https://simplex.giss.nasa.gov/gcm/doc/HOWTO/newio.html}
\tablenotetext{6}{Name of the model configuration rundeck file.}
\tablenotetext{7}{COND in name indicates ``convective condensate precipitates"
in this particular  run. See Section \ref{cloudparam}.}
\tablenotetext{8}{S at end of this name indicates soil is 100\% loam/silt.}
\tablenotetext{9}{S at end of this name indicates soil is 100\% sand.}
\end{deluxetable}

\startlongtable
\begin{deluxetable}{|l|l|c|c|c|l|l|}
\tabletypesize{\scriptsize}
\tablecaption{Q-flux=0 Ocean Simulations \label{tab:qflux:simulations}}
\tablehead{
\multicolumn{1}{|l|}{ID} & \multicolumn{1}{l|}{S0X} & \multicolumn{1}{c|}{SDL} &
\multicolumn{1}{c|}{Mean} & \multicolumn{1}{c|}{Years} & \multicolumn{1}{l|}{Mean model output name} & \multicolumn{1}{l|}{Rundeck name}
}
\startdata
01Z & 1.0 & X001B\tablenotemark{1}& 10  & 200   & ANN0190-0199.accP211eoZoht100\_X001B\_O30.nc & P211eoZoht100\_X001B\_O30.R \\
02Z & 1.1 & X001B&  10 & 200   & ANN0190-0199.accP212eoZoht100\_X001B\_O30.nc & P212eoZoht100\_X001B\_O30.R \\
03Z & 1.2 & X001B&  10 & 200   & ANN0190-0199.accP213eoZoht100\_X001B\_O30.nc & P213eoZoht100\_X001B\_O30.R \\
04Z & 1.3 & X001B&  10 & 200   & ANN0190-0199.accP214eoZoht100\_X001B\_O30.nc & P214eoZoht100\_X001B\_O30.R \\
\hline
05Z & 1.0 & X001 &  10 & 200   & ANN0190-0199.accP211eoZoht100\_X001\_O30.nc & P211eoZoht100\_X001B\_O30.R \\
06Z & 1.0 & X001 &  10 & 200   & ANN0190-0199.accP211eoZoht100COND\_X001\_O30.nc & P211eoZoht100COND\_X001B\_O30.R \\
07Z & 1.1 & X001 &  10 & 200   & ANN0190-0199.accP212eoZoht100\_X001\_O30.nc & P212eoZoht100\_X001B\_O30.R \\
08Z & 1.2 & X001 &  10 & 200   & ANN0190-0199.accP213eoZoht100\_X001\_O30.nc & P213eoZoht100\_X001B\_O30.R \\
\hline
109Z & 1.0 & X002 &  10 & 200   & ANN0190-0199.accP211eoZoht100\_X002\_O30.nc & P211eoZoht100\_X002\_O30.R \\
10Z & 1.1 & X002 &  10 & 200   & ANN0190-0199.accP212eoZoht100\_X002\_O30.nc & P212eoZoht100\_X002\_O30.R \\
11Z & 1.2 & X002 &  10 & 200   & ANN0190-0199.accP213eoZoht100\_X002\_O30.nc & P213eoZoht100\_X002\_O30.R \\
12Z & 1.3 & X002 &  10 & 200   & ANN0190-0199.accP214eoZoht100\_X002\_O30.nc & P214eoZoht100\_X002\_O30.R \\
13Z & 1.4 & X002 &  10 & 200   & ANN0190-0199.accP215eoZoht100\_X002\_O30.nc & P215eoZoht100\_X002\_O30.R \\
\hline
14Z & 1.0 & X004 &  10 & 200   & ANN0190-0199.accP211eoZoht100\_X004\_O30.nc & P211eoZoht100\_X004\_O30.R \\
15Z & 1.1 & X004 &  10 & 200   & ANN0190-0199.accP212eoZoht100\_X004\_O30.nc & P212eoZoht100\_X004\_O30.R \\
16Z & 1.2 & X004 &  10 & 200   & ANN0190-0199.accP213eoZoht100\_X004\_O30.nc & P213eoZoht100\_X004\_O30.R \\
17Z & 1.3 & X004 &  10 & 200   & ANN0190-0199.accP214eoZoht100\_X004\_O30.nc & P214eoZoht100\_X004\_O30.R \\
\hline
18Z & 1.0 & X008 &  10 & 200   & ANN0190-0199.accP211eoZoht100\_X008\_O30.nc & P211eoZoht100\_X008\_O30.R \\
19Z & 1.1 & X008 &  10 & 200   & ANN0190-0199.accP212eoZoht100\_X008\_O30.nc & P212eoZoht100\_X008\_O30.R \\
20Z & 1.2 & X008 &  10 & 200   & ANN0190-0199.accP213eoZoht100\_X008\_O30.nc & P213eoZoht100\_X008\_O30.R \\
21Z & 1.3 & X008 &  10 & 200   & ANN0190-0199.accP214eoZoht100\_X008\_O30.nc & P214eoZoht100\_X008\_O30.R \\
22Z & 1.4 & X008 &  10 & 200   & ANN0190-0199.accP215eoZoht100\_X008\_O30.nc & P215eoZoht100\_X008\_O30.R \\
\hline
23Z & 1.0 & X016 &  10 & 200   & ANN0190-0199.accP211eoZoht100\_X016\_O30.nc & P211eoZoht100\_X016\_O30.R \\
24Z & 1.1 & X016 &  10 & 200   & ANN0190-0199.accP212eoZoht100\_X016\_O30.nc & P212eoZoht100\_X016\_O30.R \\
25Z & 1.1 & X016 &  10 & 200   & ANN0190-0199.accP212eoZoht100COND\_X016\_O30.nc & P212eoZoht100COND\_X016\_O30.R \\
26Z & 1.2 & X016 &  10 & 200   & ANN0190-0199.accP213eoZoht100\_X016\_O30.nc & P213eoZoht100\_X016\_O30.R \\
27Z & 1.3 & X016 &  10 & 200   & ANN0190-0199.accP214eoZoht100\_X016\_O30.nc & P214eoZoht100\_X016\_O30.R \\
28Z & 1.4 & X016 &  10 & 200   & ANN0190-0199.accP215eoZoht100\_X016\_O30.nc & P215eoZoht100\_X016\_O30.R \\
29Z & 1.5 & X016 &  10 & 200   & ANN0190-0199.accP216eoZoht100\_X016\_O30.nc & P216eoZoht100\_X016\_O30.R \\
\hline
30Z & 1.0 & X032 &  10 & 200   & ANN0190-0190.accP211eoZoht100\_X032\_O30.nc & P211eoZoht100\_X032\_O30.R \\
31Z & 1.1 & X032 &  10 & 200   & ANN0190-0190.accP212eoZoht100\_X032\_O30.nc & P212eoZoht100\_X032\_O30.R \\
32Z & 1.2 & X032 &  10 & 200   & ANN0190-0190.accP213eoZoht100\_X032\_O30.nc & P213eoZoht100\_X032\_O30.R \\
33Z & 1.3 & X032 &  10 & 200   & ANN0190-0199.accP214eoZoht100\_X032\_O30.nc & P214eoZoht100\_X032\_O30.R \\
34Z & 1.4 & X032 &  10 & 200   & ANN0190-0199.accP215eoZoht100\_X032\_O30.nc & P215eoZoht100\_X032\_O30.R \\
35Z & 1.5 & X032 &  10 & 200   & ANN0190-0199.accP216eoZoht100\_X032\_O30.nc & P216eoZoht100\_X032\_O30.R \\
36Z & 1.7 & X032 &  10 & 200   & ANN0190-0199.accP217eoZoht100\_X032\_O30.nc & P217eoZoht100\_X032\_O30.R \\
37Z & 1.9 & X032 &  10 & 400   & ANN0390-0399.accP219eoZoht100\_X032\_O30.nc & P219eoZoht100\_X032\_O30.R \\
\hline
38Z & 1.0 & X064 &  10 & 200   & ANN0190-0190.accP211eoZoht100\_X064\_O30.nc & P211eoZoht100\_X064\_O30.R \\
39Z & 1.1 & X064 &  10 & 200   & ANN0190-0190.accP212eoZoht100\_X064\_O30.nc & P212eoZoht100\_X064\_O30.R \\
40Z & 1.2 & X064 &  10 & 200   & ANN0190-0190.accP213eoZoht100\_X064\_O30.nc & P213eoZoht100\_X064\_O30.R \\
41Z & 1.2 & X064 &  10 & 200   & ANN0190-0190.accP213eoZoht100COND\_X064\_O30.nc & P213eoZoht100COND\_X064\_O30.R \\
42Z & 1.3 & X064 &  10 & 200   & ANN0190-0190.accP214eoZoht100\_X064\_O30.nc & P214eoZoht100\_X064\_O30.R \\
43Z & 1.4 & X064 &  10 & 200   & ANN0190-0190.accP215eoZoht100\_X064\_O30.nc & P215eoZoht100\_X064\_O30.R \\
44Z & 1.5 & X064 &  10 & 200   & ANN0190-0190.accP216eoZoht100\_X064\_O30.nc & P216eoZoht100\_X064\_O30.R \\
45Z & 1.7 & X064 &  10 & 200   & ANN0190-0190.accP217eoZoht100\_X064\_O30.nc & P217eoZoht100\_X064\_O30.R \\
46Z & 1.9 & X064 &  10 & 200   & ANN0190-0190.accP219eoZoht100\_X064\_O30.nc & P219eoZoht100\_X064\_O30.R \\
47Z & 2.1 & X064 &  10 & 200   & ANN0190-0190.accP221eoZoht100\_X064\_O30.nc & P221eoZoht100\_X064\_O30.R \\
48Z & 2.3 & X064 &  10 & 200   & ANN0190-0199.accP223eoZoht100\_X064\_O30.nc & P223eoZoht100\_X064\_O30.R \\
49Z & 2.5 & X064 &  10 & 200   & ANN0190-0199.accP225eoZoht100\_X064\_O30.nc & P225eoZoht100\_X064\_O30.R \\
50Z & 2.7 & X064 &  10 & 200   & ANN0190-0199.accP227eoZoht100\_X064\_O30.nc & P227eoZoht100\_X064\_O30.R \\
51Z & 2.9 & X064 &  10 & 200   & ANN0190-0199.accP229eoZoht100\_X064\_O30.nc & P229eoZoht100\_X064\_O30.R \\
\hline
52Z & 1.0 & X128 &  50 & 200   & ANN0150-0199.accP211eoZoht100\_X128\_O30.nc & P211eoZoht100\_X128\_O30.R \\
53Z & 1.1 & X128 &  50 & 200   & ANN0150-0199.accP212eoZoht100\_X128\_O30.nc & P212eoZoht100\_X128\_O30.R \\
54Z & 1.2 & X128 &  50 & 200   & ANN0150-0199.accP213eoZoht100\_X128\_O30.nc & P213eoZoht100\_X128\_O30.R \\
55Z & 1.3 & X128 &  50 & 200   & ANN0150-0199.accP214eoZoht100\_X128\_O30.nc & P214eoZoht100\_X128\_O30.R \\
56Z & 1.3 & X128 &  50 & 200   & ANN0150-0199.accP214eoZoht100COND\_X128\_O30.nc & P214eoZoht100COND\_X128\_O30.R \\
57Z & 1.4 & X128 &  50 & 200   & ANN0150-0199.accP215eoZoht100\_X128\_O30.nc & P215eoZoht100\_X128\_O30.R \\
58Z & 1.5 & X128 &  50 & 200   & ANN0150-0199.accP216eoZoht100\_X128\_O30.nc & P216eoZoht100\_X128\_O30.R \\
59Z & 1.7 & X128 &  50 & 200   & ANN0150-0199.accP217eoZoht100\_X128\_O30.nc & P217eoZoht100\_X128\_O30.R \\
60Z & 1.9 & X128 &  50 & 200   & ANN0150-0199.accP219eoZoht100\_X128\_O30.nc & P219eoZoht100\_X128\_O30.R \\
61Z & 2.1 & X128 &  50 & 200   & ANN0150-0199.accP221eoZoht100\_X128\_O30.nc & P221eoZoht100\_X128\_O30.R \\
62Z & 2.3 & X128 &  50 & 200   & ANN0150-0199.accP223eoZoht100\_X128\_O30.nc & P223eoZoht100\_X128\_O30.R \\
63Z & 2.5 & X128 &  50 & 200   & ANN0150-0199.accP225eoZoht100\_X128\_O30.nc & P225eoZoht100\_X128\_O30.R \\
64Z & 2.7 & X128 &  50 & 200   & ANN0150-0199.accP227eoZoht100\_X128\_O30.nc & P227eoZoht100\_X128\_O30.R \\
\hline
65Z & 1.0 & X256 & 100 & 200   & ANN0100-0199.accP211eoZoht100\_X256\_O30.nc & P211eoZoht100\_X256\_O30.R \\
66Z & 1.1 & X256 & 100 & 200   & ANN0100-0199.accP212eoZoht100\_X256\_O30.nc & P212eoZoht100\_X256\_O30.R \\
67Z & 1.2 & X256 & 100 & 200   & ANN0100-0199.accP213eoZoht100\_X256\_O30.nc & P213eoZoht100\_X256\_O30.R \\
68Z & 1.3 & X256 & 100 & 200   & ANN0100-0199.accP214eoZoht100\_X256\_O30.nc & P214eoZoht100\_X256\_O30.R \\
69Z & 1.4 & X256 & 100 & 200   & ANN0100-0199.accP215eoZoht100\_X256\_O30.nc & P215eoZoht100\_X256\_O30.R \\
70Z & 1.5 & X256 & 100 & 200   & ANN0100-0199.accP216eoZoht100\_X256\_O30.nc & P216eoZoht100\_X256\_O30.R \\
71Z & 1.7 & X256 & 100 & 200   & ANN0100-0199.accP217eoZoht100\_X256\_O30.nc & P217eoZoht100\_X256\_O30.R \\
72Z & 1.9 & X256 & 100 & 200   & ANN0100-0199.accP219eoZoht100\_X256\_O30.nc & P219eoZoht100\_X256\_O30.R \\
73Z & 2.1 & X256 & 100 & 200   & ANN0100-0199.accP221eoZoht100\_X256\_O30.nc & P221eoZoht100\_X256\_O30.R \\
74Z & 2.3 & X256 & 100 & 200   & ANN0100-0199.accP223eoZoht100\_X256\_O30.nc & P223eoZoht100\_X256\_O30.R \\
75Z & 2.5 & X256 & 100 & 200   & ANN0100-0199.accP225eoZoht100\_X256\_O30.nc & P225eoZoht100\_X256\_O30.R \\
\enddata
\tablenotetext{1}{B at end of name indicates radiation balance set to colder
temperatures via cloud tuning. See Section \ref{cloudparam} for details.}
\end{deluxetable}

\begin{deluxetable}{|l|l|l|}[!htb]
\tabletypesize{\scriptsize}
\tablecaption{Boundary Condition Files\tablenotemark{1}\label{tab:boundarycondfiles}}
\tablehead{
\multicolumn{1}{|l|}{Name\tablenotemark{2}} & \multicolumn{1}{l|}{Input filename\tablenotemark{3}} & \multicolumn{1}{l|}{Description}
}
\startdata
AIC.RES\_M20A.D771201\_40L.nc & AIC & Atmosphere Initial Conditions\\
GIC.E046D3M20A.1DEC1955.ext\_1.nc & GIC & Ground Initial Conditions\\
Z72X46N\_gas.1\_nocasp\_btub005.nc & TOPO & Land Topography \\
OZ72X46N\_gas.1\_nocasp\_btub005.nc & TOPO\_OC & Dynamic Ocean Topography\tablenotemark{4}\\
OIC4X5LD.Z12.gas1.CLEV94.DEC01\_btub00.nc & OIC & Ocean Initial Conditions\\
OSTRAITS\_72x46btub0.nml & OSTRAITS & Ocean Straits\tablenotemark{4}\\
RD\_modelE\_M\_btub004D.nc & RVR & River Directions\tablenotemark{4}\\
RD\_modelE\_M.names\_btub0.txt & NAMERVR & River Names\tablenotemark{4}\\
zero\_OHT\_4x5\_100m.nc & OHT & Q-flux=0 Ocean Heat Transport\tablenotemark{5}\\
zero\_OCNML\_4x5\_100m.nc & OCNML & Q-flux=0 initial ocean layer depth\tablenotemark{5}\\
V72X46.1.cor2\_no\_crops02.ext.nc& VEG & Uniform surface albedo=0.2 \\
S4X50093SANDCLAY.ext.nc & SOIL & Soil Type (50/50 sand/clay) \\
S4X50093LOAM.ext.nc & SOIL & Soil Type (100\% loam/silt) \\
S4X50093SAND.ext.nc & SOIL & Soil Type (100\% sand)\\
GLMELT\_4X5.OCN\_MWAY0.nc & GLMELT & Glacial Melt Areas \\
GHG.MWAY201412B.txt & GHG & Green House Gas concentrations\\
\hline
\enddata
\tablenotetext{1}{These files may be downloaded from
https://portal.nccs.nasa.gov/GISS\_modelE/ROCKE-3D and/or The Internet Archive
https://archive.org/details/Climates\_of\_Warm\_Earth\_like\_Planets}
\tablenotetext{2}{Files ending in nc are formatted in NetCDF
(https://www.unidata.ucar.edu/software/netcdf/), NetCDF is software developed
by UCAR/Unidata http://doi.org/10.5065/D6H70CW6. Non-nc files are simple flat
text files.}
\tablenotetext{3}{The name for the assignment variable in \rocke{}.}
\tablenotetext{4}{Fully Coupled Dynamic Ocean input files.}
\tablenotetext{5}{This file is used for Q-flux=0 oceans of 100m depth.}
\end{deluxetable}

\acknowledgements
This work was supported by the NASA Astrobiology Program through collaborations
arising from our participation in the Nexus for Exoplanet System Science, and
by the NASA Planetary Atmospheres Program. Computing resources for this work
were provided by the NASA High-End Computing (HEC) Program through the NASA
Center for Climate Simulation (NCCS) at Goddard Space Flight Center
(http://www.nccs.nasa.gov).  Thanks to Tiffany Jansen for suggesting additional
simulations at sidereal days from X002 to X008 needed for future papers in this
series. We would also like to thank the anonymous referee for their insightful
and useful comments that improved the quality of this work.

\bibliographystyle{apj}
\bibliography{bibliography}

\end{document}